\pdfoutput=1

\documentclass[a4paper,11pt]{article}
\usepackage{jheppub}

\usepackage{bm}

\usepackage{color}

\usepackage{amsmath}
\usepackage{amssymb}
\usepackage{graphicx}
\usepackage{slashed}
\usepackage{soul}
\usepackage{lscape}
\usepackage{xcolor}
\usepackage{multirow}
\usepackage{placeins,hyperref}
\usepackage[separate-uncertainty=true]{siunitx}
\usepackage{silence}
\usepackage{caption}
\usepackage{subcaption}
\usepackage{mathrsfs}
\usepackage[version=3]{mhchem}
\usepackage{amsmath, amssymb, amscd, amsthm, amsfonts}
\usepackage{wrapfig}

\usepackage[italic]{hepnames}
\usepackage{gensymb}
\usepackage{textcomp}
\usepackage{booktabs}
\usepackage{float}
\usepackage{adjustbox}

\counterwithin{figure}{section}

\usepackage{lineno}

\begin{document}

\title{\centering Mineral Detection of Neutrinos and Dark Matter 2025 \\ Proceedings}


\author[1]{Shigenobu~Hirose,}
\affiliation[1]{Center for Mathematical Science and Advanced Technology (MAT), Japan Agency for Marine-Earth Science and Technology (JAMSTEC), Yokohama, Kanagawa 236-0001, Japan}

\author[2]{Patrick~Stengel;}
\affiliation[2]{Jo\v{z}ef Stefan Institute, Jamova 39, 1000 Ljubljana, Slovenia}

\author[1]{Natsue~Abe,}

\author[3]{Daniel~Ang,}
\affiliation[3]{Quantum Technology Center, University of Maryland, College Park, MD 20742, USA}

\author[4]{Lorenzo~Apollonio,}
\affiliation[4]{INFN Milano, via Celoria 16 20133, Milano, Italy}

\author[5]{Gabriela~R.~Araujo,}
\affiliation[5]{Physik-Institut, University of Zurich, Switzerland}

\author[6]{Yoshihiro~Asahara,}
\affiliation[6]{Department of Earth and Planetary Sciences, Graduate School of Environmental Studies, Nagoya University, Furo-cho, Chikusaku, Nagoya, Aichi, 464-8601, Japan}

\author[5]{Laura~Baudis,}

\author[7]{Pranshu~Bhaumik,}
\affiliation[7]{College of William and Mary, Williamsburg, VA 23187, USA}

\author[8]{Nathaniel~Bowden,}
\affiliation[8]{Lawrence Livermore National Laboratory, 7000 East Ave, Livermore, CA 94536, USA}

\author[9]{Joseph~Bramante,}
\affiliation[9]{Department of Physics, Engineering Physics and Astronomy, Queen's University, 99 University Avenue, Kingston, Canada}

\author[4]{Lorenzo~Caccianiga,}

\author[3]{Mason~Camp,}

\author[10]{Qing~Chang,}
\affiliation[10]{Volcanoes and Earth's Interior Research Center, Japan Agency for Marine-Earth Science and Technology (JAMSTEC), Yokosuka, Kanagawa 237-0061, Japan}

\author[11]{Jordan~Chapman,}
\affiliation[11]{Virginia Tech National Security Institute, Blacksburg, Virginia 24060, USA}

\author[12,13]{Reza~Ebadi,}
\affiliation[12]{Department of Physics and Astronomy, The Johns Hopkins University, Baltimore, MD 21218, USA}
\affiliation[13]{Department of Physics and Astronomy, University of Delaware, Newark, DE 19716, USA}

\author[14]{Alexey~Elykov,}
\affiliation[14]{Institute for Astroparticle Physics, Karlsruhe Institute of Technology, 76021 Karlsruhe, Germany}

\author[15]{Anna~Erickson,}
\affiliation[15]{George W. Woodruff School of Mechanical Engineering, Georgia Institute of Technology, Atlanta, GA, USA}

\author[16]{Valentin~Fondement,}
\affiliation[16]{Department of Nuclear Engineering and Radiological Sciences, University of Michigan, Ann Arbor, MI 48103, USA}

\author[17,18,19]{Katherine~Freese,}
\affiliation[17]{Department of Physics, The University of Texas at Austin, Austin, TX 78712, USA}
\affiliation[18]{The Oskar Klein Centre, Department of Physics, Stockholm University, AlbaNova, SE-106 91 Stockholm, Sweden}
\affiliation[19]{Nordita, Stockholm University and KTH Royal Institute of Technology, Hannes Alfvéns väg 12, SE-106 91 Stockholm, Sweden}

\author[20]{Shota~Futamura,}
\affiliation[20]{Institute for Space-Earth Environmental Research, Nagoya University, Furo-cho, Chikusaku, Nagoya, Aichi, 464-8601, Japan}

\author[4]{Claudio~Galelli,}

\author[3]{Andrew~Gilpin,}

\author[10]{Takeshi~Hanyu,}

\author[21]{Noriko~Hasebe,}
\affiliation[21]{Kanazawa University, Kanazawa, Ishikawa 920-1192, Japan}

\author[22]{Adam~A.~Hecht,}
\affiliation[22]{Department of Nuclear Engineering, University of New Mexico, Albuquerque, NM 87131, USA}

\author[23]{Samuel~C.~Hedges,}
\affiliation[23]{Center for Neutrino Physics, Virginia Tech, Blacksburg, VA 24061, USA}


\author[23,24,25]{Shunsaku~Horiuchi,}
\affiliation[24]{Department of Physics, Institute of Science Tokyo, 2-12-1 Ookayama, Meguro-ku, Tokyo 152-8551, Japan}
\affiliation[25]{Kavli IPMU (WPI), UTIAS, The University of Tokyo, Kashiwa, Chiba 277-8583, Japan}

\author[26]{Yasushi~Hoshino,}
\affiliation[26]{Department of Physics, Kanagawa University, Yokohama, Kanagawa 221-8686, Japan}

\author[23]{Patrick~Huber,}

\author[20]{Yuki~Ido,}

\author[27]{Yohei~Igami,}
\affiliation[27]{Department of Geology and Mineralogy, Kyoto University, Kyoto, Japan}

\author[28]{Yuto~Iinuma,}
\affiliation[28]{Integrated Radiation and Nuclear Science, Kyoto University, Kumatori, Osaka 590-0494 Japan}

\author[11,29,30]{Vsevolod~Ivanov,}
\affiliation[29]{Department of Physics, Virginia Tech, Blacksburg, Virginia 24061, USA}
\affiliation[30]{Center for Quantum Information Science and Engineering, Virginia Tech, Blacksburg, VA 24061, USA}

\author[16]{Igor~Jovanovic,}

\author[31]{Ayuki~Kamada,}
\affiliation[31]{Institute of Theoretical Physics, Faculty of Physics, University of Warsaw, ul. Pasteura 5, PL-02-093 Warsaw, Poland}

\author[32]{Takashi~Kamiyama,}
\affiliation[32]{Faculty of Engineering, Hokkaido University, Sapporo, Hokkaido 060-0808, Japan}

\author[20]{Takenori~Kato,}

\author[1]{Yoji~Kawamura,}

\author[29]{Giti~A.~Khodaparast,}

\author[6]{Yui~Kouketsu,}

\author[20,33]{Yukiko~Kozaka,}
\affiliation[33]{Faculty of Geosciences and Civil Engineering, Institute of Science and Engineering, Kanazawa University, Kakuma-machi, Kanazawa, Ishikawa, 920-1192, Japan}

\author[34]{Emilie~M.~LaVoie-Ingram,}
\affiliation[34]{Department of Physics, University of Michigan, Ann Arbor, MI 48103, USA}

\author[35]{Matthew~Leybourne,}
\affiliation[35]{Department of Geological Sciences and Geological Engineering, Queen's University, Kingston, Canada}

\author[3]{Gavishta~Liyanage,}

\author[29]{Brenden~A.~Magill,}

\author[4]{Paolo~Magnani,}

\author[36,37]{William~F.~McDonough,}
\affiliation[36]{Advanced Institute for Marine Ecosystem Change, Department of Earth Sciences and Research Center for Neutrino Science, Tohoku University, Sendai, Miyagi 980-8578, Japan}
\affiliation[37]{Department of Geology, University of Maryland, College Park, MD 20742, USA}

\author[6]{Katsuyoshi~Michibayashi,}

\author[38]{Naoki~Mizutani,}
\affiliation[38]{Department of Physics, Toho University, 2-2-1 Miyama, Funabashi, Chiba, Japan}

\author[39,40]{Kohta~Murase,}
\affiliation[39]{Department of Physics, Department of Astronomy and Astrophysics, The Pennsylvania State University, 104 Davey Lab, University Park, PA 16802, USA}
\affiliation[40]{Yukawa Institute for Theoretical Physics, Kyoto University, Kyoto, Kyoto 606-8267, Japan}

\author[38]{Tatsuhiro~Naka,}

\author[21]{Taiki~Nakashima,}

\author[1]{Kenji~Oguni,}

\author[29]{Mariano~ Guerrero~Perez,}

\author[41]{Noriaki~Sakurai,}
\affiliation[41]{Institute for Marine-Earth Exploration and Engineering (MarE3), Japan Agency for Marine-Earth Science and Technology (JAMSTEC), Yokosuka, Kanagawa 237-0061, Japan}

\author[14]{Lukas~Scherne,}

\author[3]{Maximilian~Shen,}

\author[34]{Joshua~Spitz,}


\author[42]{Kai~Sun,}
\affiliation[42]{Department of Materials Science and Engineering, University of Michigan, Ann Arbor, MI 48103, USA}

\author[43]{Katsuhiko~Suzuki,}
\affiliation[43]{Submarine Resources Research Center, Japan Agency for Marine-Earth Science and Technology (JAMSTEC), Yokosuka, Kanagawa 237-0061, Japan}

\author[28]{Koichi~Takamiya,}

\author[3]{Jiashen~Tang,}

\author[44]{Erwin~H.~Tanin,}
\affiliation[44]{Leinweber Institute for Theoretical Physics at Stanford, 382 Via Pueblo, Stanford, CA 94305, USA}

\author[34]{Ethan~Todd,}

\author[45]{Atsuhiro~Umemoto,}
\affiliation[45]{International Center for Quantum-field Measurement Systems for Studies of the Universe and Particles (QUP), High Energy Accelerator Research Organization (KEK), 1-1 Oho, Tsukuba, Ibaraki 305-0801, Japan}

\author[29]{Keegan~Walkup,}

\author[3]{Ronald~Walsworth,}

\author[9]{Alexis~M.~Willson,}

\author[21]{Norihiro~Yamada,}

\author[46]{Seiko~Yamasaki,}
\affiliation[46]{Geological Survey of Japan, AIST, 1-1-1 Higashi Tsukuba Ibaraki, Japan}

\author[47]{Wen~Yin,}
\affiliation[47]{Department of Physics, Tokyo Metropolitan University, Minami-Osawa, Hachioji-shi, Tokyo 192-0397, Japan}

\author[21]{Akihiko~Yokoyama}

\emailAdd{shirose@jamstec.go.jp}
\emailAdd{patrick.stengel@ijs.si}


\abstract{The third ``Mineral Detection of Neutrinos and Dark Matter'' (MD$\nu$DM'25) meeting was held May 20-23, 2025 in Yokohama, Japan, hosted by the Yokohama Institute for Earth Sciences, Japan Agency for Marine-Earth Science and Technology (JAMSTEC). These proceedings compile contributions from the workshop and update the progress of mineral detector research. MD$\nu$DM'25 was the third such meeting, following the first in October of 2022 held at the IFPU in Trieste, Italy and the second in January of 2024 hosted by the Center for Neutrino Physics at Virginia Tech in Arlington, USA. Mineral detectors record and retain damage induced by nuclear recoils in synthetic or natural mineral samples. The damage features can then be read out by a variety of nano- and micro-scale imaging techniques. Applications of mineral detectors on timescales relevant for laboratory experiments include reactor neutrino monitoring and dark matter detection, with the potential to measure the directions as well as the energies of the induced nuclear recoils. For natural mineral detectors which record nuclear recoils over geological timescales, reading out even small mineral samples could be sensitive to rare interactions induced by astrophysical neutrinos, cosmic rays, dark matter and heavy exotic particles. A series of mineral detectors of different ages could measure the time evolution of these fluxes, offering a unique window into the history of our solar system and the Milky Way. Mineral detector research is highly multidisciplinary, incorporating aspects of high energy physics, condensed matter physics, materials science, geoscience, and AI/ML for data analysis. Although realizing the scientific potential of mineral detectors poses many challenges, the MD$\nu$DM community looks forward to the continued development of mineral detector experiments and the possible discoveries that mineral detectors could reveal.}

\maketitle
\flushbottom

\section*{Preface}
\addcontentsline{toc}{section}{Preface}

Particle interactions can induce a variety of damage features in synthetic and natural minerals. Mineral detector (MD) experiments implement one or a combination of imaging techniques capable of reading out damage features with characteristic sizes ranging from the nano- to micro-scale~\cite{Fleischer:1964,Fleischer383,Fleischer:1965yv,GUO2012233}. As passive nuclear recoil detectors which can deployed at room temperature in almost any environment, laboratory-prepared MDs could be sensitive to the damage induced by neutrinos produced in nuclear reactors~\cite{Cogswell:2021qlq,Alfonso:2022meh}. MDs could also be sensitive to nuclear recoils induced by dark matter particles in a controlled laboratory experiment, with the potential to measure the direction as well as energy of the nuclear recoil~\cite{Rajendran:2017ynw,Marshall:2020azl,ebadi_directional_2022}. These applications demonstrate the potential for Mineral Detection of Neutrinos and Dark Matter (MD$\nu$DM) on timescales up to several years. The design and execution of MD experiments in laboratory environments also provides a potential foundation for the measurement of particle interactions in MDs over geological timescales.   

While significantly more complicated to realize than a laboratory experiment, there are significant potential applications for MDs using geological samples. MDs have long been used as detectors of fission~\cite{Wagner:1992,Malusa:2018} and alpha-recoil~\cite{Goegen:2000,Glasmacher:2003} tracks induced by the heavy nuclear remnants produced in the decays of heavy radioisotopes such as $^{238}$U. For certain minerals in which such tracks can be retained over geological timescales and be read-out by optical microscopes after the preparation of the mineral samples using chemical etchants, MDs can reveal geochronological information concerning the age, ambient temperature and depth of the samples. Subsequent to these geological applications of MDs, similar techniques have also been used to search for damage features induced by exotic particles hypothesized to exist beyond the Standard Model of particle physics. With exposure times spanning over geological timescales, MDs were used to probe the rare interactions of dark matter~\cite{Snowden-Ifft:1995zgn} and magnetic monopoles~\cite{physrevlett56} with ordinary matter. 

More recently there has been renewed interest in applications of MDs for both synthetic and natural samples. In addition to facilitating MD$\nu$DM applications with laboratory-prepared samples above, advances in the imaging and data analysis techniques necessary to search for rare events in geological samples have sparked theoretical and experimental efforts to realize the scientific potential of MDs~\cite{Baum:2023cct,baum:2024eyr}. Phenomenological studies suggest that the sensitivity of MD searches for weakly interacting massive particle (WIMP) dark matter interactions with nuclei could be competitive with current and next generation conventional direct detection experiments~\cite{Baum:2018tfw,Drukier:2018pdy,Edwards:2018hcf,Baum:2021jak,Fung:2025cub}. As our solar system has rotated around the Milky Way every $\sim 250 \,$Myr, a series of MD samples of different ages could be sensitive to substructure in the dark matter halo of the galaxy~\cite{baum:2021chx,Bramante:2021dyx,Zhang:2025xzi}. Following earlier searches for magnetic monopoles, recent studies have proposed MDs as probes of additional exotic phenomena, such as heavy composite dark matter~\cite{Sidhu:2019qoa,Ebadi:2021cte,Acevedo:2021tbl}, charged black hole remnants~\cite{Lehmann:2019zgt} and proton decay~\cite{Baum:2024sst}.   

Although a background for WIMP searches, MDs can also probe the fluxes of astrophysical neutrinos and how those fluxes evolve over time. A series of MDs of different ages could measure the evolution of solar neutrino fluxes, potentially differentiating between solar metallically models~\cite{Tapia-Arellano:2021cml}. MDs could also probe galactic core collapse supernova neutrinos, potentially tracing the star formation history of the Milky Way~\cite{Baum:2019fqm} and determining the heavy flavor component of the supernova neutrino flux~\cite{Baum:2022wfc}. Atmospheric neutrinos and muons are produced in the interactions of cosmic rays with the Earth's atmosphere. A series of MDs could measure how the atmospheric neutrino and muon fluxes have changed over geological timescales, which is sensitive to the evolution of the galactic cosmic ray flux, as well as the Earth's atmosphere and magnetic field~\cite{Jordan:2020gxx}. Atmospheric muons are shielded by the overburden of the Earth and, thus, MD measurements of the atmospheric muon flux can also be sensitive to changes in Earth's crust and surface~\cite{caccianiga_2024}. 

Realizing this broad scientific potential, however, hinges on overcoming the longstanding central obstacle: the efficient readout of nano-scale damage tracks from large volumes of mineral. Specifically, the etching technique used in the initial MD works could only reveal portions of the tracks near the etched surface, which severely limited the analyzable mineral volume~\cite{physrevlett56,Snowden-Ifft:1995zgn}. Therefore, the development of high-throughput readout techniques has become the primary focus of current experimental efforts. Table~\ref{tab:experiments} summarizes the experimental studies addressing this challenge presented at this workshop, including ongoing projects and those in developmental or conceptual stages.

The experimental efforts listed in Table~\ref{tab:experiments} fall into two main categories: those that use etching and those that do not. The former group now employs innovative techniques to overcome the inherent throughput limitations of etching. For instance, DMICA (Sec.~\ref{sec:DMICA}) replaces slow AFM scanning with white-light interferometry to achieve the high scan rates needed for ton-year exposures. Similarly, Toho University (Sec.~\ref{sec:Tatsuhiro_Naka}) is developing a high-speed scanning technology, QTS, designed for speeds of approximately 50 cm$^2$/h at 20× magnification. Meanwhile, Queen's University (Sec.~\ref{sec:Queens}) is exploring Scanning Electron Microscopy (SEM) and Polarized Light Microscopy (PLM) to enhance readout sensitivity.

For direct, etching-free measurements, two primary approaches are being pursued. The first involves bulk scanning of synthetic crystals—such as diamond (University of Maryland: Sec.~\ref{sec:UMD}) and LiF (PALEOCCENE: Sec.~\ref{sec:PALEOCCENE})—using light-sheet microscopy to visualize fluorescent defects associated with tracks. These methods currently focus on detecting real-time events, much like conventional detectors, and offer the key advantage of identifying tracks within the bulk material, enabling directional sensitivity. The second approach is a hybrid strategy that combines high-resolution track observation using Transmission Electron Microscopy (TEM) with larger-scale scanning via X-ray Microscopy (University of Michigan: Secs.~\ref{sec:KaiSun}, \ref{sec:U-M_LaVoie-Ingram}; KIT: Sec.~\ref{sec:KIT_IAP}).

While these readout advances are promising, a significant challenge for MDs is their inherent, uncontrollable background in natural mineral samples. Unlike modern detectors that can be shielded, MDs have been continuously exposed over geological timescales—cosmic rays are reduced by overburden at depth, but nuclear recoils from radioactivity in both the minerals and the surrounding rock have been ever-present. To overcome this, some efforts target phenomena that are intrinsically background-free. For example, charged Q-balls or magnetic monopoles would create unique, penetrating tracks, providing an effectively background-free signal (Toho U: Sec.~\ref{sec:Tatsuhiro_Naka}; Queen's U: Sec.~\ref{sec:Queens}). In a novel approach, the PRImuS project (Sec.~\ref{sec:PRImuS}) re-frames cosmic rays—typically considered a background—as its primary signal of interest.

\begin{table}[htbp]
\centering
\caption{Summary of experimental studies of paleo-detectors. The content is kept compact for ease of reference.}
\label{tab:experiments}
\adjustbox{center}{  
\setlength{\tabcolsep}{3pt}  
\begin{tabular}{rlllll}
\hline
\S &\textbf{Collab/Inst} &  \textbf{Mineral} & \textbf{Etching} & \textbf{Readout} & \textbf{Target} \\
\hline
\hline
\ref{sec:PALEOCCENE} & PALEOCCENE &LiF (synth) & N/A & mesoSPIM & $n$, DM, $\nu$ \\
\hline
\ref{sec:UMD} &U of Maryland & Diamond (synth)  & N/A & LS-QDM & DM \\
\hline
\ref{sec:KIT_IAP} &KIT & Mica & N/A & nano-CT, TEM & DM, $\nu$ \\
\hline
\ref{sec:KaiSun}, \ref{sec:U-M_LaVoie-Ingram} &U of Michigan & Quartz, Olivine & N/A & TXM, TEM & $\nu$ (atms), DM \\
\hline
\ref{sec:DMICA} & DMICA &Mica & chemical & WLI & DM\\
\hline
\ref{sec:Tatsuhiro_Naka} &Toho U & Mica, Olivine & chemical & OM & UHDM, Q-balls \\
\hline
\ref{sec:Queens} &Queen's U & Mica & chemical & OM, SEM, PLM & UHDM, monopoles \\
\hline
\ref{sec:PRImuS} &PRImuS & Halite & plasma & OM & Cosmic rays \\
\hline
\end{tabular}
}
\begin{flushleft}
\footnotesize
\textbf{Abbreviations:} mesoSPIM = mesoscale Selective Plane-Illumination Microscopy; LS-QDM = Light Sheet Quantum Diamond Microscopy; nano-CT = nano-Computed Tomography; TEM = Transmission Electron Microscopy; TXM = Transmission X-ray Microscopy; WLI = White-Light Interferometry; OM = Optical Microscopy; SEM = Scanning Electron Microscopy; PLM = Polarized Light Microscopy; UHDM = Ultra-Heavy Dark Matter; $n$ = neutrons; $\nu$ = neutrinos. 
\end{flushleft}
\end{table}

The MD$\nu$DM'25 workshop at the Yokohama Institute for Earth Sciences JAMSTEC in May of 2025\footnote{\href{https://indico.ijs.si/event/2583/}{\url{https://indico.ijs.si/event/2583/}}} was the third such meeting, following the first in October of 2022 held at the Institute for Fundamental Physics of the Universe (IFPU) in Trieste, Italy\footnote{\href{https://agenda.infn.it/event/32181/}{\url{https://agenda.infn.it/event/32181/}}} and the second in January of 2024 hosted by the Center for Neutrino Physics at Virginia Tech in Arlington, USA\footnote{\href{https://indico.phys.vt.edu/event/62/}{\url{https://indico.phys.vt.edu/event/62/}}}. A whitepaper summarizing the history and current state of the field was published after the first workshop~\cite{Baum:2023cct} and a proceedings collected contributions following the second workshop~\cite{baum:2024eyr}, similar to this proceedings for the third meeting. All contributions to this and previous proceedings are intended to focus on the progress of various groups between successive workshops. 

As can be seen from the contributions below, many exciting advances are being made in MD research. The MD$\nu$DM community is international, with several groups actively working on MDs in Asia, Europe and North America. The MD$\nu$DM community is also highly multidisciplinary. In addition to aspects of MD research focused on the materials science challenge of reading out microscopic damage features in macroscopic mineral samples, the MD$\nu$DM community includes experts from high energy physics, condensed matter physics, geoscience, and AI/ML for data analysis. The application of MDs to the challenge of detecting dark matter and neutrino interactions will require years of research and development. With the active experimental and theoretical efforts summarized below, the MD$\nu$DM community aims to make significant progress before the next workshop scheduled for April 2026 and hosted by the Karlsruhe Institute of Technology (KIT) in Germany.

\acknowledgments
We thank the Yokohama Institute for Earth Sciences JAMSTEC for hosting the MD$\nu$DM'25 workshop in May 2025. We also thank Yoji Kawamura for the careful review of these proceedings.

\clearpage

\section{Current Status of DMICA: Exploring Dark Matter in Natural MICA}
\label{sec:DMICA}

Authors: {\it Shigenobu~Hirose$^1$, Natsue~Abe$^1$, Qing~Chang$^1$, Takeshi~Hanyu$^1$, Noriko~Hasebe$^2$, Yasushi~Hoshino$^3$, Takashi~Kamiyama$^4$, Yoji~Kawamura$^1$, Kohta~Murase$^5$, Tatsuhiro~Naka$^6$, Kenji~Oguni$^1$, Katsuhiko~Suzuki$^1$, and Seiko~Yamasaki$^7$}
\vspace{0.1cm} \\
$^1$Japan~Agency~for~Marine-Earth~Science~and~Technology, $^2$Kanazawa~University, \\
$^3$Kanagawa~University, $^4$Hokkaido~University, $^5$The~Pennsylvania~State~University,\\ \
$^6$Toho~University, $^7$National~Institute~of~Advanced~Industrial~Science~and~Technology
\vspace{0.3cm}

\subsection{Mica as dark matter detector}

\begin{figure}[ht]
   \centering
   \includegraphics[width=0.49\textwidth]{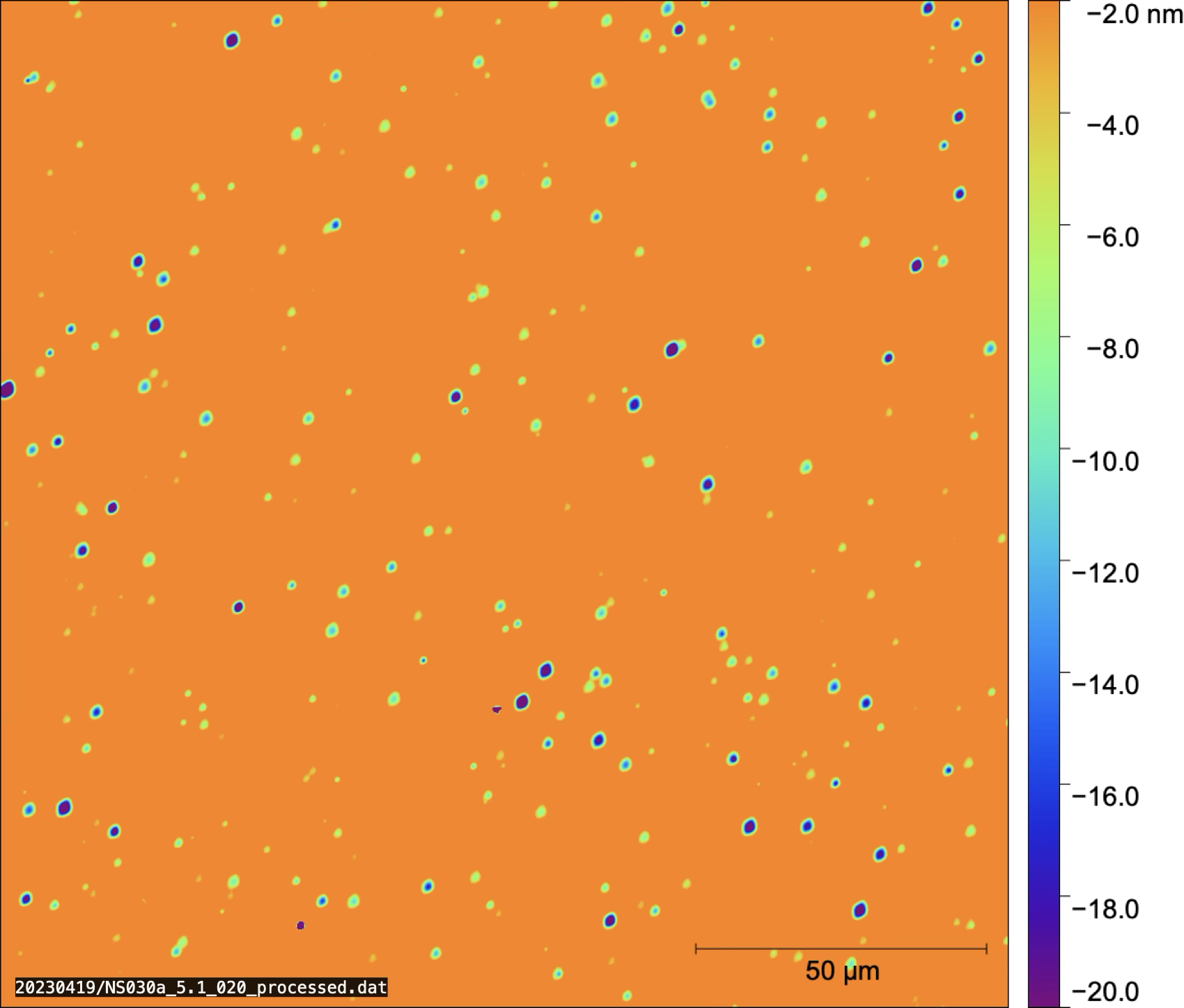}
   \includegraphics[width=0.49\textwidth]{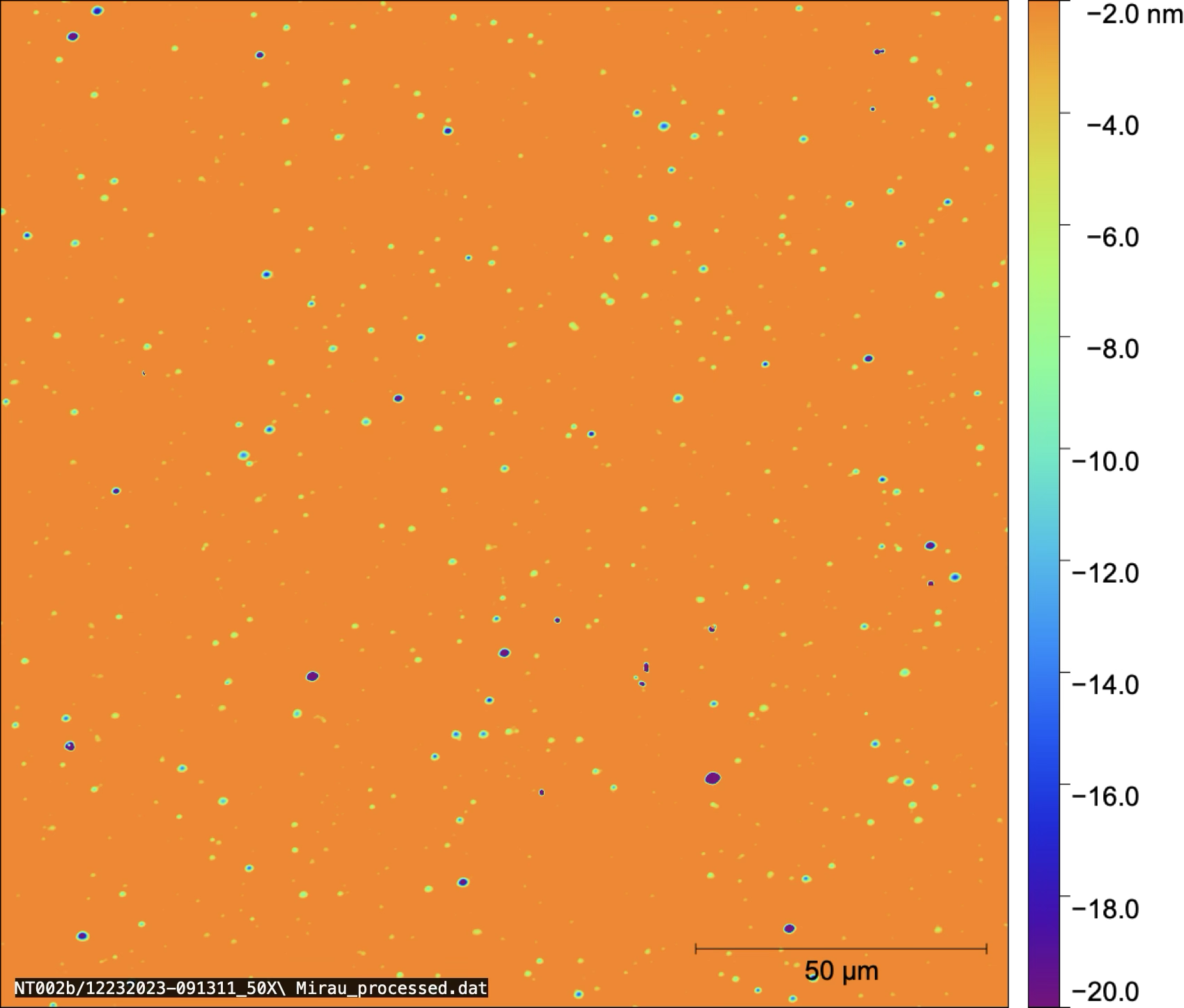}
   \caption{Optical profiler images of the etched, cleaved surface of natural mica (left) and of fast-neutron-irradiated mica after annealing (right). }
   \label{fig:DMICA_etched_surfaces}
\end{figure}

Muscovite mica is an efficient solid-state track detector.
Figure~\ref{fig:DMICA_etched_surfaces} displays optical-profiler (white-light interferometer) images of surfaces that were first cleaved and then etched in HF.
In the right panel, the mica was annealed to erase radiogenic defects (cf. the left panel), irradiated with $\mathcal{O}(1)$ MeV neutrons at the Hokkaido University Neutron Source, and subsequently processed; each pit corresponds to a single neutron-recoil track.
Because a dark-matter (DM) scatter is likewise expected to leave an isolated recoil track with a recoil energy in the $\sim$keV/amu range, the right-hand image demonstrates mica’s suitability as a DM detector.

The first mica-based DM search was performed by Ref.~\cite{Snowden-Ifft:1995zgn}.
Building on that pioneering study, the DMICA project is now in an R\&D phase and aims for an exposure several orders of magnitude greater.


\subsection{Restoration of surface topography obtained by an optical profiler}

\begin{figure}[ht]
   \centering
   \includegraphics[width=0.49\textwidth]{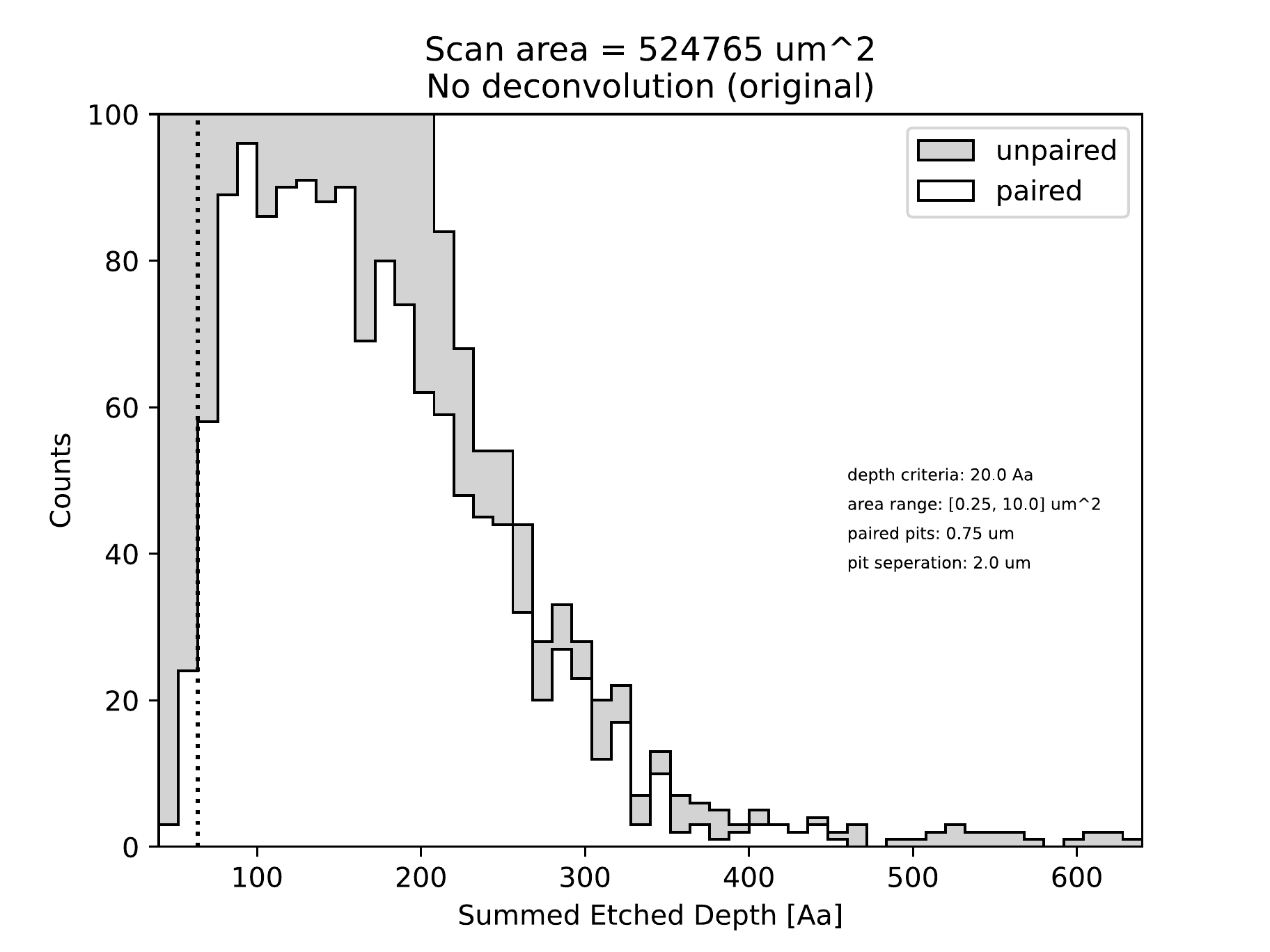}
   \includegraphics[width=0.49\textwidth]{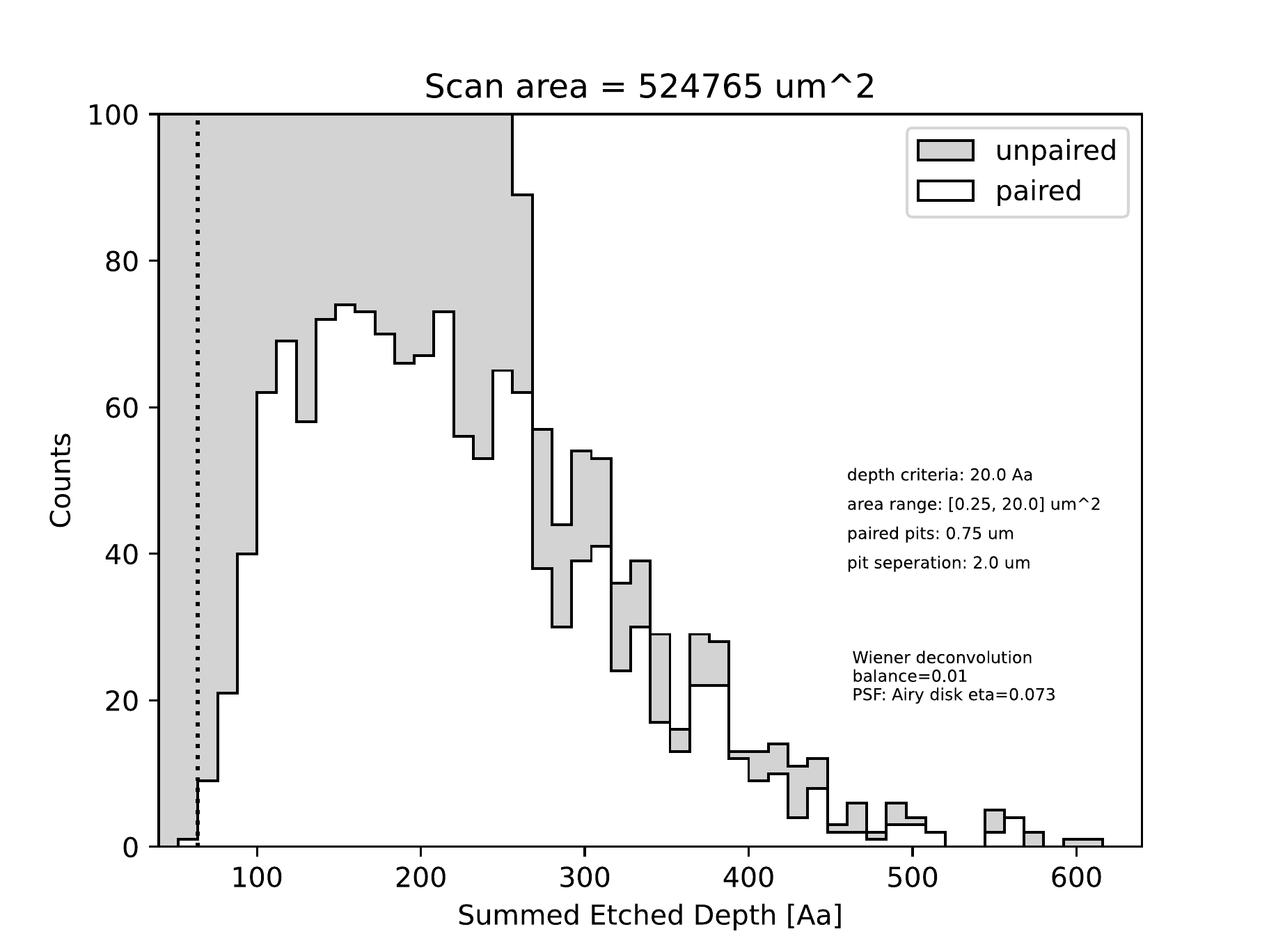}
\caption{Pit-depth histograms summed over processed mica surfaces with a total area of 524,765~$\mu\mathrm{m}^{2}$.  
The histogram derived from the restored topography (right) recovers pit depths that were underestimated in the original measurement~\cite{hirose2024} (left).}
   \label{fig:DMICA_histogram}
\end{figure}

The throughput in Ref.~\cite{Snowden-Ifft:1995zgn} was limited by the slow scanning rate of the atomic-force microscope used to inspect mica surfaces. The DMICA experiment overcomes this constraint by employing an optical profiler, enabling the scan rate required to achieve an exposure of 1 ton yr. Optical profiling is, however, subject to the diffraction limit of the objective lens, and thus the recorded topography is blurred and pit depths are systematically underestimated. We mitigate this limitation with a restoration procedure based on the profiler’s instrument transfer function (ITF)~\cite{Hirose:25}. The resulting restored pit-depth distribution is compared with the original one in Fig.~\ref{fig:DMICA_histogram}.

\subsection{Measurement of the bulk etch rate}\label{sec:bulk_etch_rate}

\begin{figure}[ht]
   \centering
   \includegraphics[width=0.52\textwidth]{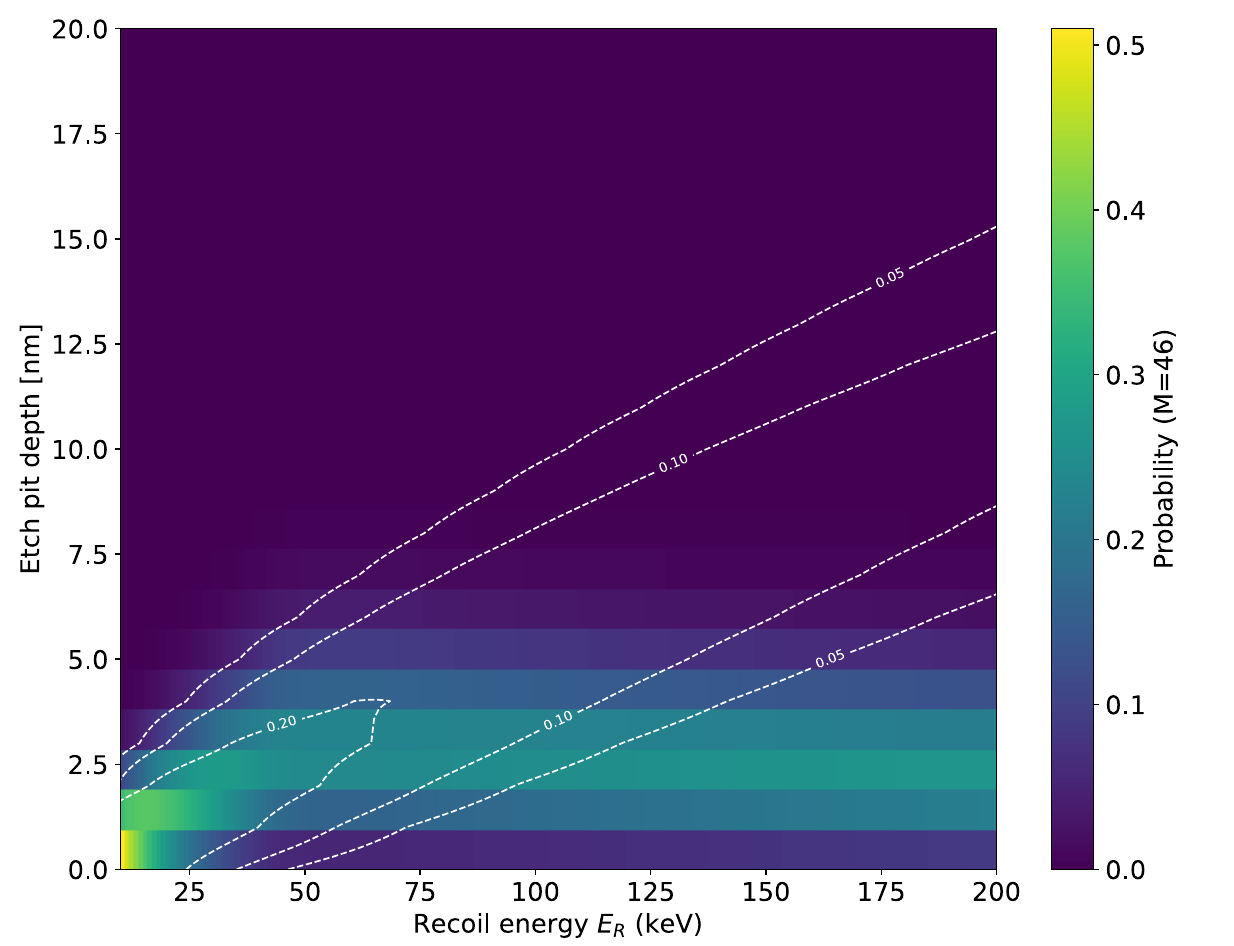}
   \includegraphics[width=0.47\textwidth]{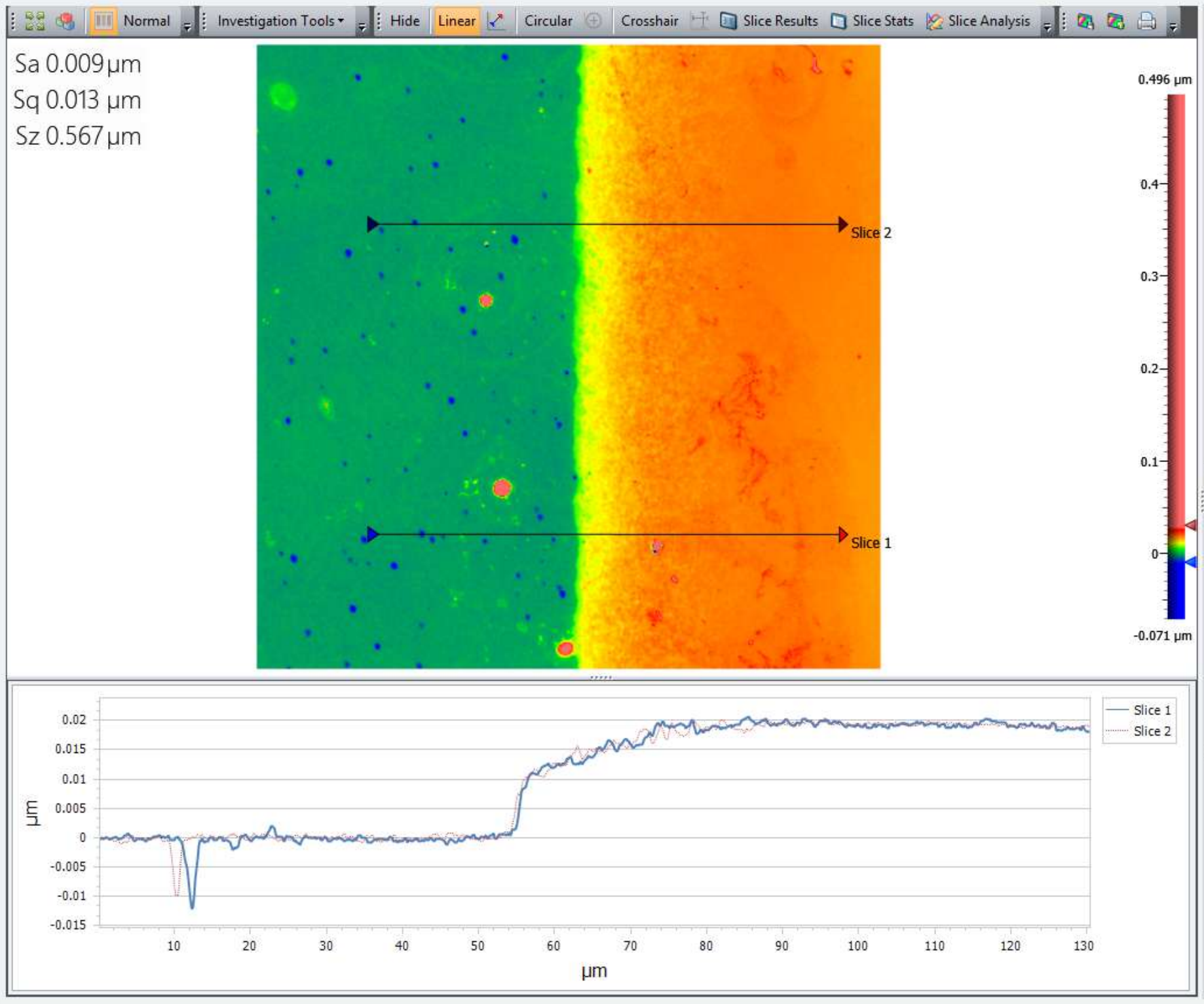}
    \caption{Left: color map showing the relation between etched pit depth and recoil energy, after Ref.~\cite{Snowden-Ifft:1995rip}.  Right: after a 1~h HF etch, the step height between the HF-resistant coated region (right) and the uncoated region (left) corresponds to a bulk etch rate of $20~\mu\mathrm{m/h}$~\cite{freeman:1996}.}
   \label{fig:bulk_etch_rate}
\end{figure}

The etching process exposes, in fact, only a portion of each latent track. To use mica as an energy-resolving detector, Ref.~\cite{Snowden-Ifft:1995rip} established a relation between pit depth and recoil energy from ion-irradiation experiments (Fig.~\ref{fig:bulk_etch_rate}, left). This calibration depends on the bulk etch rate of the mica, which therefore must be measured. We have devised a new direct method that employs an HF-resistant coating and have obtained a bulk etch rate that is consistent with the only direct measurement reported to date (Fig.~\ref{fig:bulk_etch_rate}, right).

\acknowledgments
We are grateful to Prof. Snowden-Ifft for his invaluable insights and discussions, which have significantly shaped the development of the DMICA project. This work was supported by JSPS KAKENHI Grant Number JP25K07350.

\clearpage

\section{Q-ball Interaction with Matter}

Authors: {\it Ayuki Kamada}
\vspace{0.1cm} \\
University of Warsaw \\
\vspace{0.3cm}

\subsection{Q-ball dark matter}
A Q-ball is a non-topological soliton made of scalar particles carrying a conserved charge (differentiate it with electromagnetic charge)~\cite{Coleman:1985ki}.
It is (quasi-)stable when the chemical potential (energy cost per charge) is smaller than the mass of the scalar particle.
The stability condition requires a certain shape of the scalar-field potential (in brief, it should be shallower than the quadratic potential, at some field value).
Meanwhile, in the standard model (SM) of particle physics, baryon number is known not only to be approximately conserved, but also to be generated somehow in the early Universe to account for baryon asymmetry of the Universe (BAU).
Therefore, it is well-motivated to think about a Q-ball made of some scalar field carrying a baryon number (some literature calls it B-ball to be more specific).

Though there is no such a field in the SM (quarks are fermions), it appears in the well-motivated extension of the SM (remember that we need an extension to generate BAU, anyway): supersymmetry (SUSY), which explains why the energy scale of the SM (electroweak scale) is smaller than the energy scale of the more fundamental theory (Planck scale) by many orders of magnitude~\cite{Martin:1997ns}.
In a supersymmetric extension of the SM, there are supersymmetric partners of each SM particle, whose quantum number (or charge) is the same as the SM particles, except for quantum statistics: quark (spin 1/2) $\leftrightarrow$ squark (spin 0).
Such squarks carrying a baryon number also satisfy the potential condition mentioned above and thus can form a Q-ball.
In particular, in a certain class of supersymmetric extensions (gauge mediation scenarios), the chemical potential per baryon number is larger than the mass of the nucleon~\cite{Kusenko:1997si, Kusenko:1997vp}.
Then, a Q-ball is stable over the age of the Universe and thus a good candidate of dark matter.
(See, e.g., Ref.~\cite{Kasuya:2015uka} for more about Q-ball dark matter and its relic abundance.)

\subsection{Q-ball interaction with matter}

Though a Q-ball has a large geometrical cross section $\gtrsim 1 \, {\rm barn}$, it evades the constraints from conventional direct-detection searches.
This is because its large mass $\gtrsim 1 \, {\rm g}$ limits the number of Q-balls coming to the detector of $\sim 1 \, {\rm t}$ during the exposure of $\sim 1 \, {\rm yr}$.
Here come large-size experiments such as Kamiokande~\cite{Super-Kamiokande:2006sdq} and IceCube~\cite{Kasuya:2015uka} or longer-exposure experiments such as paleo detectors~\cite{Baum:2018tfw, Baum:2023cct}.
Despite its phenomenological importance, interaction of a Q-ball with ordinary matter has not been well-understood yet.
This is mainly because a Q-ball is a composite object rather than a single particle (more is different).
What we can study robustly is so far limited with test free quarks on a Q-ball background, which ignores any electromagnetic/color forces on quarks and any backreaction to a Q-ball.
Their results suggest that the in-coming quarks are reflected as anti-quarks with order-unity branching ratio~\cite{Kusenko:2004yw}.
This conversion is expected to (remember that it is not automatically incorporated in the computation) add $2/3$ baryon numbers to the Q-ball and thus costs the chemical potential.
It is actually shown that when initial quark does not have a sufficient energy to pay the energy cost, anti-conversion actually does not occur~\cite{Kamada:2025mji}.

Translating such results into a nucleon (or hadron) is quite challenging because of the non-perturbative nature of strong interaction.
Even worse, strong interaction becomes very weak inside a Q-ball (where a sfermion has a large field value).
There still exist qualitative best guesses about what happens when a nucleon collides with a Q-ball.
An in-coming nucleon would be dissolved into quarks on the surface of the Q-ball.
The quarks cannot go away from the the Q-ball, since strong interaction is very strong outside a Q-ball.
Let us call such a quasi-excited state with quarks as a Q$^{*}$-ball.
There would be three possibilities in the fate of the Q$^{*}$-ball:
1) emitting one (or more) pion (called KKST process~\cite{Kusenko:1997vp});
2) emitting one nucleon (suggested by elastic scattering of quark);
3) emitting one anti-nucleon (suggested by conversion of quark into anti-quark).
Actually the last possibility adds two baryon numbers to the Q-ball and thus costs the chemical potential.
With the typical collision velocity in our Milky Way, such an energy requirement is not satisfied~\cite{Kamada:2025mji}. (The situation is different, e.g., for a collision of a Q-ball and a nucleus in the neutron star~\cite{Kusenko:2005du}, where typical collision velocity is semi-relativistic.)
Therefore, one may focus on the first two possibilities.
Assuming the branching ratios are comparable, the first possibility would be a more striking signal in actual experiments.

The above basic picture is further modified when a Q-ball carries an electromagnetic charge.
Any nucleus is positively charged and thus cannot collide with a positively-charged Q-ball because of Coulomb repulsion.
This is why an early literature~\cite{Bakari:2000dq, Arafune:2000yv} studies two cases with Supersymmetric Electrically Neutral Solitons (SENS) and Supersymmetric Electrically Charged Solitons (SECS) separately.
Even if Q-balls are charge neutral in the early Universe, there are two mechanisms for them to obtain a charge:
1) Q-ball carries a lepton number in addition to a baryon number, and emitting charged leptons~\cite{Hong:2016ict, Hong:2017qvx};
2) Q-ball obtains a charge by absorbing a nucleon and emitting pions/nucleon successively. 
In the latter case, a positively-charged Q-ball would be an attractor, since further absorption of a nucleus is suppressed by Coulomb repulsion, unless there exists some other discharge mechanism like electron conversion into positron.
In addition, to place a more robust constraint, one needs to take account of (dis)charging a Q-ball in the passage to and through the detector, which is ignored in the early literature~\cite{Bakari:2000dq, Arafune:2000yv}.

\acknowledgments
A. K. thanks Takumi Kuwahara and Keiichi Watanabe for the collaboration, and Kohta Murase and Shigenobu Hirose for encouraging us to work on this study.

\clearpage

\section{ Probing Ancient Cosmic Ray Flux with Paleo-Detectors and the Launch of the PRImuS Project}\label{sec:PRImuS}

Authors: {\it Claudio~Galelli$^1$, Lorenzo~Caccianiga$^1$, Lorenzo~Apollonio$^2$, and Paolo~Magnani$^1$}
\vspace{0.1cm} \\
$^1$INFN~Milano\\
$^2$Università~Statale~di~Milano \\
\vspace{0.3cm}

\subsection{Paleodetectors for cosmic rays and the Messinian Salinity Crisis}

For the emerging field of paleo-detectors, one of the primary challenges in searching for rare events like dark matter or neutrino interactions is the overwhelming background from cosmic rays (CRs)~\cite{Baum:2023cct}. In particular, this necessitates sourcing mineral samples from deep underground sites to ensure adequate shielding. Our work inverts this paradigm by treating the CR-induced nuclear recoil tracks not as a background, but as the scientific signal of interest. By analyzing minerals with a known history of successive surface exposure and shielding, we can recover the ``frozen'' track record to open a new observational window on the past of the high-energy universe, possibly allowing us to probe the evolution and history of the CR flux over geological timescales. This approach offers a new channel for paleo-astroparticle physics and aims at providing the first direct experimental characterization of the CR background, an important input for the entire mineral detection community.

\begin{figure}[!ht]
    \centering
    \includegraphics[width = 0.7\textwidth]{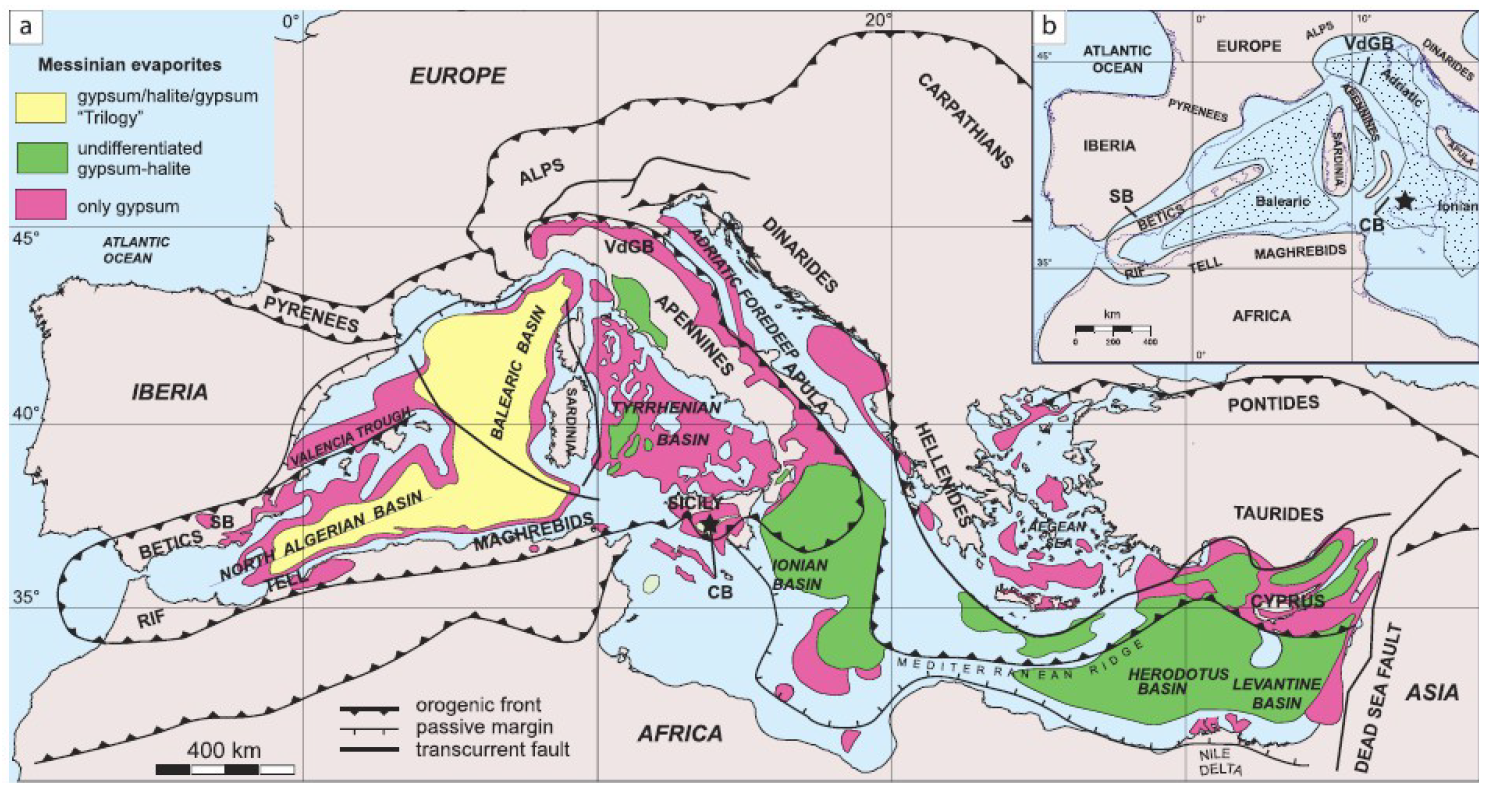}
    \caption{a) Map of the Messinian evaporites in the Mediterranean. b) Paleocenaographic map of the Western Mediterranean basins during the Messinian salinity crisis, showing the main evaporite depocentres (dotted areas). Emerged areas are in gray. Dotted line is the modern coastline. From~\cite{Meilijson:2019}.}
    \label{fig:PRIMUS_msc}
\end{figure}

To make such a measurement feasible, a mineral sample must have a well-defined ``exposure window'' followed by a rapid and permanent shielding event. Our phenomenological work has identified a very interesting target that meets these criteria: evaporite minerals, specifically halite (NaCl), formed during the Messinian Salinity Crisis (MSC) approximately 6 million years ago~\cite{Meilijson:2019}. This unique geological event saw the near-complete desiccation of the Mediterranean Sea, during which thick layers of halite were deposited and exposed to the surface CR flux for a period of about 500,000 years (Fig.~\ref{fig:PRIMUS_msc}). The crisis ended abruptly with the catastrophic Zanclean Flood, which refilled the basin and buried the evaporites under kilometers of water and sediment. This event acted as a perfect natural ``off-switch'', preserving the integrated record of CR interactions from that specific epoch.

\begin{figure}[ht!]
    \centering
    \includegraphics[width=0.7\columnwidth]{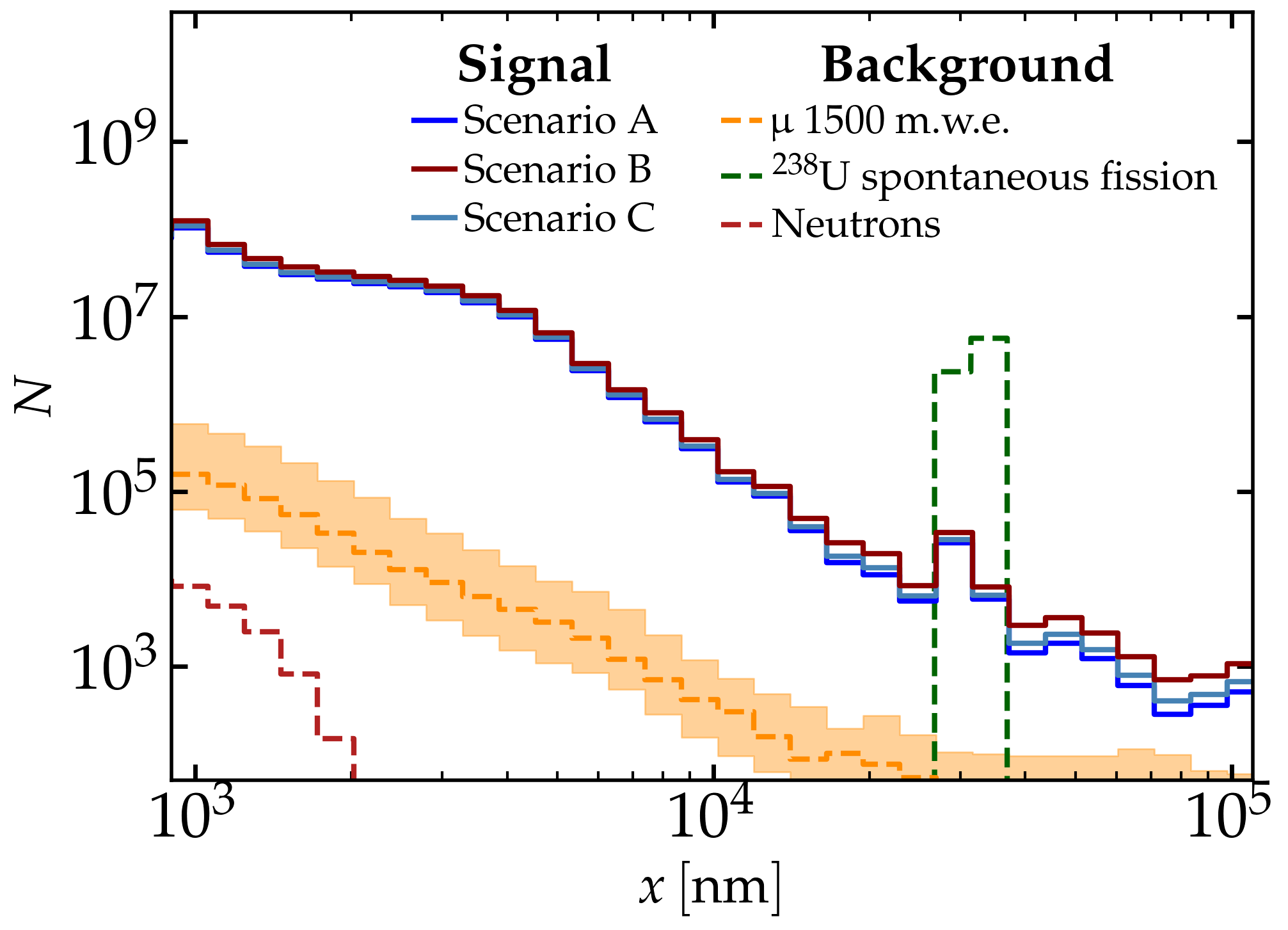} 
    \caption{Expected total number of tracks in a $10\,$g sample of halite which was created $5.6\,$Myr ago, exposed for $270\,$kyr to muons and then covered by a $1.5\,$km overburden of water. The backgrounds are integrated for the whole age of the sample with the exception of the underwater $\mu$, which is integrated for the period when the sample is sea-covered, i.e. $5330\,$kyr.}
    \label{fig:PRIMUS_histhalite}
\end{figure}

We performed a simulation of this scenario, modeling the propagation of CR-induced muons through the atmosphere and into the halite deposits, and calculating the resulting spectrum of nuclear recoil tracks. We selected muons as our primary particle of interest as they are abundant in CR showers and easily penetrate small shields of water or sediment. The results confirm that the expected signal from CR muons is orders of magnitude greater than any anticipated backgrounds, such as from internal radioactivity or atmospheric neutrinos, as visible in Fig.~\ref{fig:PRIMUS_histhalite}. Most importantly, our simulations demonstrate that the track length spectrum is sensitive to percent-level variations in the primary CR flux. This sensitivity could be sufficient, depending on track counting and recovery efficiency and instrumental effect, to probe for the effects of significant astrophysical events, such as a nearby supernova explosion within the local bubble, which are hypothesized to have occurred during that geological period. These results were published in~\cite{caccianiga_2024}.

\subsection{The launch of PRImuS}

\begin{figure}[!ht]
\centering
\begin{minipage}[b]{0.45\linewidth}
\centering
\includegraphics[width=\linewidth]{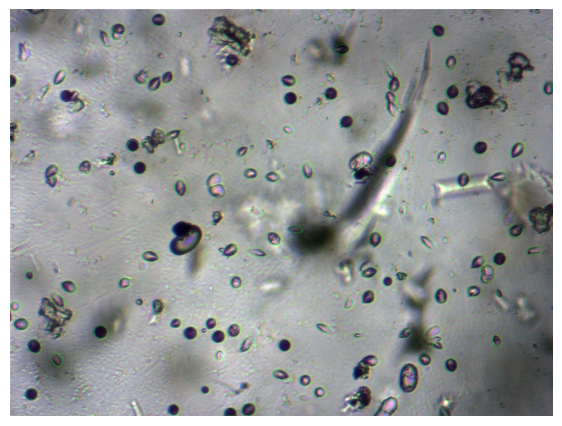}
\end{minipage}
\hfill
\begin{minipage}[b]{0.54\linewidth}
\centering
\includegraphics[width=\linewidth]{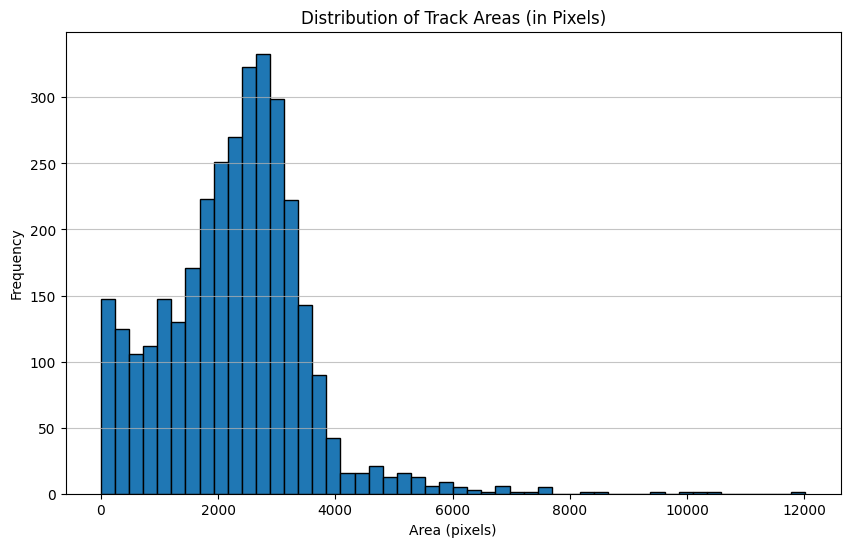}
\end{minipage}
\caption{Test results of the analysis pipeline applied to enlargements on chemically treated irradiated obsidian containing induced fission tracks. Left: detected track mask (in green) for an image overlaid on the original image. Right: histogram of pixel size for all the tracks detected over 60 test images.}
\label{fig:PRIMUS_pipeline}
\end{figure}

Building on these promising phenomenological results, we have launched the PRImuS (Paleo-astroparticles Reconstructed with the Interactions of MUons in Stone) project. PRImuS is an INFN-funded experimental program based in Milan, designed to move from simulation to the first direct detection and measurement of these ancient CR-induced tracks. The experimental strategy is centered on a high-throughput analysis pipeline. We are developing protocols for plasma etching to enlarge the latent tracks, making them visible with optical microscopy. We have acquired an automated microscopy station capable of automatically scanning large, cm$^2$-scale areas to gather the necessary statistics.

The immense volume of image data produced will be analyzed by a dedicated software pipeline with a machine learning model (a U-Net for semantic segmentation) at its core. This model is being trained for the automated identification of tracks in optical microscopy images. The pipeline then can analyze the detected tracks and produce the scientific output i.e. size and density histograms. The pipeline is being tested on chemically treated irradiated obsidian samples, which contain induced fission tracks. The results of one of these tests are shown in Fig.~\ref{fig:PRIMUS_pipeline}.

PRImuS is in its setup phase, and, once fully active, it could be a cornerstone in the understanding of CR-induced tracks in paleodetectors, characterizing the backgrounds for other mineral detection experiments and potentially opening the possibility of studying ancient cosmic ray fluxes.

\acknowledgments
We thank the INFN CSN5 Young Researchers' Grant for providing the funding for PRImuS. CG thanks Vincent Breton for the interest and the discussions on samples.

\clearpage

\section{Probing Cosmic Walls with Paleo Detectors}

Authors: {\it Wen~Yin} \\
\vspace{0.1cm}
Department~of~Physics, Tokyo~Metropolitan~University
\vspace{0.3cm}

\subsection{Direct detection of cosmic walls using ancient minerals}

Paleo detectors---ancient minerals containing fossilized damage tracks from nuclear recoils---offer a unique opportunity to probe rare cosmological phenomena.\footnote{This talk is based on Ref.~\cite{Yin:2025wuv}.} In this talk, we propose utilizing these detectors for the direct detection of \textbf{cosmic walls}: either bubble walls from late-time first-order phase transitions or domain walls in a scaling regime. 

Cosmic walls are predicted in various extensions of the Standard Model and have been discussed in connection with phenomena such as cosmic birefringence~\cite{Takahashi:2020tqv}, gravitational waves~\cite{Kitajima:2023cek}, and potentially with time variations in fundamental constants that may help alleviate cosmological tensions~\cite{Sekiguchi:2020teg}. Although it is well known that late-time cosmic walls can be indirectly probed via distortions of the cosmic microwave background due to their gravitational potentials~\cite{Zeldovich:1974uw}, direct detection has rarely been considered (see, e.g.,~\cite{GNOME:2023rpz} for non-scaling domain walls).

Since such walls are expected to traverse Earth only $\mathcal{O}(1)$ time throughout cosmic history, continuous, passive, and time-integrated detection is essential-making paleo detectors an ideal platform. A wall-nucleus or wall-electron coupling can induce detectable nuclear recoils in minerals such as muscovite, olivine, or apatite, forming parallel damage tracks in samples older than the wall-crossing epoch, distributed globally.

\subsection{From theory to observables: signatures and constraints}

Assuming a scalar field $\phi$ associated with the wall, we consider interactions of the form $f(\phi)\,\bar{N}N$, where $N$ denotes a nucleon or nucleus. We compute wall-induced recoil processes, including $N \to N + \phi$, using perturbative WKB techniques. These calculations show that even extremely feeble couplings can lead to observable damage patterns, provided the wall passed within the last $\mathcal{O}(1)$\,Gyr. Our estimates follow the approach used in studies of bubble wall velocities and particle productions~\cite{Bodeker:2009qy,Azatov:2020ufh,Azatov:2021ifm,Azatov:2024crd}.

\begin{figure}
   \centering
   \includegraphics[width=0.7\textwidth]{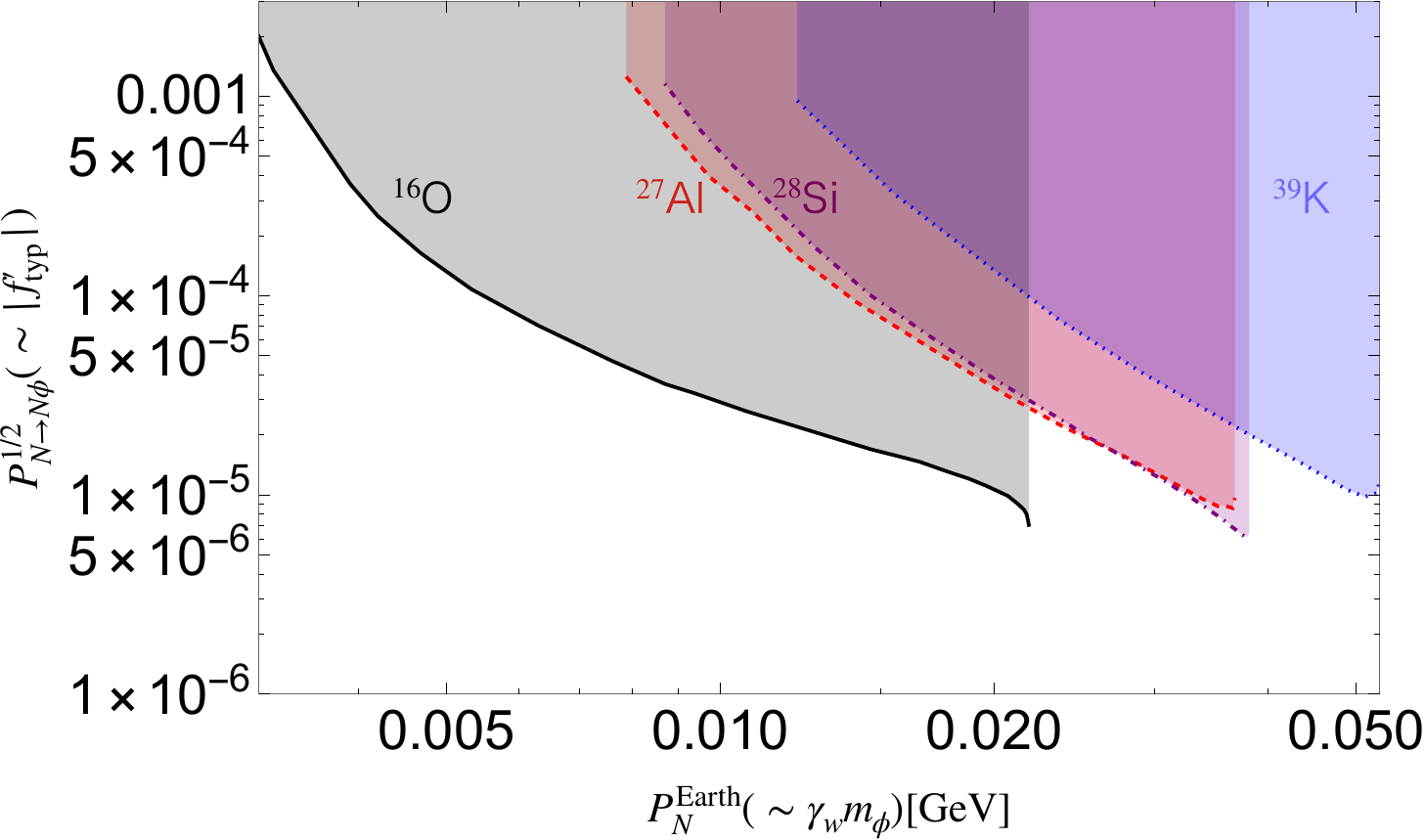}
   \caption{Direct detection limits on cosmic walls derived from muscovite mica data in Ref.~\cite{Snowden-Ifft:1995zgn}. The limit applies to a wall passage occurring within the last 0.5\,Gyr.}
   \label{fig:CW}
   \vspace{1cm}
\end{figure}

Recasting data from the ancient mica search in Ref.~\cite{Snowden-Ifft:1995zgn}, we derived the first experimental constraints on cosmic wall interactions under the assumption that a wall traversed Earth within the past 0.5\,Gyr. These limits depend on the target nucleus and the wall width (which is inversely related to the scalar mass $m_\phi$ for semi-relativistic walls). The transition probability scales as $P_{N \to N \phi} \sim |f'(\phi)|^2$, where $f'(\phi)$ denotes a typical value of the effective coupling, not necessarily the vacuum value. Figure~\ref{fig:CW} shows the resulting bounds for O, Al, Si, and K nuclei.

Importantly, if the wall velocity is sub-relativistic in the Earth's frame, the resulting recoil directions are nearly parallel across the globe, offering a smoking-gun signature. With future datasets involving larger mineral exposures and improved track resolution, such anisotropic patterns could provide decisive evidence for a wall-crossing event. Moreover, the distribution of track orientations and the length encode information about the wall’s velocity, width, and even whether it corresponds to a bubble wall or a domain wall.

Paleo detectors may therefore offer the only viable means of directly detecting cosmic walls and could reveal valuable insights into fundamental physics through their encoded wall profiles.

\acknowledgments
I am grateful to the organizers of the ``Mineral Detection of Neutrinos and Dark Matter 2025'' workshop for the invitation that inspired this study. This work is supported by JSPS KAKENHI Grants No. 22K14029 and 22H01215, and by the Selective Research Fund for Young Researchers from Tokyo Metropolitan University.%
\clearpage

\section{Olivines from Archean Komatiites}

Authors: {\it William~F~McDonough$^{1,2}$ and Emilie~LaVoie-Ingram$^3$}
\vspace{0.1cm} \\
$^1$Advanced Institute for Marine Ecosystem Change, Department of Earth Sciences and Research Center for Neutrino Science, Tohoku University, Sendai, Miyagi 980-8578, Japan
\\$^2$Department of Geology, University of Maryland, College Park, MD 20742 USA
\\$^3$Department of Physics, University of Michigan, Ann Arbor, MI 48103 USA \\
\vspace{0.3cm}

\subsection{Preservation and potential of ancient olivines as particle detectors}
Olivines from komatiites represent some of the best available mineral detectors for neutrino and dark matter searches. Komatiites are Mg-rich lavas that are almost exclusive to the Archean ($>$2.5 billion years old) time period of Earth's history. The high MgO contents of these lavas (typically 25 wt\% MgO~\citep{arndt2023komatiite}, cf., 10 wt\% MgO for modern basalts) reveal that their eruption temperatures (about 1700$\degree$C) were much higher than what is observed today for basalts erupted at mid-ocean ridges and places like Hawaii (1100 to 1250$\degree$C). 

Ref.~\cite{nisbet1987uniquely} was the first paper to report melt inclusions (10-50 micron diameter) in fresh olivines from 2.7 billion-year-old Belingwe komatiites in Zimbabwe. This amazing discovery revealed that these lavas are incredibly well preserved. Following this discovery, Ref.~\cite{mcdonough1993intraplate} reported ionprobe trace element data for these melt inclusions documenting that fluid mobile elements (e.g., K, Sr, and Ba) were still present in their original relative abundances. These data also provided the opportunity to constrain the tectonic environment for komatiite genesis. Since then, other locations of unaltered komatiites hosting olivine-bearing melt inclusions have been identified~\citep{sobolev2016komatiites,asafov2018belingwe}, including locations that are up to 3.3 billion years old~\citep{sobolev2019deep,vezinet2025growth}.

Komatiitic olivines are excellent detector minerals given their natural abundances in the Earth and their low content of radioactive elements. The olivine content of basalts and komatiites is approximately 20 to 30\% by mode, but can reach 70\% in some komatiites~\citep{arndt2023komatiite}. Olivine is the most common mineral in the upper mantle, typically estimated to be about 60\% by volume~\citep{mcdonough1998mineralogy}. The U and Th contents of these olivines are typically $\leq10^{-10}$ g/g, with Th / U of approximately 4~\citep{vezinet2025growth} and densities of 3300 kg/m$^3$, so that they will likely have only a few radioactive atoms per cubic micron. 

\subsection{Estimating the overburden of ancient olivines from South African Komatiites}
An unknown in the history of these komatiites is the depth for which they have remained during their multibillion-year residence in the continental crust. We have quantitative constraints ($\pm\sim$1\% uncertainties from radiometric ages) on the eruption and emplacement ages of these lavas~\citep{connolly2011highly,hofmann2021layered}. Currently, we do not have quantitative constraints on the depth of burial versus time. Their freshness tells us that these lavas were not at the surface where there is abundant water, which drives almost all alteration processes. The fact that these rocks are $>$2.5 billion years old and still fresh means that they have been held in some ``special place'' for a long time to preserve them. A standard crustal geotherm is typically between 10$\degree$C and 20$\degree$C/km. Therefore, the maximum temperature should only be order 100$\degree$C at 5 km depth. This condition is usually not enough to anneal tracks, but might alter these minerals. However, their preservation tells us that the system has been relatively dry over their history and does not have active alteration processes.

Although there is no quantitative constraint on the history of crustal depth of these lavas, a reconstruction of the average depth of burial may be possible with strict constraints on their age, location, and U-Th concentrations. The muon flux and, correspondingly, the differential cosmogenic neutron flux vary as a function of the depth and composition of the rock overburden~\citep{Fedynitch_2022, Woodley:2024eln, MARINO2007611, MeiHime2005}. In addition, given the global transport of continents over geological time and due to the action of plate tectonics, these lithologies probably spent time both at the magnetic poles and equatorial regions. Therefore, the paleopole position and variation in the magnetic-field intensity must be taken into account in addressing the total muon flux.

The relative shape of the induced recoil spectrum is expected to remain constant with depth, varying only in amplitude. This spectrum can be fitted, and thus the average depth or cosmogenic signal can be `reconstructed', with experimental data and simulation. There are open-source software applications available that we are experimenting with to build a successful workflow for modeling cosmogenic neutron-induced backgrounds. CRY (the ``Cosmic-Ray Shower Generator'') is one example of how we can model cosmic ray, or more specifically muon flux, through some given area and location on Earth's surface. The resulting flux and energy spectrum of muons and other primary particles can be fed into Geant4~\citep{collaboration2003geant4} for calculation of muon-spallation of nuclei in and around our target material, producing a flux of fast neutrons that induce nuclear recoils in our target. Modeling the true flux of muons and neutrons on the surface of Earth can be complicated. To achieve great precision in our simulation, we must take into account fast neutrons produced in the atmosphere that can shower down onto our target on the surface and model the affects of atmospheric changes like composition and humidity over the timescales of surface exposure. At great depths into Earth's crust, however, these surface effects are neglected, and thus only the composition and density of the overburden, and possible transient overburden and depth effects, must be taken into account. To model the induced flux at depth, the Monte Carlo code PROPOSAL (PRopagator with Optimal Precision and Optimized Speed for All Leptons)~\cite{koehne_proposal_2013} can be used to calculate survival matrices of muons through some given depth and overburden, and the resulting parameters can be integrated into Geant4 for the same calculation of neutron flux mentioned before. Our algorithm in Geant4 will report the number and energy of recoils induced on specific elements in our target mineral, and calculation of physical damage - or the length of the recoil track induced - can be done with SRIM and TRIM (Stopping and Range of Ions in Matter, and Transport of Ions in Matter)~\cite{ZieglerSRIM2010}.

\subsection{Additional background considerations}
To reach the sensitivity necessary to detect rare particle signals like those induced by neutrinos and dark matter in these ancient rocks, we must carefully take into account all other background signals. The Earth's crust has a natural flux of neutrons due to alpha decays and alpha-nucleon interactions. On average, the upper crust has a neutron flux of approximately 10$^4$ neutrons / kg / year (or $\sim10^{-8}$ neutrons per cm$^2$ s$^{-1}$)~\citep{vsramek2017subterranean}. The flux from the associated komatiite lithologies is likely to be lower than the average crust, given that komatiites and associated basaltic lava are likely to have lower Th and U contents, and thus fewer alpha emitters. Consequently, there will be low concentrations of heavy nucleus recoil tracks from alpha-emission and spontaneous fission tracks.

Finally, when considering the detectable flux of cosmic particles, that is dark matter and neutrinos, we can consider what might be the long-term flux of supernovae (extra bright neutrino emitters) in this region of the galaxy. Supernovae are estimated to emit $>$10$^{50}$ $\nu$ and $\bar \nu$  in all lepton flavors. Observations of SN 1987A (51 kiloparsecs away) had a neutrino luminosity of 10$^{45}$ W and neutrino energies of up to a few tens of MeV~\citep{arnett1989supernova}. Recently, it has been documented that a few\;\% of fresh $^{60}$Fe was captured in dust and deposited on Earth 1.5 to 3.2 million years and 6.5 to 8.7 million years ago~\citep{wallner2016recent}. This and other lines of evidence point to multiple supernova and massive star events that have occurred during the last 10 million years at up to 100 parsecs from Earth~\citep{fields2019near}. If a ``typical'' distance for a Milky Way supernova is 10 kpc, detectors like Super-Kamiokande expect to observe about 5000 neutrino events (\url{http://hep.bu.edu/~superk/gc.html}), given 22.5 ktons of fiducial volume and 1 neutrino interaction per 4500 kg of target mass. Overall, it is expected that there would be approximately 10$^{4}$ supernovae events/Myrs, or 20 million supernovae events/billion years. This flux is equivalent to the local galactic ($\leq$ 100 pc) flux of 1/every 2 to 4 million years total rate in the Milky Way (2.0 $\pm$ 0.7 per century)~\citep{wallner2016recent}.

It is also interesting to note that when using these ancient minerals to detect atmospheric neutrinos, the sensitivity of the required paleo-detector is much less influenced by depth. Tracks induced by nuclear recoils from atmospheric neutrinos can be up to a millimeter in length, exceeding track lengths caused by cosmogenic neutrons and many other backgrounds. Komatiitic olivines from ages ancient to recent (3300 to 90 million years ago)~\citep{walker2023182w} provide the best possible exposure and purity for atmospheric neutrino detection, allowing us to map changes in the cosmic ray rate throughout Earth's history. 

These numbers and komatiitic olivines give us much hope for particle track discoveries in the near future.

\acknowledgments
WFM gratefully acknowledges the NSF for support (grant EAR-2050374). ELI is grateful to Dr. Sam Hedges for his invaluable guidance in helping build a simulation workflow, as well as to many undergraduates working at the University of Michigan in helping get this project started. ELI is also incredibly grateful for the donation of ancient samples by Dr. Igor S. Puchtel at the University of Maryland and Dr. Alexander Sobolev at the Université Grenoble Alpes-CNRS. 

\clearpage

\section{Progress Toward a Solid-State Directional Dark Matter Detector}\label{sec:UMD}

Authors: {\it Daniel~Ang, Jiashen Tang, Maximilian Shen, Mason Camp, Andrew Gilpin, Gavishta Liyanage, and Ronald Walsworth}
\vspace{0.1cm} \\
Quantum Technology Center, University of Maryland
\vspace{0.3cm}

\subsection{Introduction}
Next-generation Weakly Interacting Massive Particle (WIMP) dark matter (DM) detectors will soon encounter the “neutrino fog,” where solar neutrino backgrounds obscure dark matter signals~\cite{ohareNewDefinitionNeutrino2021}. Diamond-based directional detectors offer a potential solution by exploiting the differing angular distributions of neutrinos and WIMPs, allowing background rejection through detailed imaging of nuclear recoil damage tracks paired with conventional diamond-based real-time event detection technology~\cite{Rajendran:2017ynw,Marshall:2020azl,ebadi_directional_2022}. Building upon our contribution to MD$\nu$DM 2024~\cite{Ang2024}, we outline recent progress at the University of Maryland Quantum Technology Center (QTC) towards demonstrating the technical capabilities required for such a large-scale directional DM detector, focusing on three critical areas: experimental validation with ion-implanted damage tracks, particle track simulation, and advances in imaging methodologies, specifically the light sheet quantum diamond microscope (LS-QDM).

\subsection{Experimental studies of ion-induced tracks in nitrogen-rich diamond}
\begin{figure}
   \centering
   \includegraphics[width=1.0\textwidth]{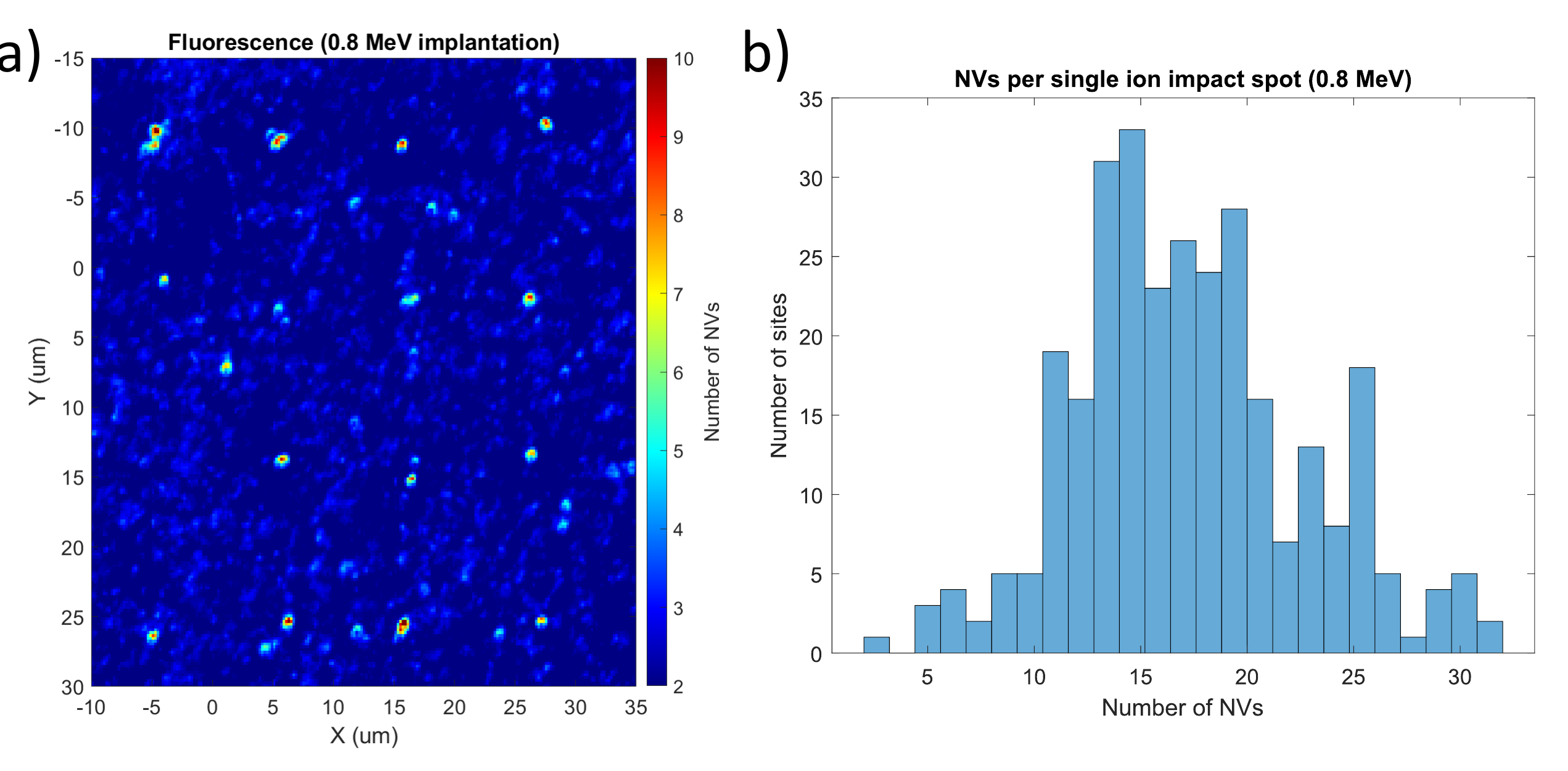}
   \caption{Preliminary confocal scanning results of ion-implanted mm-scale diamond samples. a) Large-scale scan of diamond surface showing a grid of single-ion impact sites from 0.8 MeV implantation with 10 $\mu$m spacing. Finer scans are subsequently performed on identified sites to deduce number of NVs produced per single-ion impact spot. b) Histogram of number of NVs produced from $\sim$300 sites, showing a mean of 17(5) NVs per site.}
   \label{fig:sandiaimplant}
\end{figure}

As a first step toward establishing the detectability of individual nuclear recoil damage tracks, we conduct experimental investigations of single-ion-induced damage tracks in mm-scale diamond samples. Precision implantation of $\sim$1 MeV carbon ions into nitrogen-rich Type 1b HPHT diamond chips ($\sim$200 ppm nitrogen) was carried out at the Sandia National Laboratories ion microbeam facility~\cite{titzeFocusedIonBeam2022}. Implantation was performed at a low flux to ensure production of single ion impact sites. After implantation, samples were annealed at 800°C, inducing the resulting vacancy clusters to pair with nitrogen atoms and form bright nitrogen-vacancy (NV) centers. Confocal microscopy of approximately 300 implantation sites reveal an average of 17(5) NVs per ion impact for 800 keV ion energy, compared to a background of approximately 3 NVs (Fig.~\ref{fig:sandiaimplant}). 

To interpret these observations, we perform Kinetic Monte Carlo (KMC) annealing simulations using the SPPARKS software package~\cite{mitchell2023spparks}, incorporating realistic vacancy diffusion, recombination, clustering, and NV formation~\cite{oliveira2017}. The initial vacancy distributions are provided by SIIMPL, a binary collision approximation (BCA) code to calculate ion damage distributions in crystalline materials~\cite{janson2003hydrogen}. Preliminary simulations indicate that about 5-10\% of the initial $\sim$300 vacancies generated per ion impact are converted into NVs, which is consistent with our experimental results. Detailed simulations incorporating local nitrogen concentrations measured with secondary ion mass spectroscopy (SIMS) are ongoing. Concurrently, we are performing $T_2$ electronic spin coherence measurements on the resulting NV clusters to assess their suitability towards nanoscale imaging with quantum sensing techniques such as Fourier gradient imaging~\cite{zhang_selective_2017,arai_fourier_2015}. 

Ion implantation at lower energies (50-400 keV) is scheduled for Fall 2025 at the University of Michigan. Samples will be prepared using plasma etching procedures developed in our group to minimize pre-existing NV backgrounds and enhance signal-to-noise ratio. Implantation will also be carried out on NV-rich, quantum-grade CVD diamonds to enable track detection via our previously established quantum strain imaging methods~\cite{marshall_high-precision_2022}. 

\subsection{Particle track simulations}
Precise and accurate simulations are essential for understanding the damage tracks produced from neutrino and DM-induced nuclear recoils in diamond. Initial simulations used the widely popular TRIM BCA code~\cite{ZieglerSRIM2010}, which is limited by its neglect of lattice structure effects such as channeling. Recent work employs SIIMPL~\cite{janson2003hydrogen}, a lattice-aware BCA code, to more realistically model track formation including channeling effects. A fully atomistic simulation, however, requires molecular dynamics (MD). Using LAMMPS~\cite{LAMMPS2022}, we perform MD simulations of nuclear recoil events in diamond, incorporating the EDIP interatomic potential~\cite{MarksEDIP2001,MarksEDIP2002} and the ESPNN model for electronic stopping~\cite{Haiek2022espnn}. This work extends earlier simulations by the Marks group~\cite{Buchan2015}. The results agree with TRIM within a factor of two, but demand significant computational resources, limiting the maximum achievable initial recoil energy to about 10 keV.

To overcome these constraints, we have transitioned towards machine learning (ML)-driven MD, using the GPUMD package~\cite{Fan2022GPUMD} with neuroevolution potentials (NEP)~\cite{Fan2021nep} trained on DFT data~\cite{DerringerCsanyi2017carbonMLpotential}. Due to GPU parallelization, this approach enables much faster simulations with improved accuracy from DFT-informed interatomic forces. Preliminary benchmarks indicate a substantial ($\sim$100×) speedup, which would allow simulation of higher energy recoils up to 60-80 keV (limited only by GPU memory). Work is ongoing to refine the NEP potential with newly generated close-range DFT training data, which is required to accurately model nuclear recoil damage cascades. In the future, similar ML-driven methods could be applied to model track production in other materials of interest for mineral detection of dark matter and neutrinos~\cite{Baum:2023cct}.

\begin{figure}[h]
   \centering
   \vspace{0.5cm}
   \includegraphics[width=0.7\textwidth]{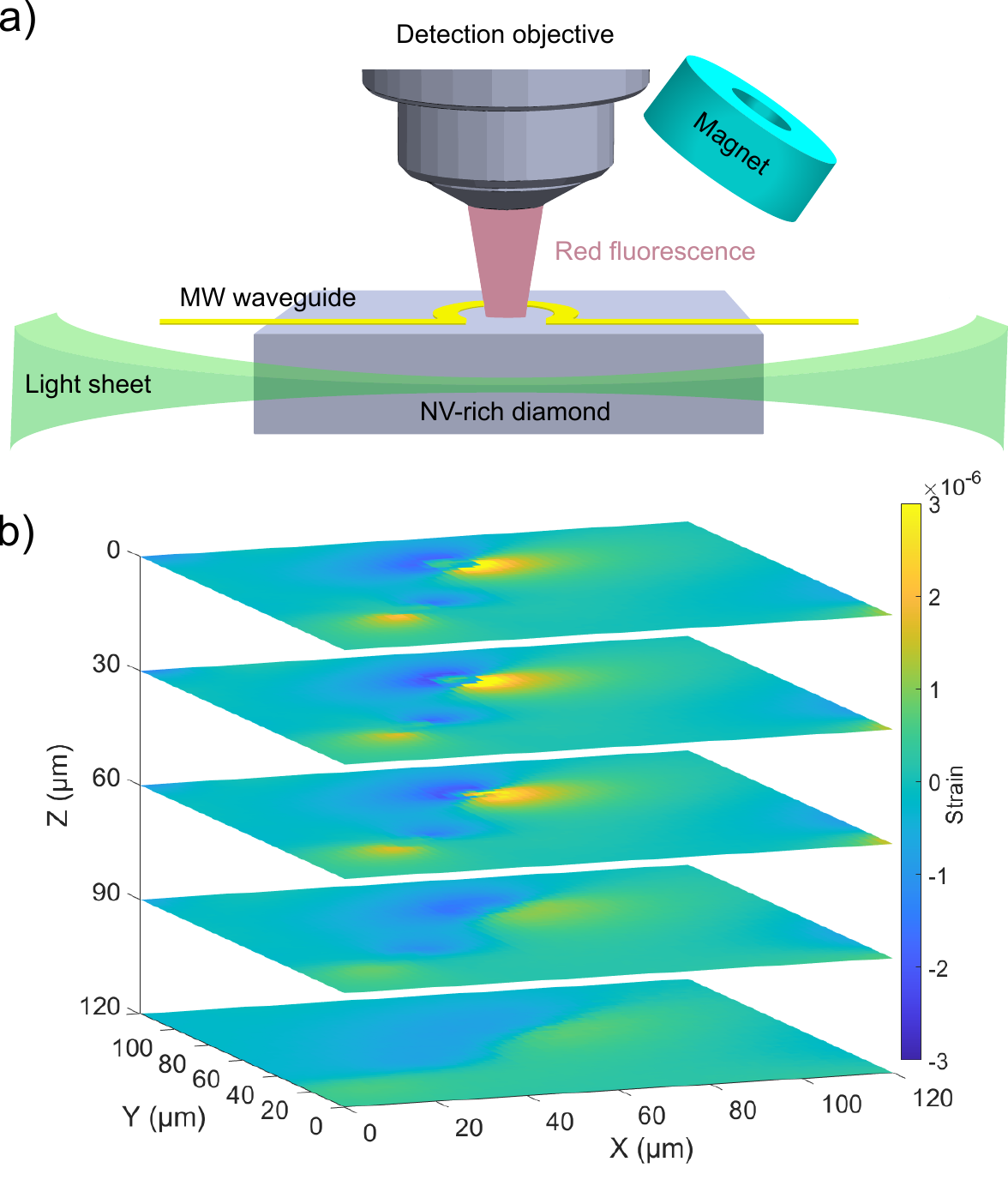}
   \caption{a) Schematic of light-sheet quantum diamond microscope (LS-QDM), which combines light-sheet laser excitation with conventional QDM components (magnets and microwave delivery). b) 3D widefield strain imaging of a quantum-grade diamond using the LS-QDM and the strain-CPMG protocol~\cite{marshall_high-precision_2022}, revealing a large-scale strain feature across multiple Z-stack planes.}
   \label{fig:lsqdm}
\end{figure}

\subsection{New imaging technique: light-sheet quantum diamond microscope}

A full-scale diamond-based directional DM detector will require the capability to perform rapid, high-precision 3D strain imaging of mm-scale diamond chips. Building upon our previous work in NV-based strain sensing~\cite{marshall_high-precision_2022}, we are developing a light-sheet quantum diamond microscope (LS-QDM), which utilizes a sheet of light to excite a thin but wide slice of diamond, inspired by similar setups used for biological imaging~\cite{Vladimirov_MesoSPIM2023} (Fig.~\ref{fig:lsqdm}a). Our current setup has a Gaussian light sheet with 9 $\mu$m thickness and $\sim$100 $\mu$m$^2$ field of view. Preliminary measurements with the strain-CPMG quantum interferometric protocol demonstrate the capability of the LS-QDM to perform widefield 3D-imaging of natural strain features in a mm-scale diamond chip (Fig.~\ref{fig:lsqdm}b). Ongoing engineering efforts focus on delivering uniform microwave fields across the entire sensing region of the diamond; and compensating for displacement of the light sheet induced by the high refractive index of diamond, which leads to systematic spatial shifts during volumetric scanning. Future LS-QDM iterations will utilize non-Gaussian beams to achieve $\sim$1 $\mu$m light sheet thicknesses, required for timely DM-induced track detection~\cite{ebadi_directional_2022}. 

In addition, a second confocal setup is being built for development of super-resolution quantum sensing techniques such as spin-RESOLFT~\cite{jaskula_superresolution_spinresolft_2017} and Fourier gradient imaging~\cite{zhang_selective_2017,arai_fourier_2015,guo_wide-field_2024}, which are required to perform detailed nanoscale NV strain imaging of the shape and orientation of a damage track once it is located with the LS-QDM. To achieve Fourier gradient imaging at depths beyond 10 $\mu$m from the diamond surface (required to resolve tracks deeply embedded in mm-scale chips), we are collaborating with the Laboratory of Physical Sciences (LPS) at UMD to design and fabricate microcoil structures capable of delivering 1–10 G/$\mu$m gradients at the required imaging depths.

\subsection{Conclusion and outlook}

In summary, we successfully detected $\sim$1 MeV single nuclear recoil-induced tracks in diamond via precision ion implantation and confocal NV imaging; and we developed a computational model that reproduces our data. Progress in developing ML-driven molecular dynamics simulations will enable even more accurate and precise modeling of DM and neutrino-induced tracks. Finally, our LS-QDM setup exhibits promising preliminary results towards widefield, fully 3D strain scanning of mm-scale diamond chips to enable particle track mapping. Overall, these developments illustrate substantial progress in key sectors towards diamond-based directional dark matter detection. 

\acknowledgments
This work is supported by, or in part by, the DOE QuANTISED program under Award No. DE-SC0019396; the U.S. Army Research Laboratory under Contract No. W911NF2420143; the LPS/QTC Jumping Electron Fellowship through award H9823022C0029; and the University of Maryland Quantum Technology Center.

\clearpage

\section{Preliminary ICP-MS Analysis of Uranium and Thorium in Olivine and Muscovite}

Authors: {\it Yukiko~Kozaka$^{1,2}$, Takenori~Kato$^{1*}$\note[*]{Corresponding author}, Katsuyoshi~Michibayashi$^3$, Yui~Kouketsu$^3$, and Yoshihiro Asahara$^3$}
\vspace{0.1cm}\\
$^1$Institute~for~Space--Earth~Environmental~Research, Nagoya~University\\
$^2$Faculty~of~Geosciences~and~Civil~Engineering, Institute~of~Science~and~Engineering, Kanazawa~University\\
$^3$Department~of~Earth~and~Planetary~Sciences, Graduate~School~of~Environmental~Studies, Nagoya~University
\vspace{0.3cm}

\subsection{Introduction}

The paleo-detector concept proposes the use of ancient minerals as long-exposure detectors for dark matter (DM) interactions, potentially extending sensitivity beyond existing techniques. The idea originated in early studies that explored track preservation in natural minerals, with muscovite mica used in searches for massive monopoles~\cite{physrevlett56} and WIMP-type dark matter~\cite{Snowden-Ifft:1995zgn} during the 1980s and 1990s. These studies demonstrated the feasibility of solid-state track detection over geological timescales.

More recently, theoretical proposals have revitalized the paleo-detector concept by evaluating a broader range of candidate minerals, such as olivine, rutile, and apatite, based on track retention capabilities and background rejection~\cite{Baum:2021jak,Drukier:2018pdy}. Notably, these proposals emphasized olivine’s potential due to its ubiquity in the mantle, low porosity, and mechanical durability. These renewed efforts have prompted experimental work to assess the radiopurity and background characteristics of candidate minerals.

A critical parameter for evaluating the suitability of paleo-detectors is the concentration of naturally occurring radioactive nuclides, such as uranium (U) and thorium (Th), in the mineral matrix. These nuclides generate nuclear recoil backgrounds through alpha decay and the recoil of daughter products. Accurate quantification is essential for estimating background track densities and signal-to-noise ratios.

Haines and Zartman~\cite{Haines1973} analyzed peridotite inclusions of probable mantle origin from the Adirondack Mountains and found uranium concentrations in olivine ranging from 3 to 200~ng/g. These results provide further evidence for the variability and generally low concentrations of uranium in mantle olivine.

De Hoog et al.~\cite{DeHoog2010} reported uranium concentrations in mantle olivine ranging from sub-ng/g to approximately 30~ng/g, indicating high variability depending on geological context. For crustal muscovite, Carignan et al.~\cite{Carignan1996} reported uranium concentrations ranging from 60 to 300~ng/g and thorium concentrations between 50 and 500~ng/g, illustrating the elevated trace element content in crustal micas.

Despite these advances, systematic solution-based analysis of U and Th in minerals relevant to paleo-detectors has been limited. In this work, we present preliminary results from solution ICP-MS analysis of olivine and muscovite samples, representing mantle-derived and crustal environments, respectively. Our data provide a basis for evaluating their suitability for low-background dark matter detection.

\subsection{Samples and analytical procedure}

The olivine samples include four types: peridotite from the Tonga Trench, peridotite from the Mariana Trench, a peridotite xenolith in basalt from Damaping, China, and olivine from a pallasite meteorite. The muscovite sample was derived from a pegmatite from Minas Gerais, Brazil. 

The Tonga and Mariana peridotites were crushed using a hammer and clean olivine grains without inclusions were handpicked under a binocular microscope. The xenolith and pallasite samples consisted of separated olivine grains without further treatment. Muscovite was peeled along its cleavage plane into thin flakes.

All mineral samples were then subjected to the same cleaning and dissolution procedure. Approximately 30--100~mg of each sample was washed in an ultrasonic cleaner with Milli-Q water and methanol to remove surface contamination. After three rinses with Milli-Q water, samples were refluxed with 1~mL of 70\% perchloric acid and 0.5~mL of 38\% hydrofluoric acid (TAMAPURE-AA-100) on a 180~\textdegree{}C hotplate for three days. The solutions were evaporated to dryness, then redissolved in 1~mL of 70\% perchloric acid and evaporated again to remove remaining HF. Finally, samples were dissolved in 2\% nitric acid.

Trace element concentrations were measured using an Agilent 7700x ICP-MS at Nagoya University. The calibration standard was the SPEX multi-element solution XSTC-331 (SPEX CertiPrep Inc., USA), and In was used as the internal standard. In addition to the mineral samples, we measured uranium and thorium concentrations in the GSJ geochemical reference sample JB-1. The measured values were consistent with the recommended values reported by Imai et al.~\cite{Imai1995} within analytical uncertainty.

\subsection{Results}

Table~\ref{tab:results} summarizes the preliminary U and Th concentrations measured in the five olivine and one muscovite samples. The Tonga and Mariana Trench olivine exhibit elevated U concentrations (20.9 and 38.2~ppb, respectively) with relatively low Th levels. The Damaping xenolith sample shows lower U but slightly higher Th compared to Mariana. The pallasite olivine sample exhibits extremely low U and Th. The muscovite sample shows moderate U and Th values consistent with a crustal environment.

\begin{table}[htbp]
\centering
\caption{Preliminary ICP-MS results for U and Th concentrations in mineral samples.}
\label{tab:results}
\begin{tabular}{lcc}
\hline
Sample & U (ppb) & Th (ppb) \\
\hline
Tonga Trench olivine & 20.9 & 0.41 \\
Mariana Trench olivine & 38.2 & 0.12 \\
Damaping olivine xenolith & 0.53 & 0.73 \\
Pallasite olivine & 0.09 & 0.20 \\
Minas Gerais muscovite & 1.8 & 3.5 \\
\hline
\end{tabular}
\end{table}

\subsection{Discussion}

The U and Th concentrations in mantle-derived olivine show variation over two orders of magnitude, with values ranging from sub-ng/g levels in extraterrestrial olivine (pallasite) to tens of ng/g in subduction zone peridotites. De Hoog et al.~\cite{DeHoog2010} and Haines and Zartman~\cite{Haines1973} reported similar levels of variability, consistent with our findings from the Tonga and Mariana peridotites.

The xenolith and pallasite samples exhibit lower concentrations, consistent with earlier expectations for relatively undisturbed olivine.

The muscovite sample showed U and Th concentrations of 1.8 and 3.5~ng/g, respectively, significantly lower than the tens to hundreds of ng/g reported by Carignan et al.~\cite{Carignan1996}, possibly reflecting variability among crustal muscovite sources.

These results confirm that olivine and muscovite can contain trace amounts of U and Th at levels relevant for background estimation in paleo-detector studies. The relatively low concentrations observed in pallasite olivine and our muscovite sample suggest they are promising candidates for low-background dark matter detection.

\acknowledgments
This work was supported by the Japan Society for the Promotion of Science (JSPS) KAKENHI Grant 20K20944 and by the Director's Leadership Fund of the Institute for Space-Earth Environmental Research (ISEE), Nagoya University.


\clearpage

\section{Surrogate Sample Preparation with Neutron Irradiation}

Authors: {\it Igor~Jovanovic, Valentin~Fondement, and Ethan~Todd}
\vspace{0.1cm} \\
University of Michigan
\vspace{0.3cm}

\subsection{Motivation}

\begin{figure}[h]
   \centering
   \includegraphics[width=0.8\textwidth]{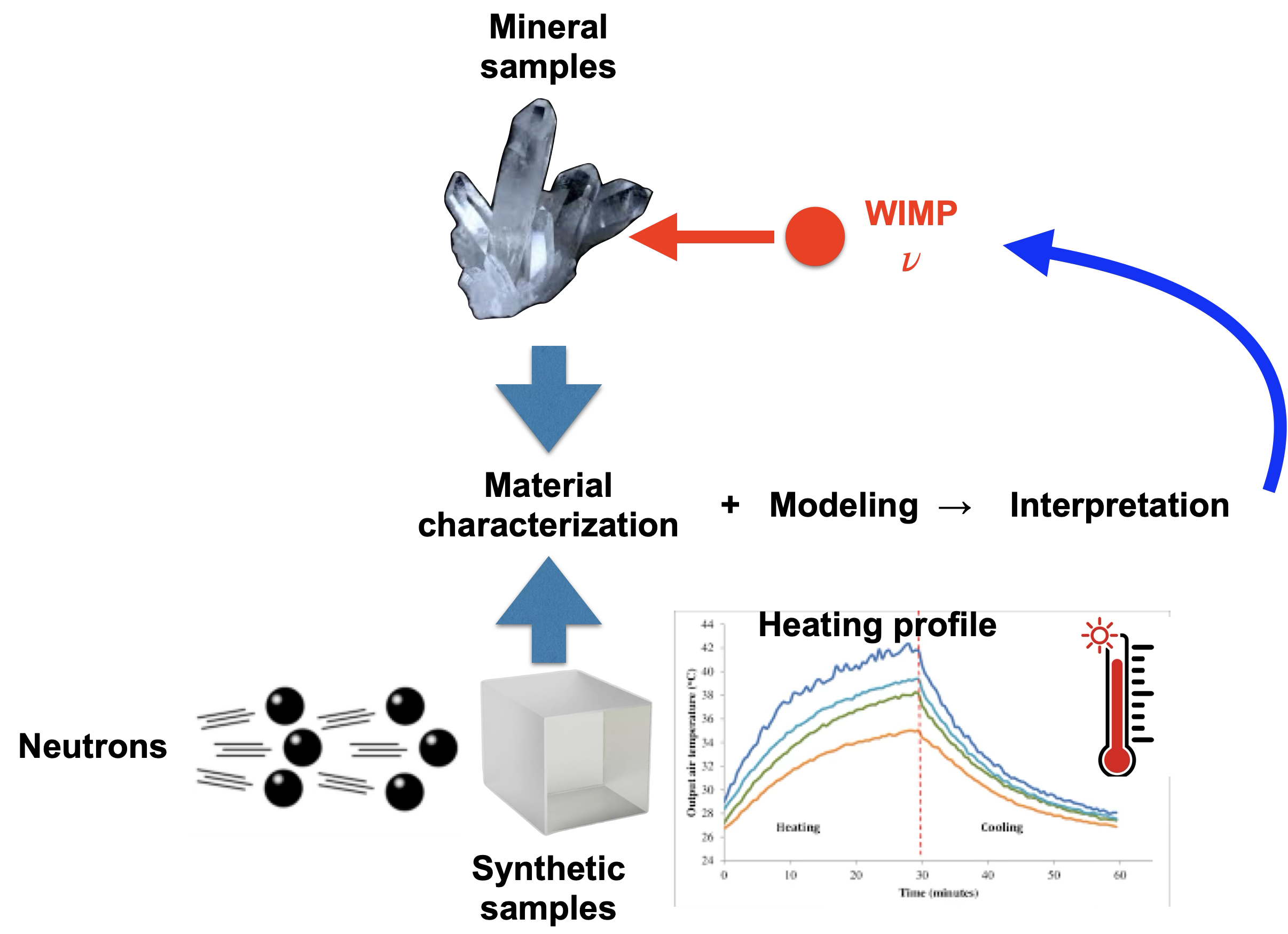}
   \caption{Illustration of the process used to develop surrogate calibration standards.}
   \label{fig:Surrogates}
\end{figure}

A key part of the effort to detect Dark Matter weakly interacting massive particles (WIMPs) and neutrinos in minerals is the development of experimental benchmarks, calibration schemes, and accompanying simulations for surrogate samples (Fig.~\ref{fig:Surrogates}). Controlled neutron irradiation can provide a well-understood energy scale, including the spectrum and spatial distribution of nuclear recoils by neutron scattering and capture. We have been developing a simulation framework and experimental infrastructure to perform such precision calibrations. Calibrations require knowledge of the neutron spectrum, flux, and directionality, which we achieve using well-characterized sources and simulations of the irradiation environment. In addition, we seek to incorporate temperature control (heating and/or cooling) at various stages of sample preparation to evaluate its effect on defect formation and evolution.

In preliminary studies discussed here, we consider the simple case of irradiating a 1 cm$^3$ sample of LiF(nat), consisting of $^{19}$F, $^6$Li (7.5 a/o of Li), and $^7$Li (92.5 a/o of Li) placed at a practical distance of 10~cm from a radiation source. We use Geant4 simulations to estimate the recoil energy range of [0,5~keV], corresponding to WIMP scatters with a mass of up to $\sim$5~GeV/c$^2$. 

\subsection{Calibration with neutron generators}

\begin{figure}[h]
   \centering
   \includegraphics[width=1\textwidth]{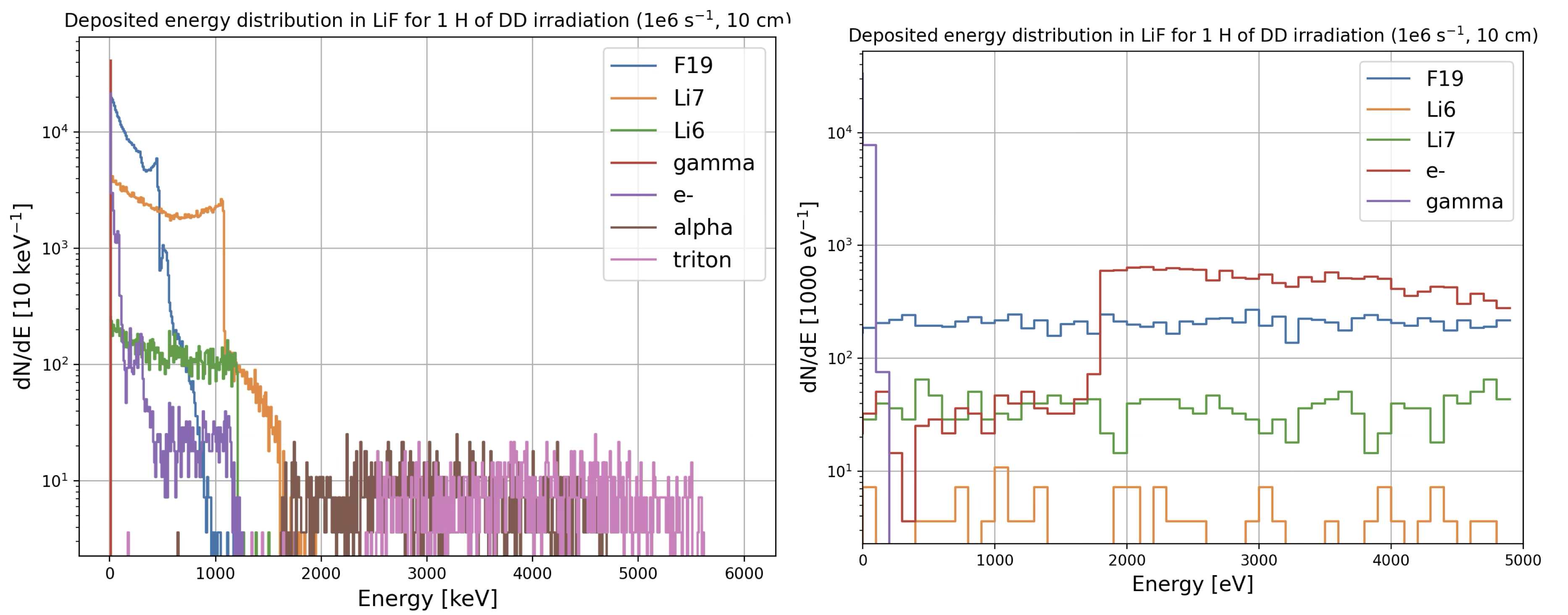}
   \caption{Nuclear recoil energy spectrum for irradiation with 2.45~MeV neutrons from a DD generator.}
   \label{fig:DD}
\end{figure}

\begin{figure}[h]
   \centering
   \includegraphics[width=1\textwidth]{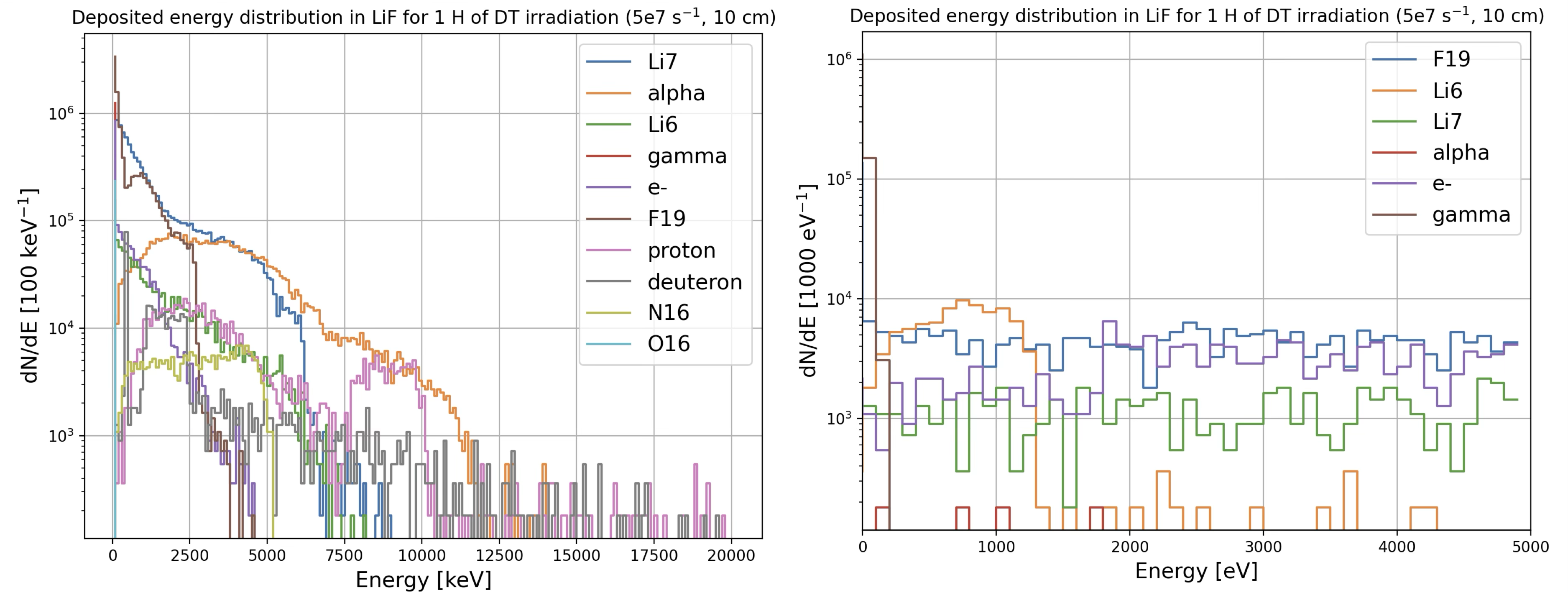}
   \caption{Nuclear recoil energy spectrum for irradiation with 14.1~MeV neutrons from a DT generator.}
   \label{fig:DT}
\end{figure}

The available neutron generators at the University of Michigan have a typical output of 10$^6$~n/s for DD (2.45~MeV) and 5.5$\times$10$^7$~n/s for DT (14.1~MeV). In Figs.~\ref{fig:DD} and~\ref{fig:DT} we show the simulated recoil spectra for those two sources, respectively. Notable is the presence of sharp cutoffs in the DD spectra, which are not as prominent with DT neutrons. The corresponding estimated nuclear rates in the [0,5~keV] energy range are $\sim$10$^3$ and $\sim$3$\times$10$^4$~s$^{-1}$. Despite the considerably higher recoil rate with the DT source, surrogate sample calibrations and understanding of various readout techniques will likely benefit more from DD irradiations because of the prominence of spectral features.

\subsection{Calibration with the \texorpdfstring{$^7$Li(p,n)}{7Li(p,n)} reaction}

\begin{figure}[h]
   \centering
   \includegraphics[width=0.5\textwidth]{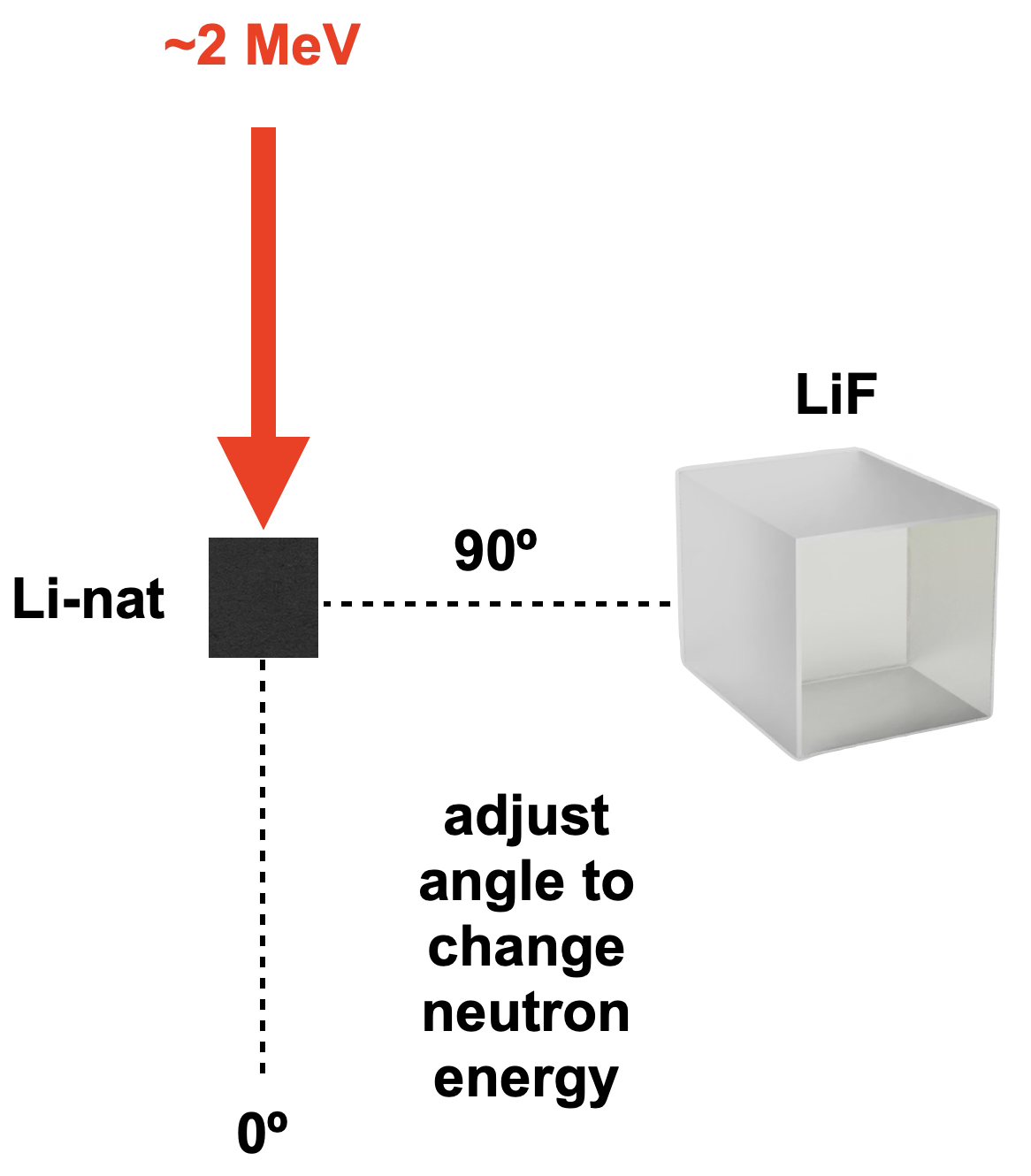}
   \caption{A scheme for production of $\sim$20~keV neutrons using the $^7$Li(p,n) reaction.}
   \label{fig:LiF_Scheme}
\end{figure}

\begin{figure}[h]
   \centering
   \includegraphics[width=1\textwidth]{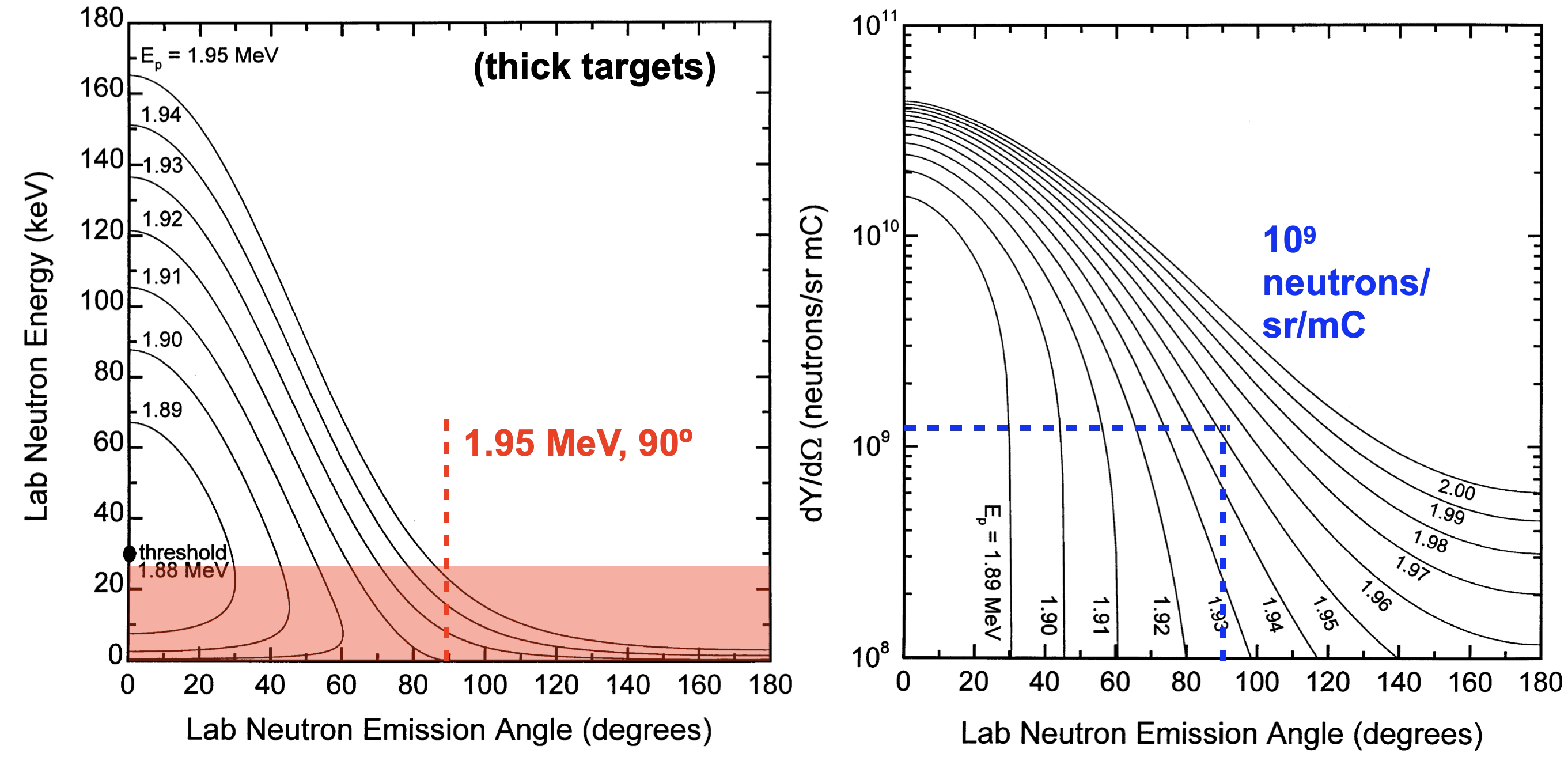}
   \caption{Neutron production characteristics in the $^7$Li(p,n) reaction~\cite{lee1999thick}.}
   \label{fig:Li_Production}
\end{figure}

We have also considered calibrations that employ the $^7$Li(p,n) reaction, which can be kinematically tuned to produce neutrons in the suitable energy range. For example, a $\sim$2~MeV proton beam (Fig.~\ref{fig:LiF_Scheme}) produces neutrons in the $\sim$20~keV range at an angle of 90$^\circ$ (Fig.~\ref{fig:Li_Production}).

We use the published data on thick-target yields for this reaction~\cite{lee1999thick} to estimate the rate of nuclear recoils $\sim$7$\times$10$^6$~hr$^{-1}$ in the [0,5~keV] range in this scheme using a 50~\textmu A proton beam. Again, this assumes that a 1~cm$^3$ LiF crystal is placed 10~cm from the source.

\subsection{Summary}

Neutron radiation is a versatile tool for surrogate sample preparation to be used in mineral detection of neutrinos and Dark Matter. Through the use of neutron sources with various spectra, the process of defect formation can be better understood, and the performance of various readout technologies under development can be evaluated. We continue to refine the methods for surrogate source preparation using accelerator-based sources discussed here, as well as radioisotope sources and nuclear reactors we can routinely access.

\acknowledgments
This work was supported by the National Science Foundation Growing Convergence Research award 242850.
\clearpage

\section{First-Principles Screening of Mineral Candidates for Dark Matter Detection with the PALEOCCENE Technique}

Authors: {\it Mariano~Guerrero~Perez$^1$, Keegan~Walkup$^1$, Jordan~Chapman$^2$, Pranshu~Bhaumik$^4$, Giti~A.~Khodaparast$^1$, Brenden~A.~Magill$^1$, Patrick~Huber$^1$, and Vsevolod~Ivanov$^{1,2,3}$}
\vspace{0.1cm} \\
$^1$Department~of~Physics,~Virginia~Tech,~Blacksburg,~Virginia~24061,~USA\\
$^2$Virginia~Tech~National~Security~Institute,~Blacksburg,~Virginia~24060,~USA\\
$^3$Virginia~Tech~Center~for~Quantum~Information~Science~and~Engineering,~Blacksburg,~Virginia~24061,~USA\\
$^4$College~of~William~and~Mary,~Williamsburg,~VA~23187,~USA
\vspace{0.3cm}

At the lowest threshold for particle detection using the PALEOCCENE technique, nuclear recoils may result in only small numbers of single vacancy or interstitial defects. When such defects are optically active \emph{color centers}, they can be used to image damage tracks down to atomic-scale resolution~\cite{araujo2025nuclear}. First principles calculations are an effective approach for screening large numbers of minerals based on their propensity to form color centers from nuclear recoils. We have recently applied a methodology for calculating the electronic and optical properties of color centers, using a properly tuned hybrid functional. Tuning the hybrid functional to simultaneously satisfy the generalized Koopman's theorem and reproduce the experimental band gap leads to the cancellation of two dominant error contribution and enables the precise prediction of emission wavelengths and formation energies. We demonstrate the accuracy of this method by comparing with experimental measurements of color center defects in lithium fluoride~\cite{perez2025firstprinciples}. Moreover, we show that for most LiF defects, reproducing the experimental bandgap is a sufficient condition for predicting absorption/emission frequencies, since they do not vary significantly with the choice of functional.

The reasonable agreement between first-principles calculations and experimental measurements of LiF defect optical properties motivates the predictive screening of materials for particle detection. Taking into account natural abundance at depths below 1km, low radioactive nuclide content, and availability of synthetic samples for testing, diamond, olivine (Mg/FeSiO$_4$), zircon (ZrSiO$_4$), scheelite (CaWO$_4$), and corundum (Al$_2$O$_3$) have previously been proposed. Color center defects in diamond have already been extensively studied, while the other materials are less explored. Using the HSE06 functional (which reproduces the bandgaps of the aforementioned compounds), we avoid the full Koopman's tuning procedure and preform a preliminary screening of simple vacancies and self-interstitial defects in these materials. Using the computed transition dipole moment as the metric for emission brightness, and the Kohn-Sham energy difference as a stand-in for emission energy, aim to identify in which materials such simple defects would result in bright emitters with emission wavelength within or near the visible band (400nm-700nm). Out of all the candidate materials, we find the largest number of bright \& visible simple defects in ZrSiO$_4$ and MgSiO$_4$, identifying these two minerals as more promising for PALEOCCENE from the perspective of bright color-center formation.

\acknowledgments
This work has been supported through the U.S.\ National Science Foundation Growing Convergence Research Grant OIA-2428507, titled ``Collaborative Research: GCR: Mineral Detection of Dark Matter'' and by the National Nuclear Security Administration Office of Defense Nuclear Nonproliferation R\&D through the ``Consortium for Monitoring, Technology and Verification'' under award number DE-NA0003920.
\clearpage


\section{Queen's Paleo Group Mica Studies}\label{sec:Queens}

Authors: {\it Joseph~Bramante$^1$, Matthew~Leybourne$^2$, and Alexis Willson$^{1,2}$}
\vspace{0.1cm} \\
$^1$Department~of Physics, Engineering Physics, and Astronomy, Queen's University, Kingston, Canada\\
$^2$Department~of Geological Sciences and Geological Engineering, Queen's University, Kingston, Canada 
\vspace{0.3cm}

\subsection{Muscovite mica for particle detection}
Muscovite mica is a mineral of ideal structure for particle detection. Its perfect basal cleavage results in thin parallel sheets easily separated, imaged, and reconstructed to display damage tracks from particle interactions with the lattice. These interactions form point defects, imperfections around a lattice point that can be etched to a visible scale under the microscope. Recently the Queen's Paleo group has begun developing etch and microscopy protocols to study how ancient muscovite mica can be used as a dark matter detector. We hope to eventually explore dark matter models with masses ranging from $10^{10}$ to $10^{28}$ GeV, including strongly interacting composite dark matter and monopoles. Similar work was undertaken by Price \& Salamon in 1986~\cite{physrevlett56}, whose primary focus was monopole searches, and who achieved a reported nuclear energy deposition sensitivity limit for as small as $\sim 7$ GeV/cm in muscovite, with a microscopy depth sensitivity of $\sim 0.1$ $\mu$m and using a hydrofluoric (HF) acid etching duration of 48 h. There is also more recent work on these techniques from Snowden-Ifft et al.~\cite{Snowden-Ifft:1995zgn} and the DMICA group~\cite{baum:2024eyr}. We have begun to re-examine the Price and Salamon experimental procedure and to investigate methods to improve on their sensitivity. We are currently pursuing three avenues for augmenting sensitivity; increasing etching duration, introducing scanning electron microscopy (SEM) to achieve increased resolution at the few nm level, and investigating sharpened 2.5D images through polarized light microscopy (PLM).  

\subsection{Etching}
The crystal lattice of muscovite is arranged in a stacked tetrahedral-octahedral-tetrahedral-cation (TOT-c) pattern of elements K, Al, Si, O, H, F, and naturally occurring impurities. Hydrofluoric acid is a strong vessel of etching, attacking the elemental bonds in the muscovite lattice, weakened by point defects induced by incoming particles, or radioisotope decay within the mineral. This induces further structure weaknesses, creating tunnels or 3D tracks along the track of a past particle, visible as rounded to angular circular holes on the surface of a single mica sheet. Based on the work of Price \& Salamon, etching a muscovite mica sheet in 40\% HF results in visibility for low energy alpha particles associated with radiogenic U and Th decays. It may be possible to increase the etching time to achieve sensitivity to lower energy particles.

\subsection{Scanning electron microscopy}
As part of our research we are also looking to better characterize etched muscovite at sub-micron scales. In recent years, SEM has become increasingly employed for microscopic mineral and material studies. Chosen for its nm level resolution, gaseous SEM (GSEM) and environmental SEM (ESEM) has been employed to image particle track features pre- and post-etching. Settings were predominantly set to 1.5 torr in the pressure chamber and 10 kV of energy. In the gaseous SEM chamber, an electron beam is directed at the muscovite sheet sample, and the secondary electrons, backscattered electrons, and X-rays are collected to image high resolution surficial features. Damage holes, feature size, collective density, and 2D geometry indicate the responsible particle of the track. After HF etching, these features get increasingly visible for effective proposed detection. SEM allows for reorientation of the sample and the ability to catch all side surfaces of the sample. This feature can be used for 3D reconstruction by volumetric rendering, and track depth calculations. Limitations of SEM include the analytical restriction to surficial features only.

\begin{figure}[h!]
   \centering
   \includegraphics[width=0.49\textwidth]{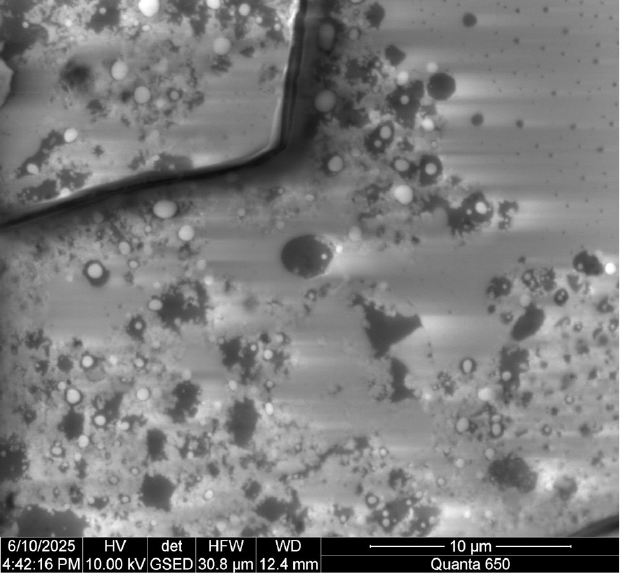}
      \includegraphics[width=0.5\textwidth]{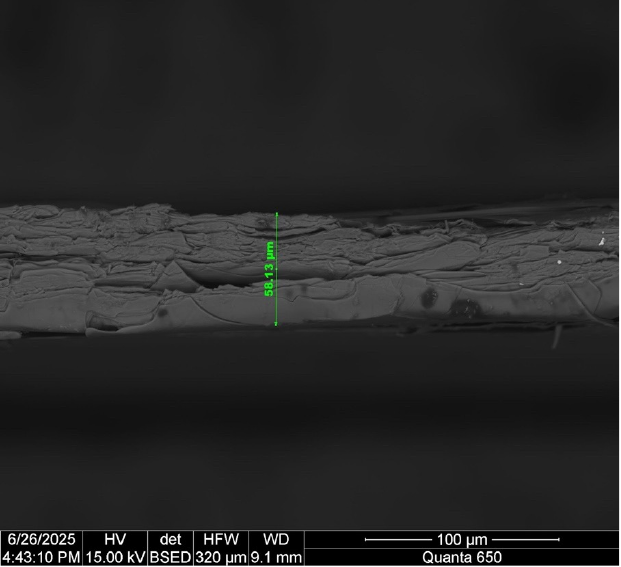}
   \caption{Left: SEM micron-scale and lower resolution in muscovite samples, enhanced with HF etching. Right: Lateral SEM imaging of a muscovite mica sheet, highlighting $\sim 50$ micron thickness and an array of subsheets.}
   \label{fig:SEM Figure 1}
\end{figure}

Figure~\ref{fig:SEM Figure 1}(Left) displays SEM ability to image at the micron level and below. Holes, precipitate, and a notable fracture line is the product of 40\% HF etching for 48 h at 25℃. Figure~\ref{fig:SEM Figure 1}(Right) highlights the SEM ability to observe a muscovite sheet at any desired orientation, including laterally to roughly measure sheet thickness.

\subsection{Polarized light microscopy}
\begin{figure}[h!]
   \centering
   \includegraphics[width=0.5\textwidth]{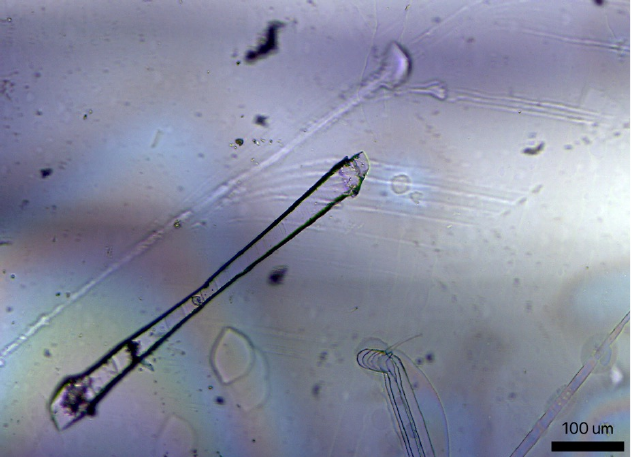}
      \includegraphics[width=0.48\textwidth]{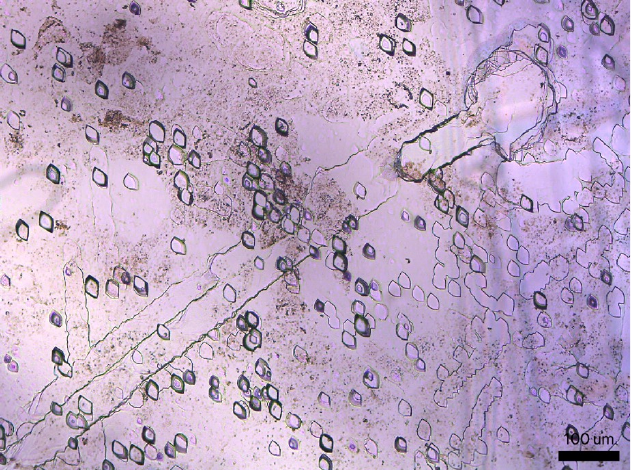}
   \caption{Left: PLM image showing micron level structures resembling tracks after the previously established 48h HF exposure duration. Right: PLM image of surficial features on 48 h HF etched muscovite mica, including an abundance of small etched holes and a tunnel geometry feature under analysis.}
   \label{fig:PLM Figure 1}
\end{figure}

PLM has two main modes, plane polarized light (PPL) for main feature observation of tracks in muscovite samples, and cross-polarized light (XPL) for subsheet differentiation and identification of foreign minerals and impurities. The study has started with PPL analytics under a black tinted thin section for contrast against the white muscovite sheet. Such a method is practiced predominantly for its ability to see within the sample, rather than strictly surficial features. This allows for sub-3D geometry, or 2.5D imagery to aid greatly in identification of the origin particle and behaviour of the particle as it interacted with the lattice. Fission tracks, alpha showers from radioactive sample Am-241, electron scattering, and more are in the process of being identified and analyzed. Limitations of PLM include lower than SEM resolution, which is less effective in the study’s goal to push the boundary of detection sensitivity. Nevertheless, PLM can observe sharp features on the micron level shown in Fig.~\ref{fig:PLM Figure 1}.

Figure~\ref{fig:PLM Figure 1} displays PLM imaging at 100 $\mu$m, such a large scale already being useful for feature observation at the same HF etching conditions. 
Our near-term objectives are as follows: after better establishing PLM microscopy, we will be turning to direct calibration of the mica samples. Specifically, we plan to use Am-241 to test sensitivity to low energy alpha tracks (reported in~\cite{physrevlett56}) and their differentiability from other tracks. We also plan to establish U/Th concentration and isotope levels using laser ablation and confirm this with direct fission track identification and differentiation.

\acknowledgments
We thank Colin Aldis, Agatha Dobosz, Steve Gillen, and Emma Scanlan for many helpful discussions, engineering, and scientific support.

\clearpage

\section{Experimental \& Simulation Studies for Mineral Detectors at KIT}\label{sec:KIT_IAP}

Authors: {\it Lukas~Scherne and Alexey~Elykov}
\vspace{0.1cm} \\
Karlsruhe Institute of Technology
\vspace{0.3cm}

\subsection{Introduction}

Key fundamental questions about the nature of our Universe still remain unanswered and are firmly tied to the properties of Dark Matter and neutrinos. 
The novel research field of mineral detectors aims to uncover the nature of these mysterious particles by using the advent of modern technologies to read out microscopic damage features produced by their nuclear recoils off nuclei of minerals.
Karlsruhe Institute of Technology (KIT) is one of the major scientific research institutes in Europe with a wide array of expertise in scientific areas ranging from geology and material sciences to particle and astroparticle physics. 
Ongoing experimental and simulation studies at the Institute for Astroparticle Physics (IAP) at KIT, in collaboration with geologists from Heidelberg University and microscopy experts from the Institute of Nanotechnology (INT) and the Laboratory for Electron Microscopy (LEM) at KIT aim to tackle key challenges associated with the realisation of the concept of mineral detectors. 

\subsection{Simulation Studies}
\label{sec:simulaitons}

To support our efforts for imaging particle-induced damage tracks in various materials, a series of GEANT4 and SRIM/TRIM simulations were performed, providing guidance for imaging studies and interpretation of observations.  

\subsubsection*{Simulation of Neutron/Alpha-induced damage in LiF with TRIM and GEANT4}
\label{sec:simulation_of_neutron_alpha_induced_damage_in_lif}

The natural lithium isotope ${}^6\text{Li}$ has a high cross section for the thermal neutron-induced process $\text{n} + {}^6\text{Li} \rightarrow {}^3\text{H} + \text{$\alpha$}$. 
This process produces a clear fission signal, since the two decay products, an alpha particle and a triton ion, have well-defined energies and therefore well-defined ranges. 
This process and the transparent nature of LiF makes it a suitable mineral for investigations with light-sheet microscopy; for more details see the efforts of the PALEOCCENE collaboration~\cite{Alfonso:2022meh}.
\par
The simulation of possible colour centers from neutron irradiation was performed using a two-step approach. 
First, neutrons were simulated in GEANT4 as primary particles, used to irradiate a LiF cube. 
All ions produced inside the LiF cube (secondary particles), regardless of their production channel, were tracked and some details needed for the second step in the simulation process were saved. 
To simulate the damage produced by recoils and fission particles in more detail, the characteristics of the produced secondary particles were written into a customizable input file for the TRIM simulation.
The crucial TRIM inputs are the position at which the particles were produced, their kinetic energy, mass, and flight direction.
To obtain the most detailed information about the vacancy distribution in the material, TRIM was run in full cascade mode with monolayer collision steps.
In this setting, every collision, either by the primary ion, or the cascade particles, was tracked and saved.
\par
\begin{figure}
    \centering
    \includegraphics[width=1\linewidth]{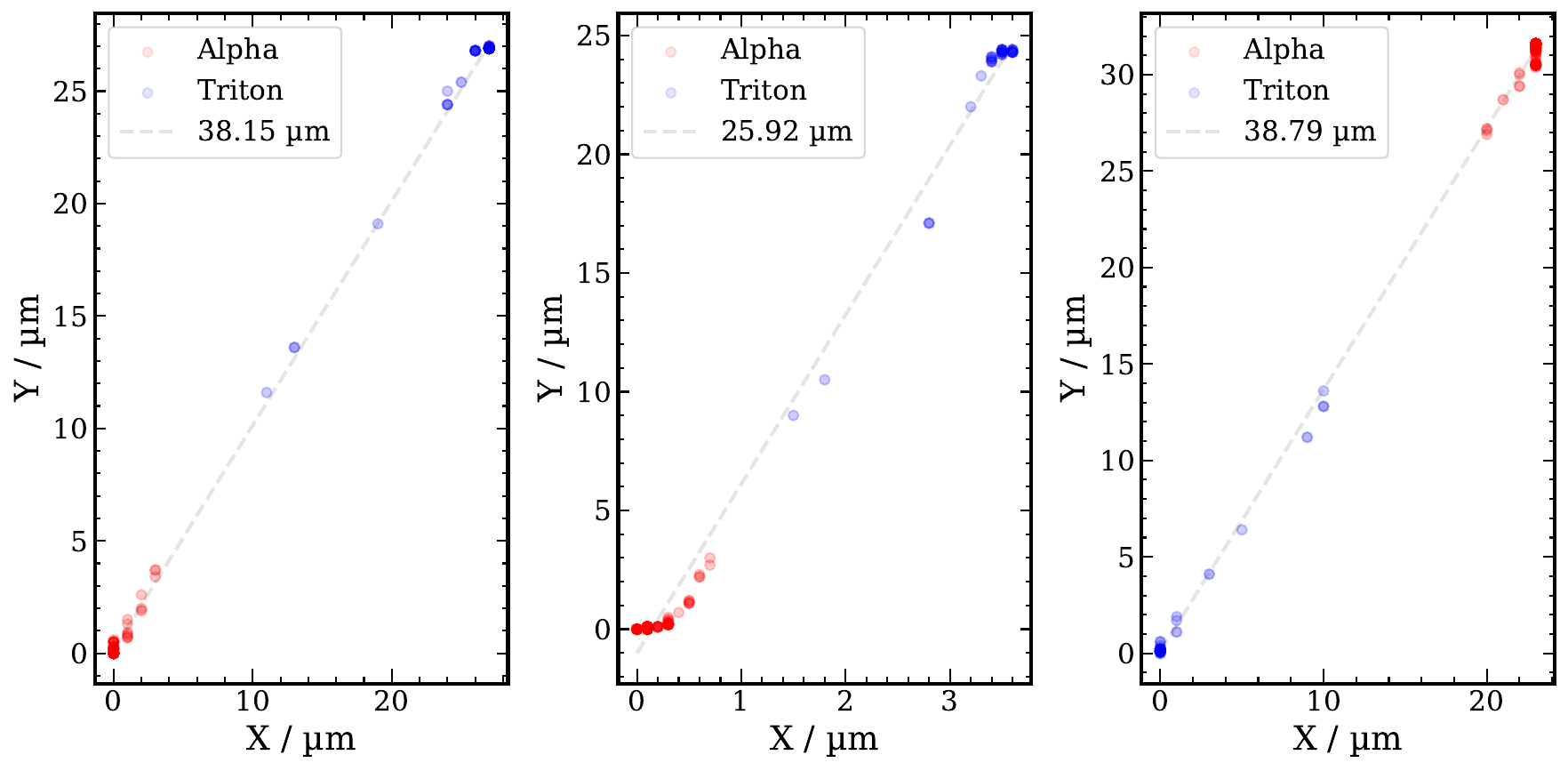}
    \caption{A selection of simulated thermal-neutron induced fission tracks. Marked in blue are the vacancies produced by the triton ion. In red, the vacancies produced by the alpha particle are shown. The intensity of the colour represents the number of vacancies in a cluster. The points are not to scale. Most of the vacancies are located at the end of the tracks.}
    \label{fig:projection_fission_track}
\end{figure}

Since the decay products of thermal-neutron induced fission are of relatively high energy ($E_{\alpha} = \SI{2.05}{MeV}, E_{\mathrm{^3H}}
 = \SI{2.73}{MeV}$), the tracks produced by them will be dominated by a line-like distribution of vacancies along their path through the sample. 
A selection of tracks with different orientations is shown in Fig.~\ref{fig:projection_fission_track}. 
Most of the vacancies are produced at the ends of the tracks---this is due to the fact that the lower the energy of a particle the more of its energy is transferred to the nuclear system of the target. 
Certainly damage will be produced at the beginning of the tracks as well, but it will be due to coupling of the ions to the electronic system of the mineral. 
That does not lead to vacancies, but could lead to amorphization of the target material. 
The amount of damage depends on the properties of the material and cannot be simulated with TRIM. 
    
\par
\begin{figure}
    \centering
    \includegraphics[width=1\linewidth]{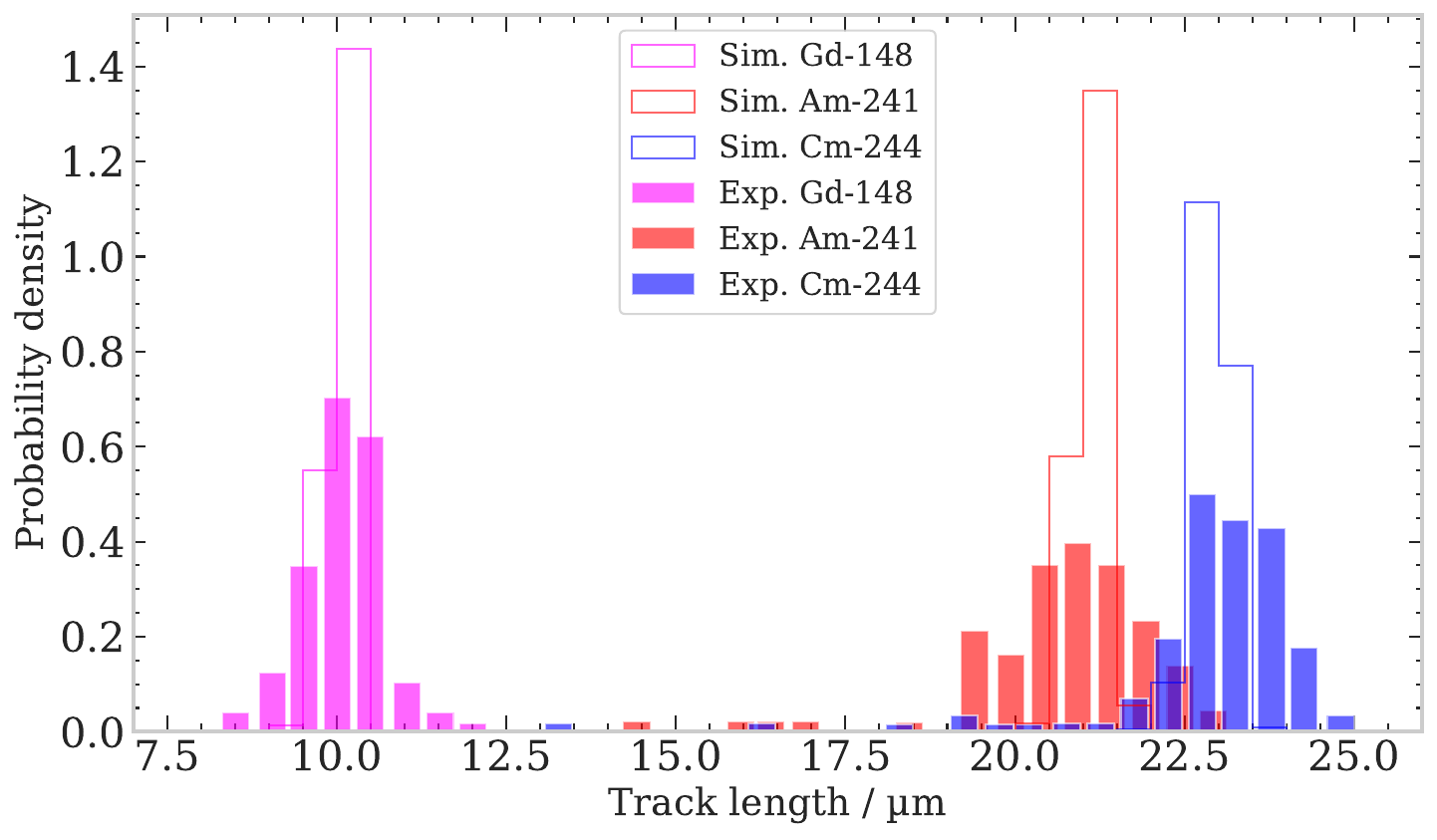}
    \caption{Comparison of simulated and measured track-length spectra from alpha particles of various energies. Solid, unfilled lines showcase simulated data, filled bars represent data that were digitized from~\cite{Bilski:2020}. The mean range values are well described by TRIM, although the distributions differ in shape. The difference in shapes is likely due to experimental uncertainties in the measurement.}
    \label{fig:comparison_alpha_spectra_sim_and_exp}
\end{figure}

To validate our simulation results we compared the resulting track length spectra to experimental data that were obtained in~\cite{Bilski:2020}.
A comparison of the track-length spectra is shown in Fig.~\ref{fig:comparison_alpha_spectra_sim_and_exp}, and the mean ranges are listed in Table \ref{tab:mean_vals_sim_exp}.
For consistency, the calculation of track lengths was performed via the method described in~\cite{Bilski:2020}. 
The track length $R^2$ is given by $R^2 = [r^2+z^2]$, where $r$ is the projection of the track into the plane, and $z$ is the depth of the extent of the damage. 
The mean range of the simulated particles is in good agreement with experimental observations, while the difference in the shape of the distributions could be potentially attributed to experimental effects.
\par
It could be concluded that the use of GEANT4 to simulate neutron irradiation of LiF, complemented with the use of TRIM for the calculation of the vacancies produced by the secondary particles, gives a reliable prediction of the distribution of vacancies in a given material. 
A comparison of simulated and experimentally observed track-length spectra of vacancies has shown a good agreement for particles with MeV-scale energies.
Future work will focus on simulation of vacancies produced by low energy particles.

\subsubsection*{Simulation of $^{132}{\text{Xe}}$ ion-induced damage in Biotite Mica with TRIM}
\label{sec:simulation_of_ion_induced_damage}

\begin{figure}
    \centering
    \includegraphics[width=1\linewidth]{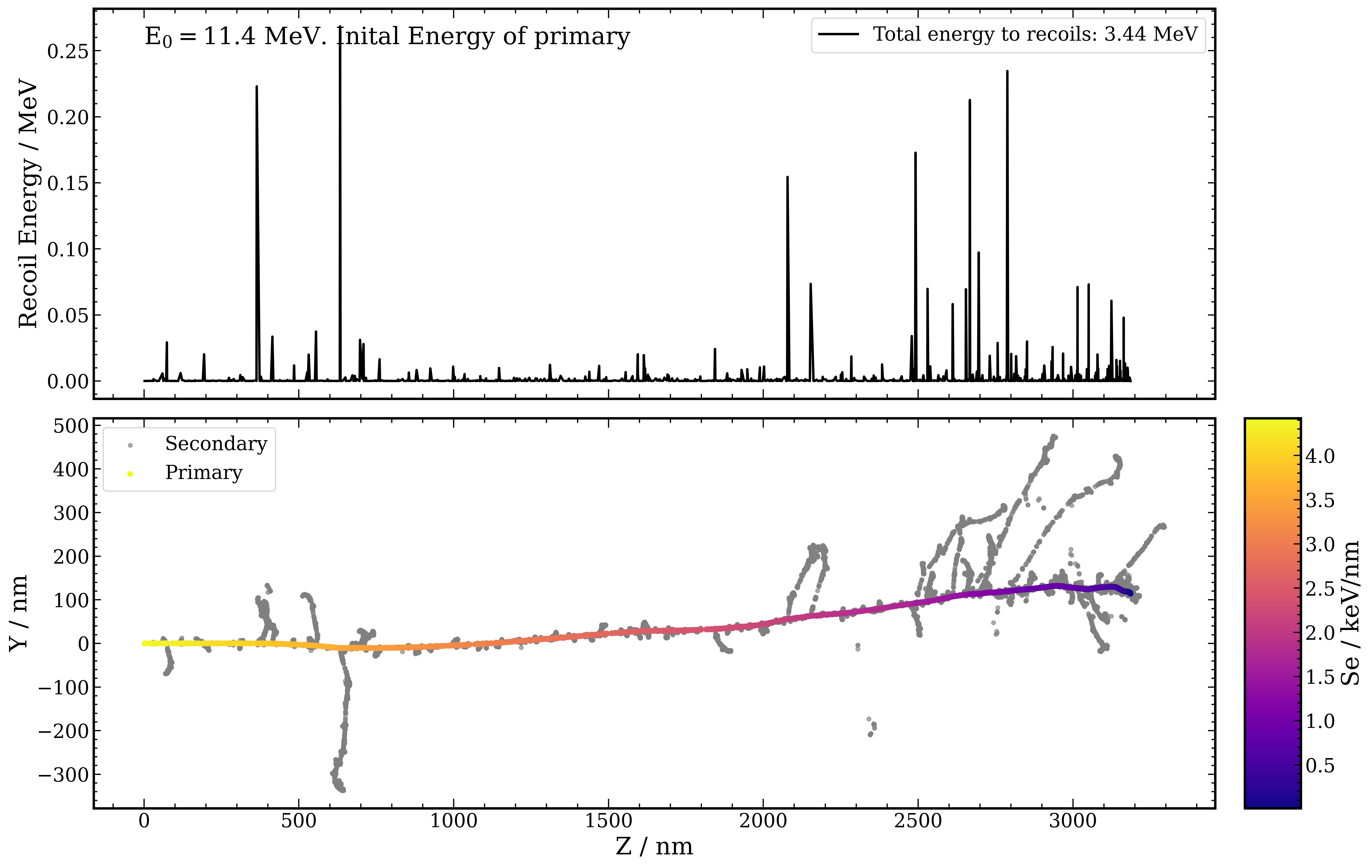}
   \caption{Simulation of a track induced by an $^{132}\text{Xe ion}$ with $E_0 = \SI{11.4}{MeV}$. Bottom: Every dot corresponds to a vacancy - the points are not to scale. The damage region can be estimated to be of about $\SI{3}{\mu m}$ in length and about $\SI{0.6}{\mu m}$ in width, although the core damage region is only about $\SI{0.2}{\mu m}$ in width. Indicated in colour is the electronic-energy loss of the ion. One can observe the expected decrease as the particle travels trough the material. Sub-cascades induced by recoiling atoms are shown in grey. Top: Energy of the recoil cascades. The number of recoil cascades increases with decreasing electronic energy loss. The total energy lost to recoils is $\SI{3.44}{MeV}$, which constitutes approximately $30\%$ of the ions initial energy.}
    \label{fig:11.4MeV_Xe_track_in_Biotite}
\end{figure}

Microscopy imaging of a Biotite Mica sample that was irradiated with $\SI{11.4}{MeV}$ $^{132}\text{Xe}$ ions is currently in preparation. 
To support this study, a detailed simulation of the expected extent of damage induced by these ions in the sample was performed. 
The simulations were performed using TRIM, operated with its most detailed settings.
In Fig.~\ref{fig:11.4MeV_Xe_track_in_Biotite} (bottom) an example of an ion track is shown in the $y$-$z$-projection, where $z$ is the flight direction of the ion. 
The path of the primary ion is plotted in a colour-map that indicates the magnitude of the electronic energy losses of the particle, and recoil-cascades are depicted in grey.
\par

The track is dominated by the primary ion's path up to a depth of $\SI{2}{\mu m}$. 
Beyond this depth, a tree-like structure made of many cascades develops. 
The upper part of Fig.~\ref{fig:11.4MeV_Xe_track_in_Biotite} shows the frequency of recoils and the energy transferred by the primary ion to the initial atom of each cascade. 
Although relatively high energetic recoils can occur along the track, most of them are concentrated around the endpoint. 
The sub-cascades produced by the primary ion are of great interest for further studies, as their typical energy is in the range of a few keV.
\par
TRIM does not allow to simulate the effects of electronic energy losses directly. 
Therefore, one needs to compare the morphology of the track with stopping power tables to estimate the extent of damage in the target material. 
The possible extent of damage induced by an ion in a given material depends on the material's track-formation threshold.
In the case of $\SI{11.4}{MeV}$ Xe ions in Biotite Mica, electronic energy losses dominate up to a depth of $\SI{2}{\mu m}$, at that point the energy of a typical particle has dropped to approximately $\SI{2}{MeV}$. 
The remaining energy is mainly lost to the nuclear system. 
The particle travels about $\SI{1}{\mu m}$ in the nuclear-loss dominated regime.
\par
Since the track-formation threshold of Biotite Mica is unknown, it is difficult to estimate the extent of damage from amorphization.
From the study of 1000 Xe ions the mean range predicted by TRIM for these primary particles is about $\SI{3.22}{\mu m}$. 
However, the observable extent of damage could be much smaller if the track-formation threshold is high. 

\begin{table}[ht]
    \centering
    \caption{A comparison of simulated mean track-length values of alpha-particles of various energies in LiF to experimental data from~\cite{Bilski:2020}. The energy shown is the mean energy of the alphas emitted by the sources.}
    \label{tab:mean_vals_sim_exp}
    \begin{tabular}{cccc}
        \toprule
        Source & $\langle E \rangle$ / MeV & $\langle R \rangle$ Simulation / $\mu$m & $\langle R \rangle$ Experiment /$\mu$m \\
        \midrule
        Gd-148  & 3.183 & 10.07 & 9.97 \\
        Am-241  & 5.465 & 21.09 & 19.65 \\
        Cm-244  & 5.783 & 22.91 & 20.80 \\
        \bottomrule
    \end{tabular}
\end{table}

\subsection{Experimental studies} \label{sec:experimental studies}

With support from microscopy experts at KIT, we are performing studies aimed at establishing techniques for preparation of sub-samples of minerals (Muscovite Mica, Biotite Mica, Silicon) for their subsequent imaging with high-resolution microscopy.
The explored and utilized techniques range from mechanical manipulation to the use of a focused ion beam (FIB). 
In the scope of these efforts we are working closely with researchers from LEM-KIT on fabrication of lamellae for subsequent imaging with transmission electron microscopy.
The aforementioned studies have also resulted in an ongoing interdisciplinary master thesis project (IAP-LEM), whose goal is to correlate expectations obtained from simulations described in \autoref{sec:simulaitons} with microscopy imagery.
\par
\begin{figure}
   \centering
   \includegraphics[width=0.7\textwidth]{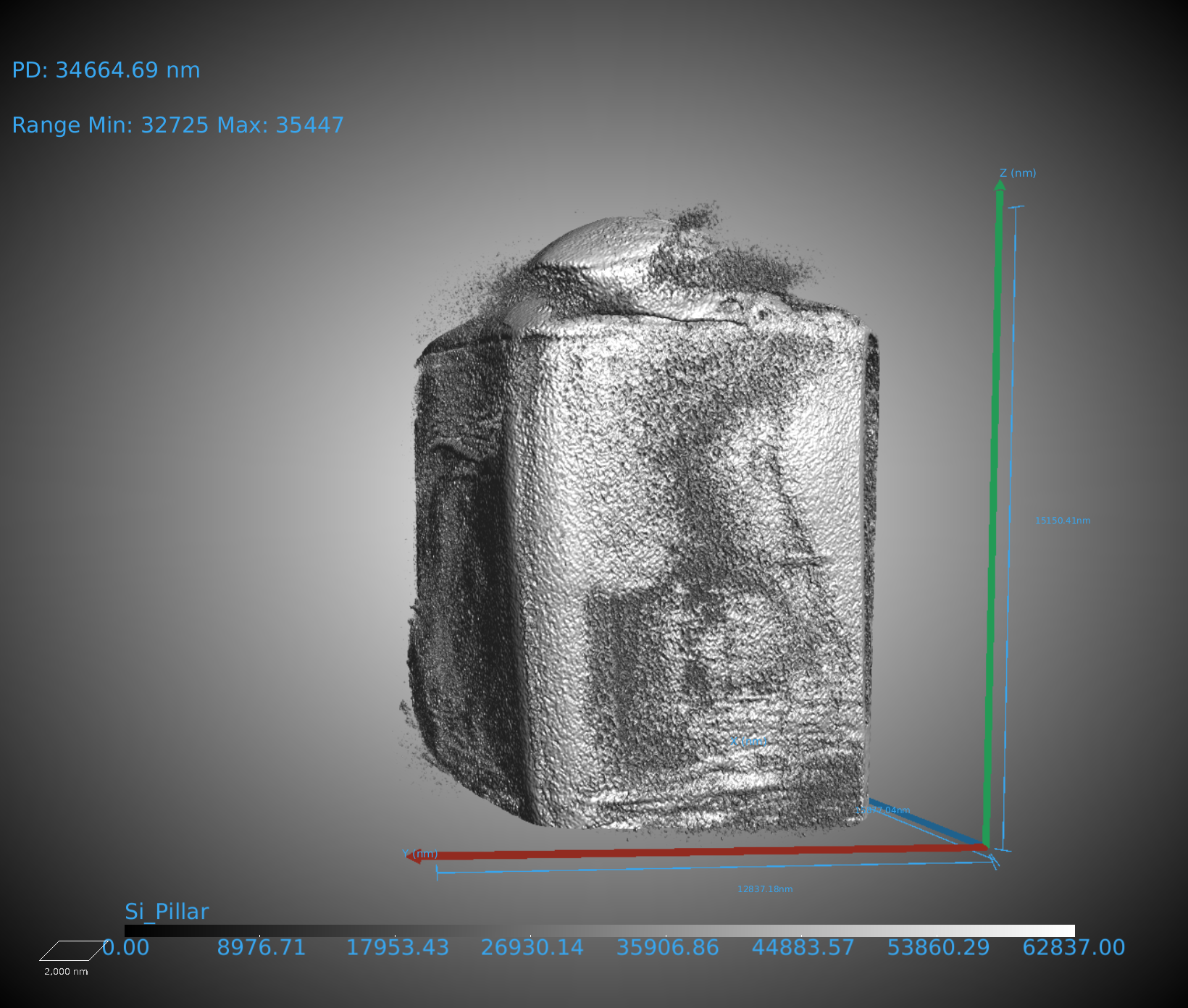}
   \caption{Reconstruction of a nanoCT scan of a Silicon pillar imaged by Dr. Rafaela Debastiani at the Institute of Nanotechnology at KIT. The 3D reconstruction was made with Dragonfly~\cite{dragonfly:2024}.}
   \label{fig:3DSiPillar}
\end{figure}

Additionally, with support from experts from INT-KIT, we have initiated the ``Dark Matter \& Neutrinos under the Microscope'' project, which strives to assess the feasibility of imaging particle-induced tracks in minerals with nanotomography (nanoCT).
As illustrated in Fig.~\ref{fig:3DSiPillar}, preliminary nanoCT imagery was already obtained for Muscovite Mica and Silicon samples.
These data are currently being used to develop image processing and analysis techniques, incorporating machine learning-based algorithms for identification of particle induced tracks.
To that end, we are exploring the use of commercially available software such as Dragonfly~\cite{dragonfly:2024}, and custom developed solutions based on packages like OpenCV.
Moreover, we have irradiated several mineral samples with a D-T neutron generator.
At present, we are also preparing for further irradiation of samples with a range of particles with energies spanning from keV to MeV, with the use of readily available neutron sources (AmBe), as well as with ion and alpha sources. 
\par
Ultimately, the ongoing studies at IAP-KIT seek to develop the simulation, microscopy and analysis techniques for imaging tracks produced by low energy nuclear recoils, bringing us one step closer to the realization of the concept of mineral detectors for neutrino and Dark Matter searches.

\acknowledgments

We thank Prof. Ulrich A. Glasmacher of the Institute of Earth Sciences,  University of Heidelberg for the continuous support for the project and for providing mineral test samples. 
We thank Dr. Christopher J. Kenney of the SLAC National Accelerator Laboratory for providing irradiated samples of mica and silicon for calibration and imaging studies.
We also thank Prof. Yolita Eggeler, Dr. Martin Peterlechner and Dr. Erich Müller of the Laboratory for Electron Microscopy at KIT, for the ongoing support of the project as well as for organizing and performing sub-sample preparation and electron microscopy imaging.
Additionally, we thank Dr. Torsten Scherer of the Institute of Nanotechnology at KIT and Dr. Rafaela Debastiani of the Institute of Nanotechnology at KIT and of the Helmholtz Institute Freiberg for Resource Technology at the Helmholtz-Zentrum Dresden-Rossendorf, for the useful discussions, and for preparation and imaging of samples with nanoCT.
This work was partly carried out with the support of the Karlsruhe Nano Micro Facility (KNMF, www.knmf.kit.edu), a Helmholtz Research Infrastructure at Karlsruhe Institute of Technology (KIT, www.kit.edu). 
The Xradia 810 Ultra (nanoCT) core facility was supported (in part) by the 3DMM2O - Cluster of Excellence (EXC-2082/1390761711).
The presented project is supported in part through the Helmholtz Initiative and Networking Fund (grant agreement no.~W2/W3-118). 
We also gratefully acknowledge the support by the KIT Center Elementary Particle and Astroparticle Physics (KCETA) for this project. 
\clearpage

\section{Advanced Microscopy Study of Tracks in Natural and Synthesized Quartz}\label{sec:KaiSun}

Authors: {\it Kai~Sun$^1$, Emilie~LaVoie-Ingram$^2$, and Joshua~Spitz$^2$}
\vspace{0.1cm} \\
$^1$Department~of~Materials~Science~and~Engineering, University~of~Michigan,~Ann~Arbor,~Michigan,~USA\\
$^2$Department~of~Physics, University~of~Michigan,~Ann~Arbor,~Michigan,~USA
\vspace{0.3cm}

This research focuses on the study of tracks in quartz, an abundant mineral in the earth. Both synthetic single crystal quartz and natural quartz samples from deep underground have been studied.

\begin{figure}[ht]
   \centering
   \includegraphics[width=0.9\textwidth]{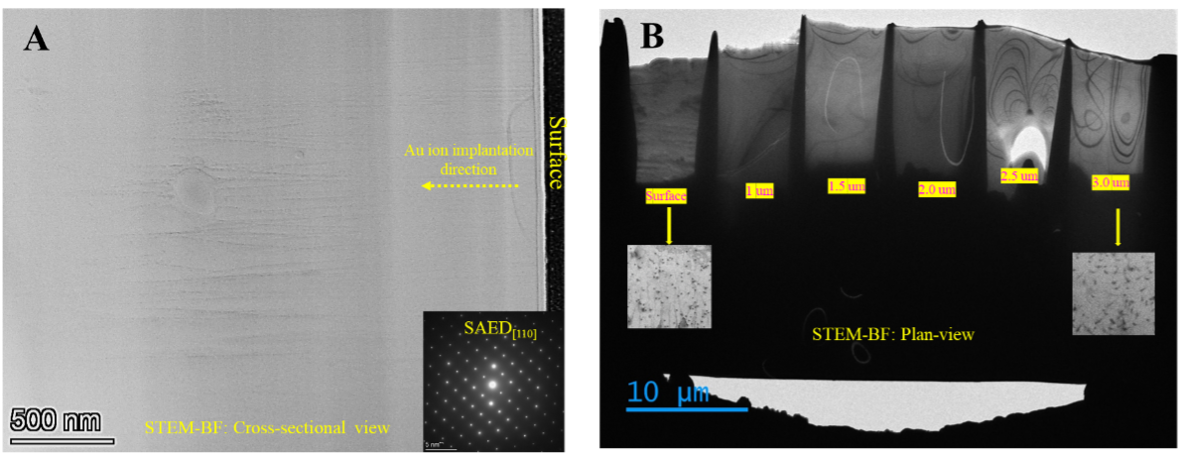}
   \caption{(A) cross-sectional view and (B) plan-view at different depths (with images showing single tracks at the surface and 3~$\mu$m depth as insets) of the Au ion tracks in Au ion irradiated synthetic single crystal quartz.}
   \label{fig:quartz_crosssection}
\end{figure}

\begin{figure}[ht]
   \centering
   \includegraphics[width=0.7\textwidth]{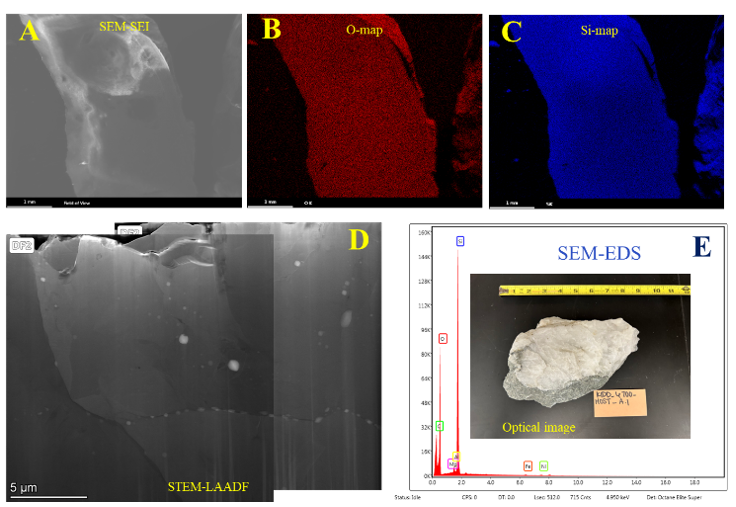}
   \caption{(A-C) SEM secondary electron image together with O- and Si-maps; (D) STEM low angle annular dark-field (LAADF) image and EDS spectrum with the optical image of the mineral piece as (E).}
   \label{fig:quartz_natural}
\end{figure}

Gold ions were introduced to the synthetic single crystal quartz by using a 15~MeV Au$^{5+}$ ion beam at current of 5~nA, $\sim$6.25$\times$10$^9$~Au$^{5+}$/s, for 15~seconds, 30~seconds and 45~seconds, respectively. Two different types of specimens for electron microscopy were prepared, plan-view along the ion beam direction (shown in Fig.~\ref{fig:quartz_crosssection}A) and cross-sectional view (Fig.~\ref{fig:quartz_crosssection}B). Natural quartz from the deep earth were also studied by SEM and TEM as shown in Fig.~\ref{fig:quartz_natural}.

\acknowledgments
The authors acknowledge the financial supports from the National Science Foundation (No: 2428508) and the Gordon and Betty Moore Foundation and the Michigan Center for Materials Characterization and Michigan Ion Beam Laboratory at University of Michigan for the advanced microscopy facilities and ion beam irradiation experiments. 

\clearpage

\section{Ultra-heavy Dark Matter Search with the Paleo Detector}\label{sec:Tatsuhiro_Naka}

Authors: {\it Tatsuhiro~Naka$^1$, Takenori~Kato$^2$, Yuki~Ido$^2$, Shota~Futamura$^2$, Naoki~Mizutani$^1$, and Yohei~Igami$^3$}
\vspace{0.1cm} \\
$^1$Toho~University\\
$^2$Nagoya~University\\
$^3$Kyoto~University
\vspace{0.3cm}

\subsection{Heavy dark matter search}
The mass parameter space for dark matter spans a very wide range. Ultra-heavy dark matter (UHDM), with masses around or exceeding the Planck scale, is also an intriguing target in astroparticle physics, motivated by baryogenesis/leptogenesis, grand unified theories (GUTs), supersymmetry (SUSY), and other beyond-standard-model frameworks.
Due to the extremely low flux of such UHDM on Earth, conventional artificial detectors lack sensitivity to it. In contrast, paleo-detectors owing to their geological timescale integration have the potential to probe UHDM candidates. 

Currently, we are investigating the search for Q-balls using muscovite mica, which is a representative candidate detector material for UHDM.
An interesting property of Q-balls is that they can carry an electric charge while remaining stable. As a result, it would behave similarly to an ultra-heavy atom by capturing electrons during the era of light element formation in the early universe~\cite{Kawasaki:gage_Q,Hong:2017qvx,Hong:2016ict}.
If such a quantum state were to pass through a paleo-detector, it would leave long, continuous tracks, making it identifiable through its highly penetrating nature.

\subsection{Target of the flux sensitivity for the Q-ball} 

\begin{wrapfigure}[16]{r}[-3mm]{65mm}
  \centering
 \raisebox{2mm}{\includegraphics[keepaspectratio,width=7cm]{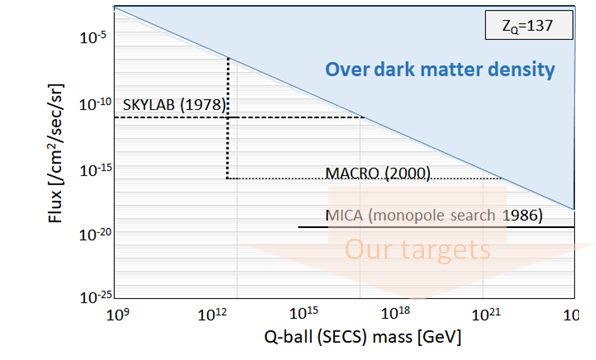}}
  \caption{Summary of limit for Q-ball flux. This search is targeting the flux lower than $10^{-16}$cm$^{-1}$sec$^{-1}$sr$^{-1}$, better than artificial detector sensitivity.}
\label{fig:sensitivity}
\end{wrapfigure}

\begin{wrapfigure}[15]{r}[2mm]{80mm}
  \centering
 \raisebox{2mm}{\includegraphics[keepaspectratio,width=4cm]{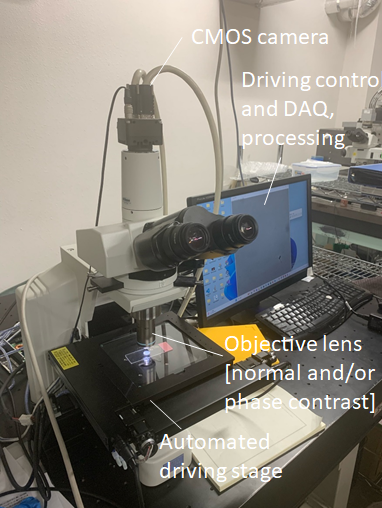}}
  \caption{Picture of constructing scanning system based on the optical microscope (QTS).}
\label{fig:QTS}
\end{wrapfigure}

Summary of flux sensitivities for the Q-ball was shown in Fig.~\ref{fig:sensitivity}. The highest sensitivity achieved so far (\textit{e.g.,} MACRO experiment~\cite{MACRO})  corresponds to a flux limit of about 
\(10^{-16} -  10^{-15}\,\mathrm{cm}^{-2}\,\mathrm{s}^{-1}\,\mathrm{sr}^{-1}\). For paleo-detectors, assuming an exposure time of 1\,Gyr, comparable sensitivity could be reached with a detection area of around \(1\,\mathrm{cm}^2\). In this context, muscovite mica is a suitable mineral due to its scalability in area. In mica-based searches, an upper limit on the Q-ball flux---converted from the magnetic monopole search by Price \textit{et al.}~\cite{physrevlett56}---was set at around \(10^{-20} -  10^{-19}\,\mathrm{cm}^{-2}\,\mathrm{s}^{-1}\,\mathrm{sr}^{-1}\). Achieving this level of sensitivity would require scanning an area of approximately \(500\,\mathrm{cm}^2\). Our goal is to extend this search to even larger areas using an automatic optical microscope scanning system. This new system, called the QTS (Fig.~\ref{fig:QTS}), is currently under development. Its designed scanning speed is expected to be \(50\,\mathrm{cm}^2/\mathrm{hour}\) using a 20\(\times\) objective lens. At this speed, scanning the area required to reach the MACRO limit would take approximately 10 minutes, and around 10 hours to reach \(10^{-16} -  10^{-15}\,\mathrm{cm}^{-2}\,\mathrm{s}^{-1}\,\mathrm{sr}^{-1}\).


\subsection{The performance for etching track in the muscovite mica}

The velocity of Q-balls is expected to be around ($\beta \sim 10^{-3}$). Heavy ions with similar energy loss characteristics, typically on the order of tens of keV, are used for calibration.

\begin{figure}[H]
   \centering
   \includegraphics[width=0.6\textwidth]{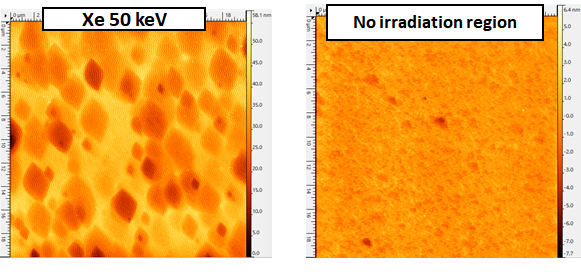}
   \caption{Surface image of the muscovite mica after chemical etching by AFM. (\textit{left}) Area of irradiated Xe 50~keV. (\textit{right}) Non-irradiation area.}
   \label{fig:Xe50keV_mica}
\end{figure}

In this study, we irradiated muscovite mica with 50~keV xenon (Xe) ions and applied chemical etching using hydrofluoric acid (HF). The feasibility for track formation were evaluated using an atomic force microscope (AFM). The AFM image for irradiated area of Xe and non-irradiated area were shown in Fig.~\ref{fig:Xe50keV_mica}.
As a result, the track formation efficiency was found to be nearly 100 $\%$. Further detailed evaluation is currently in progress. In addition, the etching velocity of the tracks was evaluated in both the lateral (width) and longitudinal (depth) directions. For such low-velocity heavy ions, the lateral etching rate was approximately 1$\mu\mathrm{m}/\mathrm{h}$, which corresponds to about 10--15\% of that for swift ions. The etching rate in the track direction was about 15 $\mathrm{nm}/\mathrm{h}$. While the typical track length of such low-velocity ions is around 50 $\mathrm{nm}$, in the case of Q-balls, the track length is expected to be significantly longer due to their ultra-heavy mass. Consequently, they exhibit reduced straggling and more uniform energy deposition. We are currently optimizing the etching conditions, taking into account a model of this unique behavior. 

\subsection{Ultra-heavy galactic cosmic-ray search with the olivine }
A time scale of 1~Gyr for paleo-detectors corresponds to several orbital cycles within the Milky Way Galaxy (MWG).
In nucleosynthesis, especially for the production of ultra-heavy elements such as transuranic nuclei, neutron star mergers (NSMs) play a crucial role through the rapid neutron capture process (r-process). However, the occurrence rate of NSMs in the MWG is estimated to be on the order of one event per 0.1 million years. Therefore, detailed investigation of such rare astrophysical phenomena is of great interest.
Olivine in meteorites has potential as a detector for extremely rare tracks induced by ultra-heavy elements produced in such events. Indeed, there are reports of tracks attributed to very high-$Z$ elements observed in meteorites~\cite{OLYMPIYA}. Taking advantage of its unique properties, our group has begun investigating track analysis in olivine to explore these possibilities further.

\subsection{Visible track ability by chemical etching in the olivine  }
\begin{figure}[H]
   \centering
   \includegraphics[width=0.65\textwidth]{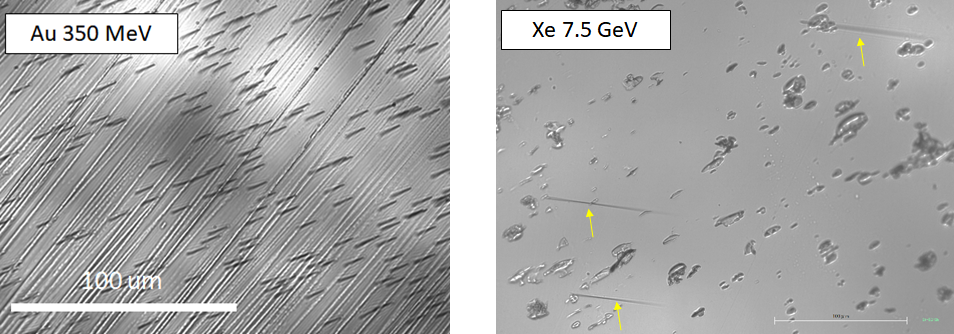}
   \caption{Optical microscope image of etched tracks in the olivine. (\textit{left}) Tracks of Au 350 MeV. Here, long diagonal lines are polishing marks, and ion tracks are observed as black lines with around 10 \textmu m. (\textit{right}) Tracks of Xe 7.5 GeV. The ion tracks are indicated by yellow allow.}
   \label{fig:Olivine_tracks}
\end{figure}

In this study, we confirmed the feasibility of an etching method based on EDTA solution. The performance of this method was evaluated using swift heavy ions. These ions were irradiated at the tandem accelerator at JAEA and at the HIMAC hadron therapy accelerator at QST, Japan.  In Fig.~\ref{fig:Olivine_tracks}, etched tracks for the Au ions of 350 MeV and Xe ions of 7.5 GeV were shown. 
Based on this evaluation, we confirmed that olivine forms tracks at an energy loss of approximately 15--20\,MeV/mg/cm\textsuperscript{2}, which is consistent with previous studies~\cite{OLYMPIYA}.
Through precise measurements of etching velocity, charge identification is expected to be achievable, as the etching rate in olivine depends on the ionization energy deposition. Based on this property, we aim to evaluate charge identification from tracks found in meteorites, and investigate new phenomena related to the history of the galaxy.

\acknowledgments
This work was supported by the Director's Leadership Grant, ISEE, Nagoya University and JSPS KAKENHI Grant Number JP25K01016.  And, experiments of the ion beam exposure were supported by the Research Project with Heavy Ions at QST-HIMAC and the Tandem Accelerator, JAEA.  

\clearpage

\section{Atmospheric Neutrino and Dark Matter
Detection with Paleo-Detectors at the
University of Michigan}\label{sec:U-M_LaVoie-Ingram}

Authors: {\it Emilie LaVoie-Ingram$^1$, Josh Spitz$^1$, Kai Sun$^2$, and Igor Jovanovic$^3$}
\vspace{0.1cm} \\
$^1$Department of Physics, University of Michigan, Ann Arbor, MI 48103 USA
\\$^2$Department of Materials Science \& Engineering, University of Michigan, Ann Arbor, MI 48103 USA
\\$^3$Department of Nuclear Engineering \& Radiological Sciences, University of Michigan, Ann Arbor, MI 48103 USA \\
\vspace{0.3cm}

\subsection{Ancient minerals as particle detectors}
Utilizing ancient minerals as paleo-detectors is an experimental technique that may revolutionize the field of direct particle detection~\cite{Baum:2019fqm, Baum:2021jak, Drukier:2018pdy, Jordan:2020gxx}. In the case of neutrinos or a theoretical dark matter particle known as a \textit{WIMP} (Weakly Interacting Massive Particle), the particle can interact with an atomic nucleus via the weak force and transfer enough energy to cause a recoil. The recoil then ionizes the lattice, creating a linear crystal defect structure constructed of a string of vacancies surrounded by a shell of interstitial atoms, which we call a \textit{nuclear recoil damage track}. These tracks are on the order of 3 nanometers in width, and up to a millimeter in length, depending on the projectile, interaction, and recoiling nucleus. In certain minerals like quartz (SiO$_2$), olivine ((Mg, Fe)$_2$ SiO$_4$), and halite (NaCl), these defects can be preserved for upwards of a billion years, providing an enormous amount of exposure to probe for these rare interactions. With the right characterization, background subtraction, and imaging technique, a 1 gram paleo-detector dated at 1 billion years old, could offer the same exposure as a 10 kiloton live direct detection experiment operating for 10 years~\cite{Edwards:2018hcf}. 

Atmospheric neutrinos are a probe into Earth's past, with their flux as a proxy for the cosmic ray rate and fluctuations in atmospheric composition and density across the last few billion years. New limits on dark matter, still one of the most intriguing mysteries of our universe, can be set with an especially old paleo-detectors and an appropriate imaging technique that would allow bulk readout of nanometer-scale tracks. Both goals require pushing the limits of microscopy and computation and rely on the careful extraction and analysis of geological samples. 

\subsection{Characterizing a natural sample}
The characterization of a natural sample for paleo-detection needs to be thoughtfully designed and carefully executed in order to properly model the background signals present in the mineral. The background and signal events induced in a target mineral are highly dependent on both the elemental composition and exposure time. When a particle interacts with an atomic nucleus and causes it to recoil, we do not detect the incoming particle, only the damage trail left by recoiling nucleus. The dimensions of this track (length and width) are dependent on the energy of the incoming particle, and the stopping power of the recoiling nucleus. Thus, a careful mapping of elemental types and composition is key in understanding the data. 

Exposure time refers to the period since a mineral last recrystallized. Geological samples are susceptible to various thermal and chemical alterations that erase crystal defects, such as annealing, hydrothermal alteration, and serpentinization. Accurate exposure times can be calculated with fission track or rare isotope dating techniques, usually done by geologists. If there is too low of a concentration of radioactive elements in the mineral itself, the surrounding host rock can be analyzed instead. In these cases, petrologists, thermochronologists, and/or geologists can infer the age by analyzing diffusion of certain elements or the rate of alteration across grain boundaries. 

\subsection{Estimating background with simulation}
At the University of Michigan, we are working to construct a simulation workflow to model backgrounds in ancient samples induced by cosmogenic sources. The bulk of this background, if samples are deep underground, is induced by fast neutrons resulting from cosmic ray muon spallation in and around our target. If the sample has been on the surface, whether after extraction or for a long geological time, we must also take into account the fast neutrons produced from cosmic ray interactions in Earth's atmosphere. The code for simulation of muon spallation is written in Geant4~\cite{collaboration2003geant4}, which is a toolkit designed to simulate the passage of particles through matter. CRY (the ``Cosmic-Ray Shower Generator'') is one example of how we can model cosmic ray, or more specifically muon flux, through some given area and location on Earth's surface, and we can then feed that flux, energy spectrum, and angular distribution into Geant4 for calculation of muon spallation into fast neutrons for sample surface exposure. If we want to model the flux induced at some point underground, we need to calculate survival matrices for the muons that will be propagating through that depth. PROPOSAL (PRopagator with Optimal Precision and Optimized Speed for All Leptons)~\cite{koehne_proposal_2013} can be used for this process, specifically through MUTE (MUon inTensity codE)~\cite{Woodley:2024eln}, which is a Python-based program made to predict muon spectra in deep underground and underwater labs. Spectra from CRY and PROPOSAL are fed into our Geant4 simulation, and the simulation will output the number of induced recoils and the corresponding energies per element in our target mineral. The calculation of physical damage can be done with SRIM and TRIM (Stopping and Range of Ions in Matter, and Transport of Ions in Matter)~\cite{ZieglerSRIM2010}. The integration of these four computational techniques is currently under progress, and our GitHub repository for the workflow can be found with the following link: \hyperlink{https://github.com/NSF-GCR-MDDM/paleo-bg-sim.git}{https://github.com/NSF-GCR-MDDM/paleo-bg-sim.git}. 

\subsection{Track imaging techniques}
\begin{figure}
   \centering
   \includegraphics[width=0.7\textwidth]{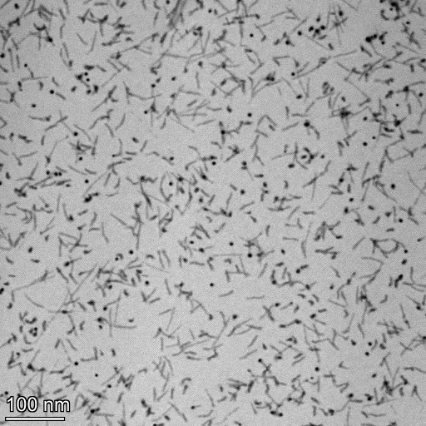}
   \caption{15 MeV Au irradiated olivine, imaged by Dr. Kai Sun at the University of Michigan, Michigan Center for Materials Characterization, with STEM BF.}
   \label{fig:TEMtracks}
\end{figure}

In addition to simulation and characterization, perhaps the most important part of this research is imaging the natural tracks. At the University of Michigan, we can easily resolve irradiation-induced ion tracks in olivine with high-resolution techniques such as transmission electron microscopy (TEM) (see Fig.~\ref{fig:TEMtracks}). However, in a matter of hours we can only image on the order of nanograms with TEM, resulting in very little throughput from this technique. Ultimately we want to image at least a tenth of a gram per sample, for competitive atmospheric neutrinos measurements in particular; thus, we need to look toward techniques that offer similar resolution but increased volumetric readout. Transmission x-ray microscopy and tomography is a technique we are determining the feasibility of for nuclear recoil damage track detection. We have conducted an initial proof of concept study using transmission x-ray microscopy at beamline 6-2c at the Stanford Linear Accelerator Center and could resolve FIB-milled tracks with radii of about 100 nanometers, but could not resolve ion tracks with radii of about 3 nanometers. We will soon have a new transmission x-ray microscope available at the University of Michigan with voxel resolution down to approximately 15 nanometers, giving us a much better chance at 3-D imaging both ion tracks and natural recoil tracks in a bulk sample on the order of milligrams to grams.

\subsection{Outlook}
At the University of Michigan, we are determining the feasibility of ancient minerals as atmospheric neutrino and dark matter detectors through investigation of how we can carefully characterize natural samples, use simulation to estimate backgrounds, and image samples with enough throughput and resolution. We anticipate the next phase of this investigation will be into various x-ray microscopy methods to allow for larger volumetric readout, but still nanometer-scale spatial resolution to resolve individual tracks. We will also investigate x-ray diffraction and strain imaging techniques that could offer higher resolution that full-field imaging techniques, like small and wide angle x-ray scattering, but at the sacrifice of indirect track measurement (i.e., not imaging a defect directly in real-space). The success of such diffraction-based methods is highly dependent on pristine crystallinity, which is quite difficult to obtain from geological samples. Nevertheless, we aim to exhaust all available imaging techniques to reach the resolution and throughput goals of the paleo-detector field.

\acknowledgments
ELI is grateful to the numerous undergraduate students and collaborators who have helped to evolve this project from a plan to experiments, as well as her advisor, Dr. Josh Spitz, for invaluable guidance on many matters of particle physics, experimental set up, and project management. This research is supported by the National Science Foundation (Grant \#EAR-2050374) and the Gordon and Betty Moore Foundation (Grant \#12234). 

\clearpage

\section{Dark Matter Search Using Synthetic Diamond}\label{sec:UmemotoDiamond}

Authors: {\it Atsuhiro~Umemoto}
\vspace{0.1cm} \\
International Center for Quantum-field Measurement Systems for Studies of the Universe and Particles (QUP), High Energy Accelerator Research Organization (KEK)
\vspace{0.3cm}

\begin{figure}[h]
   \centering
   \includegraphics[width=0.97\textwidth]{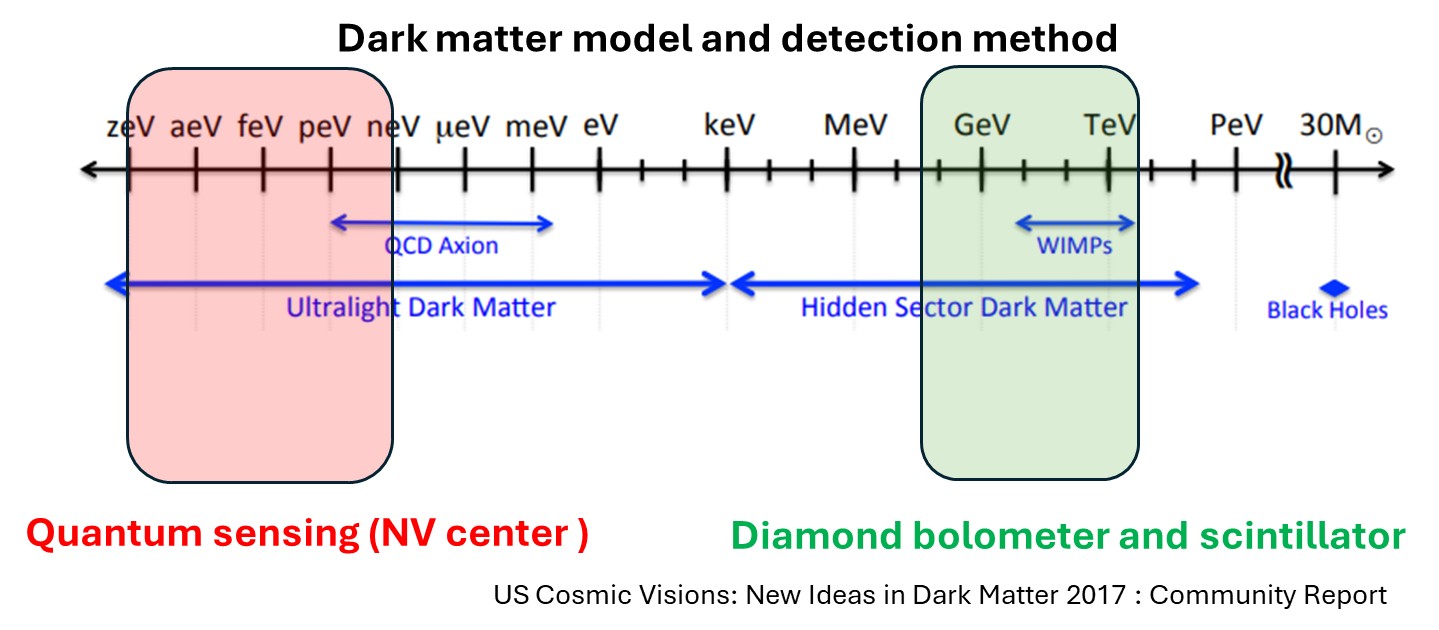}
   \caption{Dark matter candidate models and the corresponding detection methods using diamond for each model.  The figure is adapted, with minor modifications, from US Cosmic Visions: New Ideas in Dark Matter 2017 – Community Report~\cite{battaglieri2017cosmicvisionsnewideas}.}
   \label{fig:DMmodel}
\end{figure}

Diamond is a wide-bandgap ($E=5.5$ eV) semiconductor with a density of 3.52 g/cm$^{3}$, a lattice constant of 3.57 \AA, and many unique physical properties.
These properties offer advantages for dark matter searches using various detection methods.
Natural diamond, a mineral formed deep underground in the Earth over billions of years, could serve as a paleo-detector~\cite{Drukier:2018pdy}. Such diamonds contain color centers known as GR1 (general radiation) centers~\cite{GR1:2013}, originating from vacancies induced by radiation damage. Similar defects could potentially be generated as detectable signatures of dark matter interactions. 
In contrast, synthetic diamonds can be manufactured within a few days, with precise control over impurity concentrations. The impurity content strongly influences the physical properties of diamond, including its performance as a particle detector. In recent years, diamond has attracted increasing attention as a radiation detector and as a detector for GeV scale dark matter searches, with research focusing on its potential use as a diamond bolometer~\cite{DiamondBolo}, semiconductor detector~\cite{DiamondSemicon}, and scintillator~\cite{DiamondScint}.
In addition, quantum sensing techniques based on nitrogen-vacancy (NV) centers have been proposed for the detection of light bosonic dark matter~\cite{NVAxion}.
The search for dark matter using synthetic diamond is a rapidly developing field, offering promise not only for real-time event detection but also for the capability to verify dark matter models through multiple independent methods (Fig.~\ref{fig:DMmodel}).

\clearpage

\section{Supernova Neutrinos with Paleo Detectors}

Authors: {\it Shunsaku Horiuchi$^{1,2,3}$}
\vspace{0.1cm} \\
$^1$Department of Physics, Institute of Science Tokyo, Meguro-ku, Tokyo 152-8551, Japan\\
$^2$Center for Neutrino Physics, Virginia Tech, Blacksburg, VA 24061, USA\\ 
$^3$Kavli IPMU (WPI), UTIAS, The University of Tokyo, Kashiwa, Chiba 277-8583, Japan
\vspace{0.3cm}

Massive stars above $\sim 8 M_\odot$ end their evolutions spectacularly in the form of core-collapse supernovae---optical explosions so luminous they can outshine their host galaxies. However, the optical explosion is only the tip of the iceberg. Powered by the gravitational collapse of the star's iron core, the neutrino luminosity outshines the optical display by a factor of $\sim$billion. In fact, the neutrinos are predicted by numerical simulations to be crucial energy transporters responsible for energizing the optical explosion. While the copious release of neutrinos from a core-collapse supernova was confirmed by the historical supernova SN1987A~\cite{Kamiokande-II:1987idp,Bionta:1987qt}, many questions remain unanswered and future neutrinos detections are needed to test the mechanism of supernova explosions and more~\cite{Horiuchi:2018ofe}. One promising method for collecting more supernova neutrinos is to go further in distance: namely, to detect the flux of neutrinos from all core-collapse supernovae in the Universe. This is the search for the diffuse supernova neutrino background~\cite{Beacom:2010kk,Lunardini:2010ab,Ando:2023fcc}. Recently, the Super-Kamiokande has reported a $\sim 2.3 \sigma$ excess consistent with this signal~\cite{harada_2024_12726429}. Another promising method is to go further in time: namely, to detect the flux of neutrinos from all nearby core-collapse supernovae in the past. Operating over geological time scales, paleo detectors offer a unique opportunity to study such supernova neutrinos over the history of the Milky Way. 

Supernova neutrinos in paleo detectors have been explored in Refs.~\cite{Baum:2019fqm,Baum:2022wfc}. Supernova neutrinos leave damage tracks in paleo detectors over a broad range of length scales but peaking at shorter lengths. Backgrounds can be divided into two main types: neutrinos from sources other than supernovae, and non-neutrino backgrounds. For the former, solar and atmospheric neutrinos are the most relevant. Generally, supernova neutrinos dominate at some intermediate energy, with solar and atmospheric dominating at lower and higher energies~\cite{Jordan:2020gxx,Tapia-Arellano:2021cml}. Thus, a window exists for supernova neutrinos at around $\sim 100 $ nm (depending on the mineral composition). The most relevant non-neutrino background arises from neutrons. Even with the purest samples, e.g., $^{238}$U concentration of $10^{-11}$ g/g, the neutron background dominates over supernova neutrino tracks. However, the difference in their expected track length distributions can be used to extract the supernova neutrino signal. For example, it has been forecast that supernova neutrino tracks can be detected in 100 g of epsomite with $^{238}$U concentration of $10^{-11}$ g/g collecting tracks over 1 Gyr.  

Much new insight can be expected from the identification of supernova neutrino tracks in paleo detectors. For example, information regarding the past core-collapse supernova rate of the Milky Way can be obtained. While the past rate is expected to have been higher, it is a quantity of considerable uncertainty, and radiopure paleo detectors can be used to refute the constant rate scenario~\cite{Baum:2019fqm}. As another example, crucial information regarding the flavor content of supernova neutrinos can be obtained. The detection sensitivity of supernova neutrinos is strongly biased towards electron antineutrinos~\cite{Scholberg:2012id}. However, supernova neutrinos are expected to undergo complex neutrino flavor mixing between their emission and detection. Therefore, it is crucial for future supernova neutrino detections to cover all neutrino flavors. In particular, the offerings for heavy lepton type neutrinos ($\nu_\mu$, $\nu_\tau$, $\bar{\nu}_\mu$, $\bar{\nu}_\tau$) are severely limited~\cite{Suliga:2021hek}. Paleo detectors are flavor blind detectors, meaning they provide ample statistics of tracks left by heavy lepton type neutrinos which can be extracted by joint analyses with current and future detectors with mostly electron-type neutrino sensitivity~\cite{Baum:2022wfc}.

\acknowledgments
The work of SH is supported by NSF Grant No.~PHY-2209420 and JSPS KAKENHI Grant Number JP22K03630 and JP23H04899. This work was supported by World Premier International Research Center Initiative (WPI Initiative), MEXT, Japan.
\clearpage

\section{Nuclear Recoil Detection with Color Centers in Bulk Lithium Fluoride}\label{sec:PALEOCCENE}

Authors: {\it
Gabriela~R.~Araujo$^1$, Laura~Baudis$^1$, Nathaniel~Bowden$^2$, Jordan~Chapman$^3$, Anna~Erickson$^4$, Mariano~Guerrero~Perez$^5$, Adam~A.~Hecht$^6$, Samuel~C.~Hedges$^7$, Patrick~Huber$^7$, Vsevolod~Ivanov$^{3,5,8}$, Igor~Jovanovic$^9$, Giti~A.~Khodaparast$^5$, Brenden~A.~Magill$^5$, Jose~Maria~Mateos$^{10}$, Maverick~Morrison$^5$, Nicholas~ W. G.~Smith$^5$, Patrick~Stengel$^{11}$, Stuti~Surani$^{12}$, Nikita~Vladimirov$^{13,14}$, Keegan~Walkup$^7$, Christian~Wittweg$^1$, and Xianyi~Zhang$^2$
}
\vspace{0.1cm}\\
$^1$Physik-Institut, University of Zurich, Switzerland\\
$^2$Lawrence Livermore National Laboratory, Livermore, CA USA\\
$^3$Virginia Tech National Security Institute, Virginia Tech, Blacksburg VA USA\\
$^4$George W. Woodruff School of Mechanical Engineering
Georgia Institute of Technology, Atlanta, GA USA\\
$^5$Physics Department, Virginia Tech, Blacksburg VA USA\\
$^6$University of New Mexico, Albuquerque, NM USA\\
$^7$Center for Neutrino Physics, Virginia Tech, Blacksburg, VA USA\\
$^8$Virginia Tech Center for Quantum Information Science and Engineering, Blacksburg, VA USA\\
$^9$Department of Nuclear Engineering and Radiological Sciences,
University of Michigan, Ann Arbor, MI USA\\
$^{10}$Center for Microscopy and Image Analysis, University of Zurich, Switzerland\\
$^{11}$ Jo\v{z}ef Stefan Institute, Ljubljana, Slovenia\\
$^{12}$Pennsylvania State University, University Park, PA USA\\
$^{13}$URPP Adaptive Brain Circuits in Development and
Learning (AdaBD), University of Zurich, Switzerland\\
$^{14}$Center for Microscopy and Image Analysis (ZMB), University of Zurich, Zurich, Switzerland

\vspace{0.3cm}

\subsection{The PALEOCCENE concept}

Passive, persistent radiation detectors have existed since the birth
of this field, after all Becquerel discovered radioactivity using a
photographic emulsion. In the PALEOCCENE
approach~\cite{Cogswell:2021qlq} we look for persistent changes to the
crystal structure of the detector medium. Specifically, neutral
particles cause nuclear recoils and this in turn can lead to the
formation of vacancies. In many insulators vacancies form optically
active color centers and we interrogate these color centers by
fluorescence microscopy methods. Wide-field and confocal fluorescence
microscopy either lack genuine three dimensional information or have a
minuscule volume throughput. Selective plane illumination microscopy
(SPIM aka light-sheet microscopy) is a mature technology developed for
applications in biology and routinely allows for imaging large
volumes, tens of cubic centimeters, at speed and micrometer resolution.
The PALEOCCENE collaboration tries to adapt SPIM technology to the
single event detection of neutral particles via formation of color
centers in suitable materials. This opens the possibility of dark
matter, neutrino and neutron detection with passive detectors,
potentially even over geological times.

\subsection{First results from lithium fluoride}

Lithium fluoride (LiF) is a promising material with a number of
studies showing color center formation and fluorescence in response to
a wide variety of radiation types
~\cite{Bilski:2017,Bilski:2018,Bilksi:2019a,Bilski:2019b,Bilski:2024ghu}.
For our study~\cite{araujo2025nuclear} we used irradiations with thermal and fast neutrons as
well as gamma rays.  Using bulk spectroscopy we were able to show that
LiF is 50 times less sensitive to gamma rays than to neutrons and that
radiation induced fluorescence stays stable over many months.

\begin{figure}
\centering
\includegraphics[height=.36\textwidth]{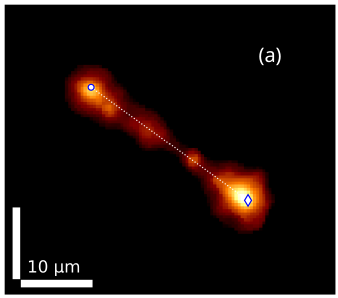}\hspace*{0.1cm}\includegraphics[height=0.35\textwidth,
  viewport = 0.00 0.00 260.00 182.0, clip =
  true]{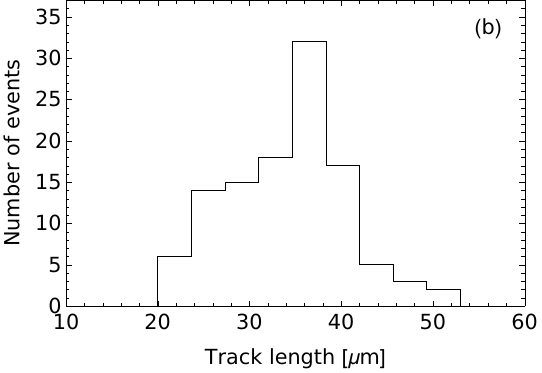}
\caption{Two dimensional projection of a thermal-neutron induced track
  candidate (a) in sample 0505 together with the axis defining the
  track length. Panel (b) shows a histogram of the three-dimensional
  track length measured for a selection of events from several scans
  of sample 0505 that meet certain quality criteria. Figure and
  caption adapted from Ref.~\cite{araujo2025nuclear}.}
\label{fig:thermal}
 \end{figure}

We used the open source mesoSPIM platform~\cite{Vladimirov:2024} to
image several samples and were able to see tracks very clearly. The
track number and morphology agreed with expectation from simulation
using SRIM. In Fig.~\ref{fig:thermal}a we show the result of a thermal
neutron interaction with $^6$Li which occurs with about 7.5\% natural
abundance. In this reaction a thermal neutron is captured and produces
an alpha particle and a triton of well defined energy: $n+
{}^6\mathrm{Li}\rightarrow \alpha + {}^3\mathrm{H}$. The alpha and
triton will move in directly opposite directions due to momentum
conservation. The shorter and brighter track corresponds to the alpha
particle whereas the longer and fainter track corresponds to the
triton. In Fig.~\ref{fig:thermal}b we show the track length
distribution for a selected group of events; the expectation from
simulation is 30--35\,\textmu m.

In a similar fashion we were able to image fast neutron recoils and
cosmic rays. We did encounter some limitations of the instrument in
particular camera noise and the resulting long exposure times resulted
in significant bleaching. We are currently building a mesoSPIM at
Virginia Tech equipped with a very-low noise camera. This will enable us
to make full use of the capabilities of the mesoSPIM and allow for
high-speed scans. Eventually, these improvements will allow us to use
$dE/dx$-information for particle identification.

A considerable effort is underway to create a custom data processing
pipeline including machine learning algorithms for track and event
identification. Combined with hardware improvements this will bring
scan times down to hours per cubic centimeter while achieving single
event sensitivity. This makes for highly efficient neutron detectors
and should pave the way for both direct dark matter detection and
reactor neutrino measurements~\cite{Cogswell:2021qlq}.

Triton tracks contain a very low density of color centers and the
fact that we can clearly see them shows that we already are sensitive
to single digit numbers of color centers per voxel. With the planned
hardware improvements we will be able to detect single color
centers. We are currently investigating underground and reactor
deployments.

A major task towards paleo-detection of dark matter is to find
materials that are naturally abundant as minerals for which the
PALEOCCENE technology is applicable.

\acknowledgments

This work was supported by the U.S. Department of Energy National
Nuclear Security Administration Office of Defense Nuclear
Nonproliferation R\&D through the Consortium for Monitoring,
Technology, and Verification under award number DE-NA0003920 and by
the National Science Foundation Growing Convergence Research award
2428507. The work of Gabriela R. Araujo was supported by the UZH
Postdoc Grant No.K-72312-14-01. The mesoSPIM project and imaging
platform were supported by the University Research Priority Program
(URPP) “Adaptive Brain Circuits in Development and Learning (AdaBD)”
of the University of Zurich. This work was supported by the Virginia
Tech College of Science Lay Nam Chang Dean's Discovery Fund. This work
was prepared by LLNL under Contract DE-AC52-07NA27344 and supported by
the LLNL-LDRD Program under Project
No. 23-FS-028. LLNL-JRNL-872914. This work is supported by the
European Union's Horizon Europe research and innovation programme
under the Marie Skłodowska-Curie Postdoctoral Fellowship Programme,
SMASH co-funded under the grant agreement No. 101081355. The SMASH
project is co-funded by the Republic of Slovenia and the European
Union from the European Regional Development Fund.

\clearpage

\section{Artificial Formation of alpha Recoil Tracks Using an Americium Source}
\label{sec:alpha_recoil_tracks}

Authors: {\it Taiki~Nakashima$^1$, Yuto~Iinuma$^2$, Koichi~Takamiya$^2$, Akihiko~Yokoyama$^1$, Norihiro~Yamada$^1$, and Noriko~Hasebe$^1$}
\vspace{0.1cm} \\
$^1$Kanazawa~University, Kanazawa, Ishikawa 920-1192, Japan \\
$^2$Integrated~Radiation~and~Nuclear~Science, Kyoto~University, Kumatori, Osaka 590-0494 Japan
\vspace{0.3cm}

\subsection{Introduction}

Uranium-238 decays to lead-206 through multiple alpha particle emissions. Uranium-238 also decays through the spontaneous nuclear fission. These decays will leave alpha recoil or fission tracks in a mineral in which uranium-238 is embedded. The observation of fission tracks in zircon using an atomic force microscope revealed the unknown waveforms~\cite{2012ohishi}. These waveforms were found to have a wavelength that did not correspond to the size of the scratches formed during the polishing of mineral surfaces or crystal structures. To seek for the possibility of these waveforms representing the surface roughness resulting from the alpha recoil damage to the crystal, the methodology to artificially form the alpha recoil track (ART) is developed~\cite{2024nakashima}.

\subsection{Experimental method and results}

A 300 Bq americium-241 source was used to irradiate muscovite, in which the ARTs were well observed in previous studies, for various time intervals after annealing treatment to erase all naturally occurring ARTs, and the samples were observed using phase contrast microscopy after chemical etching by 47\% HF at 32°C for 2 hours. The ART areal density formed on the sample surface showed a linear relationship against irradiation time, indicating the feasibility of using Am source to artificially form ARTs on a mineral surface. However, the size distribution of the artificial ARTs were larger than that the naturally observed ARTs.

\subsection{Annealing experiments and discussion}

The annealing experiment on the artificially formed ARTs showed that the size distribution of natural ARTs could be indicative of annealing at ambient temperature over a geologically long period time, or ARTs could have been annealed in the recent past at a slightly higher temperature (150--200°C) given their uniform size distribution. We cannot rule out the possibility that the observed natural ART-like tracks were formed by the movement of smaller nuclei (e.g., movement of major mineral-forming-atoms by cosmic ray irradiation), assuming that the track size reflects the energy and mass of moved nuclei. When the natural ARTs were observed by the atomic force microscope after a weaker chemical etching, we found a larger number of smaller tracks ($10^3$ times more than tracks found in the image by phase contrast microscopy). Further detailed studies of annealing behavior and track formation processes by the movement of various atoms are required to reach a definitive understanding.

\acknowledgments
The Am-241 irradiation was supported under the joint research program of the Institute for Integrated Radiation and Nuclear Science, Kyoto University.

\clearpage

\section{Multimessenger Searches for Heavy and Boosted Dark Matter}

Authors: {\it Kohta~Murase$^{1,2}$}
\vspace{0.1cm} \\
$^1$The Pennsylvania State University\\
$^2$Yukawa Institute for Theoretical Physics, Kyoto University
\vspace{0.3cm}

\subsection{Introduction}
A decade of multimessenger observations has transformed how we approach dark matter (DM) with masses beyond the weak scale. 
Recent highlights include: detection of the ``Amaterasu'' ultrahigh-energy event with an energy of $2.44\times10^{20}$~eV; evidence for high-energy neutrinos from active galactic nuclei (AGN) such as NGC 1068; measurements of diffuse sub-PeV gamma rays and neutrinos from our Galaxy; and KM3NeT's detection of a high-energy neutrino event with an energy of $220$~PeV.  
These results have been used for novel searches for heavy DM using gamma rays and neutrinos across Galactic and extragalactic targets, and they open avenues to probe boosted DM produced by beyond Standard Model (BSM) interactions of cosmic rays with ambient DM.

\subsection{Searches for superheavy DM}
Diffuse high-energy gamma-ray and neutrino data now provide comparable and complementary limits on the DM lifetime and cross section from TeV energies to the GUT and even  Planck scales. Interpreted jointly, they nearly exclude decaying-DM explanations for the diffuse IceCube signal~\cite{Murase:2015gea,Cohen:2016uyg}, and the KM3NeT event, KM3-230213A, has been discussed in light of superheavy DM~\cite{Murase:2025uwv}. Ultrahigh-energy photon and cosmic-ray observations have pushed the limits on DM lifetimes to $\tau_{\rm DM} \gtrsim 10^{29}$--$10^{30}$~s, depending on the final states~\cite{Das:2023wtk}. With ultrahigh-energy neutrino experiments including IceCube-Gen2 and lunar radio detectors, the coming decade should further probe heavy DM beyond the EeV scale~\cite{Das:2024bed}.
 
Searches for nearby dark matter halos, whose constraints are complementary to the diffuse flux limits, are also useful. Dwarf spheroidal galaxies remain gold-standard targets for DM annihilation searches thanks to low astrophysical backgrounds. Recent analyses of 14~yr Fermi-LAT data tighten constraints for heavy DM in the $10^3$--$10^{11}$~GeV range, when electromagnetic cascade emission has been taken into account~\cite{Song:2024vdc}. Galaxy clusters also provide complementary handles. For Virgo and other nearby massive clusters, both gamma-ray and neutrino observations lead to competitive limits on the lifetimes for decaying DM at PeV scales~\cite{Song:2023xdk}. 

Noble-liquid experiments probe the DM-nucleon scattering cross section, which has explored relevant parameter space of weakly interacting massive particles (WIMPs). However, in the large cross-section regime, the conventional scaling for the coherent spin-independent interaction breaks down, and the cross section would be saturated. Recent LZ analyses extend sensitivities beyond extrapolations of standard WIMP searches by leveraging multiple scattering topologies. No candidates are seen so far, yielding intriguing and competitive limits on the DM-nucleon cross section up to masses of $\sim 10^{17}$--$10^{18}$~GeV~\cite{LZCollaboration2024_MIMP}.

\subsection{Searches for boosted DM}
DM-nucleon interactions can be explored by searches for the flux of boosted DM in direct detection experiments and neutrino detectors such as Super-K. The boosted-DM limits exploit the fact that cosmic rays can upscatter ambient DM, and the yield depends primarily on the DM column that the cosmic rays traverse as well as the astrophysical cosmic-ray flux. The central black hole in a galaxy may lead to a spike in the DM distribution, although stellar heating or DM self-annihilation can smoothen it. Comparisons across TXS 0506+056, NGC 1068, and the Milky Way illustrate order-of-magnitude swings in the expected signal. The resulting boosted DM limits are also sensitive to the flux of low-energy cosmic-rays, which is highly uncertain for extragalactic sources such as AGN.

Cosmic-ray cooling limits obtained through high-energy neutrino observations provide a complementary handle. For neutrino-bright AGN such as NGC 1068, the observed IceCube flux effectively constrains the required cosmic-ray power at the source; any additional energy loss to DM-nucleon scatterings would prevent efficient neutrino production~\cite{Herrera:2023nww}. Demanding no over-cooling of cosmic rays therefore yields robust, source-anchored upper bounds on the DM-nucleon cross section, which are rather insensitive to uncertain low-energy cosmic-ray extrapolations. Not only elastic but also inelastic DM can be explored by boosted DM searches and high-energy neutrino observations~\cite{Gustafson:2024aom}. 

\subsection{Searches with paleo-detectors}
By integrating over Gyr timescales, paleo-detectors can be powerful for probing not only WIMPs but also superheavy DM~\cite{Acevedo:2021tbl}. 
In muscovite mica, etched track morphologies and depth spectra can constrain DM-nucleon scattering even at heavy masses and/or large cross sections, and the projected sensitivity of ``DMICA''~\cite{hirose2024} may be comparable or complementary to those of noble-liquid detectors, especially for composite DM (e.g., Q-balls) and/or strongly-interacting DM.
Searching for boosted DM with paleo-detectors is discussed. Because the star-formation and Galactic center activity were likely higher in the past, future samples of different ages could encode time variability and directional anisotropy.

\acknowledgments
I am grateful to the organizers of the “Mineral Detection of Neutrinos and Dark Matter 2025” workshop. The work of K.M. is supported by the NSF grants No.~AST-2108466, AST-2108467, AST-2308021, and KAKENHI No.~20H05852. 


\clearpage

\section{Rock Samples from Ocean Drilling --- Introduction of Drilling Vessel Chikyu}

Authors: {\it Noriaki~Sakurai}
\vspace{0.1cm} \\
Drilling Management Group, Chikyu Operations Department, Institute~for~Marine-Earth~Exploration~and~Engineering~(MarE3),~Japan Agency for Marine-Earth Science and Technology (JAMSTEC)
\vspace{0.3cm}


Chikyu, JAMSTEC’s deep-sea drilling vessel, is a 5th-generation drill ship built in 2005~\cite{sd-1-32-2005, Taira2014}. 
In the context of drill ship technology, the latest designs are classified as 7th generation. Although Chikyu may be considered old by its generational classification, its 20-year age does not affect its ability to safely operate in 5--6 knot currents, a state-of-the-art feat achieved by only a handful of highly skilled engineers.

\begin{figure}[ht]
   \centering
   \includegraphics[width=0.45\textwidth]{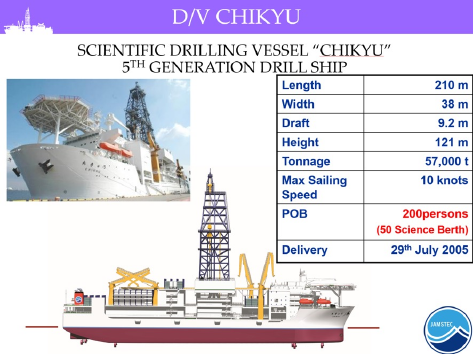}
   \includegraphics[width=0.45\textwidth]{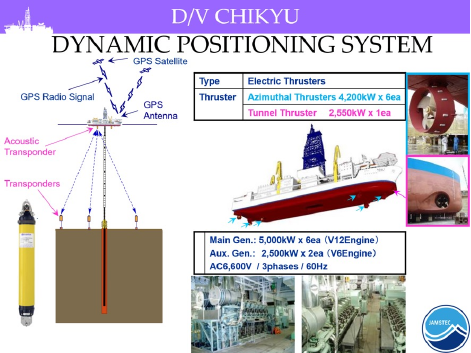}
   \caption{Left: Basic specifications of Chikyu. Right: Dynamic positioning system of Chikyu.}
   \label{fig:chikyu_overview}
\end{figure}

Chikyu measures 210 meters in length and 38 meters in width. Despite its size, the cruising speed is up to 10 knots (Fig.~\ref{fig:chikyu_overview} left). 

The Dynamic Positioning System (DPS) enables the vessel to maintain its stationary position over a target deep underwater (Fig.~\ref{fig:chikyu_overview} right).

As for the coring system used on board, Chikyu predominantly utilizes the Hydraulic Piston Coring System (HPCS), the Extended Shoe Coring System (ESCS), the Rotary Core Barrel (RCB), and the Small Diameter Rotary Core Barrel (SD-RCB) systems.

\begin{figure}[ht]
   \centering
   \includegraphics[width=0.45\textwidth]{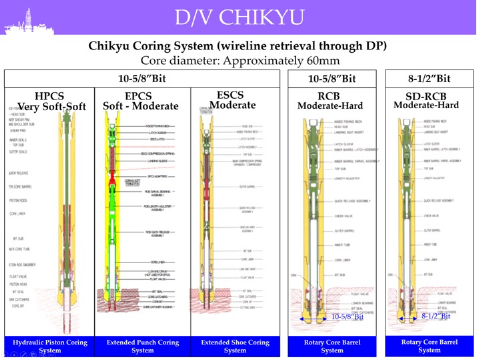}
   \includegraphics[width=0.45\textwidth]{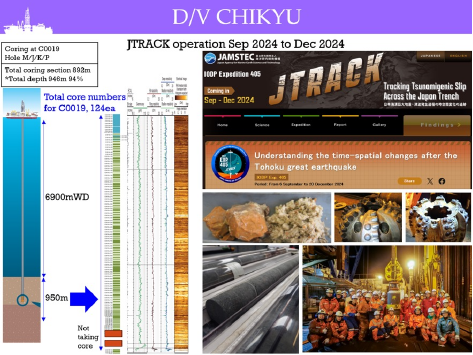}
   \caption{Left: Coring systems. Right: Latest campaign for scientific drilling.}
   \label{fig:coring_campaign}
\end{figure}

As shown in Fig.~\ref{fig:coring_campaign} (left), the HPCS is used for very soft formations down to around 100~meters below the seafloor. The ESCS is then employed for more competent and moderate rock formations. The RCB and SD-RCB systems find a use when moderately hard to hard sedimentary, igneous, and metamorphic formations are encountered.

Although the RCB and SD-RCB systems differ in the outer bit size (10-5/8 in for RCB and 8-1/2 in for SD-RCB), the actual cut core size remains the same, i.e., 59 millimeters. Some of the benefits of running the SD-RCB system rather than RCB, apart from a reduced bit size, include improved core cutting efficiency, better cut core quality, and overall better hole
stability and less vibration downhole.

Figure~\ref{fig:coring_campaign} (right) shows the latest scientific drilling campaign, i.e., the Exp 405 J-TRACK~\cite{Kodaira2023,Kodaira2024_Addendum}. 
Multiple holes were drilled at site C0019. Some holes utilized a Logging-While-Drilling (LWD) system to define and delineate substrata in real-time, while others were cored to retrieve physical cores. Comparison and integration of both yield exquisite results and allow a better understanding of the downhole processes.

\acknowledgments
The author acknowledges the support from the Institute for Marine-Earth Exploration and Engineering (MarE3), JAMSTEC.

\clearpage

%
%
%
%

\bibliographystyle{JHEP.bst}
\bibliography{theBib}

\end{document}